\documentclass[a4paper,notitlepage,12pt]{article}

\usepackage{amsfonts}
\usepackage{amsmath}
\usepackage{amssymb}
\usepackage{graphicx}

\begin{document}

\title{\textbf{Quantized Detector Networks}\\
{\emph{\Large A review of recent developments}}}
\author{\textbf{George Jaroszkiewicz} \\
School of Mathematical Sciences, University of Nottingham, \\
University Park, Nottingham NG7 2RD, UK}
\maketitle

\begin{abstract}
QDN (quantized detector networks) is a description of quantum processes in
which the principal focus is on observers and their apparatus, rather than
on states of SUOs (systems under observation). It is a realization of
Heisenberg's original instrumentalist approach to quantum physics and can
deal with time dependent apparatus, multiple observers and inter-frame
physics. QDN is most naturally expressed in the mathematical language of
quantum computation, a language ideally suited to describe quantum
experiments as processes of information exchange between observers and their
apparatus. Examples in quantum optics are given, showing how the formalism
deals with quantum interference, non-locality and entanglement. Particle
decays, relativity and non-linearity in quantum mechanics are discussed.
\end{abstract}

\

\

\begin{quote}
`` ... \textit{and all of the apparent quantum properties of light and the
existence of photons may be nothing more than the result of matter
interacting with matter directly, and according to quantum mechanical laws.}%
''

\hfill Feynman's thesis \cite{FEYNMAN-THESIS}
\end{quote}

\

\smallskip

This review is divided into four parts. Part $I$ discusses the motivation
for the QDN approach to quantum mechanics and covers some relevant
historical points. Readers more interested in mathematical details can start
with Part $II$, which outlines the formalism. Part $III$ deals with various
applications to physics, particularly quantum optics. Part $IV$ is a
discussion of some aspects related to the QDN programme.

Throughout this review we use the following acronyms many times: QDN $\equiv
$ quantized detector networks, QM $\equiv $ quantum mechanics, SQM $\equiv $
standard quantum mechanics, CM $\equiv $ classical mechanics, SUO $\equiv $
system under observation, ESD $\equiv $ elementary signal detector (or in
the case of a source, elementary signal device).

\newpage

\begin{center}
\textbf{\Large PART I: Motivation}
\end{center}

\section{Introduction and historical perspective}

The landscape of QM is littered with the debris of various interpretations
devised to explain away its strange non-classical properties such as
wave-particle duality and quantum interference. There is no space to review
any of these attempts here. Usually, they failed because their authors tried
to view too much of physical reality in terms of familiar classical
concepts, such as particles and waves moving about in a background
spacetime. Applied to everyday (large-scale) processes, such a strategy
usually works well, as evinced by the success of Newtonian and relativistic
mechanics, but when it comes to QM, it leads to various traps waiting for
the unwary theorist. A notable example is Schr\"{o}dinger, who originally
thought of his quantum waves as smeared out electronic charge, in contrast
with the Born probabilistic interpretation \cite{BORN-1926,BORN-1954}
generally accepted today.

Classical thinking in QM persists to this day in one form or another,
ranging from attempts to split electrons \cite{MARIS-2000} to paradigms such
as the Multiverse \cite{DEUTSCH-1997} and decoherence \cite{ZUREK-2002},
which in their original formulations were based on the assertion that the
Schr\"{o}dinger equation alone suffices to explain all of physics. Such
schools of thought view the quantum wavefunction as a fundamental object in
its own right. In this article we will emphasize the point that even to talk
casually about an \textquotedblleft electron wavefunction\textquotedblright
, as is commonplace amongst quantum theorists, is to risk applying classical
thinking inappropriately to quantum physics. Our view is that a quantum
wavefunction is contextual, i.e., without reference to any observer and
their apparatus, such a wavefunction is a physically meaningless concept.

Our aim in QDN is to eliminate as far as possible concepts which are
inessential, potentially misleading, or simply metaphysical (i.e., incapable
of verification). Some mathematical concepts such as labstates and Hilbert
spaces are used heavily, but in such cases, the motivation for them stems
from a desire to avoid mental imagery as much as possible. The most
important guiding principle has been to ask what exactly do experimentalists
do when they perform quantum physics experiments, and then model the answer
to that question according to established quantum mechanical principles.

\section{The importance of observers and their apparatus}

Any reasonable interpretation of QM should involve at least a rudimentary
recognition and understanding of the roles of the observer and their
apparatus in a quantum experiment. To understand what QDN says on this,
consider the analogy with cinematography. As we watch a film, we generally
tend not to think about the camera or the film crew who made the film. If we
did, not only would that spoil the film, but we would be acknowledging the
fact that the film could not have been made without the camera. We would be
reminded too of the unfortunate fact that the presence of film crews at
actual events such as riots often influences those events, and indeed,
without the presence of the film crew, those events might never have taken
place at all. The best films are those in which the presence of the camera
gets overlooked in the mind of the viewer, so that they are deceived into
believing that what is seen on screen is independent of the process of
observation, and is somehow real on that account. This is a classical world
view of reality.

Problems arise when viewers start to interpret a film, which is just a
representation of reality, as being that reality itself. Likewise, problems
arise in SQM when physical reality is thought as a wave-function. QDN
regards QM as the correct set of rules governing information exchange
between apparatus and observer, rather than any commentary on the nature of
reality. If the only things observers can ever deal with are signals from
their apparatus, then nothing need be said beyond that, and so QDN does not
comment on the existence or otherwise of SUOs.

\section{Origins of QDN}

Although many of the core principles of QDN had been developed earlier \cite%
{J2001A,EAKINS-2004,J2005A}, the catalyst for the development of QDN in the
form described here was a reading of Harry Paul's book on quantum optics
\cite{PAUL-2004}. That book emphasizes the essential non-classical
properties of the photon concept, making it clear via numerous discussions
of actual (as opposed to imagined) laboratory experiments that photons
cannot be \textquotedblleft particles\textquotedblright\ in the sense of a
cricket ball or baseball. What impressed us most were Paul's accounts of
situations where a photon appears to originate from several correlated
sources. We found ourselves trying to understand those experiments in terms
of what the observers were doing with their apparatus, rather than in terms
of the conventional quantum optics formalism, which uses photon creation and
annihilation operators. We asked the question: if photons are not actually
\textquotedblleft there\textquotedblright\ as particles \emph{per se}, then
what precisely does it mean to represent them by particle-like creation and
annihilation operators? The result was \textquotedblleft quantum register
physics\textquotedblright\ \cite{J2004C,J2006A}, which evolved into QDN in
the form described here.

Perhaps the best way to understand the core principles of QDN is to see them
as equivalent to the principles which motivated Heisenberg's approach to QM
\cite{HEISENBERG-1925}. This is now known as matrix mechanics, because the
noncommuting variables Heisenberg introduced could be regarded as transition
matrices. This allowed Schr\"{o}dinger to demonstrate the mathematical
equivalence of his wave mechanical approach to QM to the algebraic approach
of Heisenberg \cite{SCHRODINGER-1926}. However, the underlying principles on
which Heisenberg based his work were radically different to those motivating
Schr\"{o}dinger. It seems quite wrong to equate the two formulations of
quantum mechanics simply because the mathematics of one can be transformed
into the mathematics of the other.

Certainly, Heisenberg himself did not like the visual imagery associated
with Schr\"{o}dinger's equation\footnote{%
Writing to Pauli in 1926, Heisenberg wrote: \textquotedblleft The more I
think about the physical portion of Schr\"{o}dinger's theory, the more
repulsive I find it. What Schr\"{o}dinger writes about the visualizability
of his theory `is probably not quite right'...\textquotedblright .}.
Scientific theories are not just sets of mathematical rules for making
predictions, but also require specific ways of thinking about physical
situations. Sometimes, these lead to paradigms which are unphysical
dead-ends, such as the use of epicycles to describe the motion of planets,
the phlogiston theory of heat and the aether concept in Maxwellian
electrodynamics. Heisenberg's crucial idea was to focus only on those
aspects of an experiment which are accessible to observation. This seems to
be a position sufficiently far from metaphysics as to merit the status of a
central principle in physics, and we have tried to respect it as much as
possible in our development of QDN.

Not many scientists appeared to follow Heisenberg footsteps, the majority
preferring Schr\"{o}dinger's wave mechanical approach. A notable exception
was Feynman, who attempted to avoid using the electromagnetic field in a
novel formulation of electrodynamics \cite{FEYNMAN-THESIS}. Despite his
recorded disdain for the philosophy of science\footnote{%
He wrote that the philosophy of science is a disease that afflicts middle
aged scientists.} and his success in developing practical calculational
methodologies in SQM, he was not entirely disinterested in quantum
philosophy (i.e., thinking about what QM means). It led him to think of
photons as something to do with apparatus, as the quote from his thesis,
given at the start of this review, expresses in a succinct way. We shall
return to his ideas presently.

To give some balance in this review of QDN, we point out some of the issues
it does not explain at this time. It does not \textquotedblleft explain
away\textquotedblright\ intrinsic quantum phenomena such as outcome
randomness and interference, but this is true also of SQM. An important
currently unresolved issue awaiting our attention is a QDN description of
SUOs conventionally described via continuous degrees of freedom. For
example, we have not yet developed a QDN approach to the calculation of
atomic energy levels, something which the Schr\"{o}dinger equation does with
relative ease. However, that was a problem with Heisenberg's matrix
mechanics also, and in this respect QDN is no different. It is possible QDN
will not turn out to be a good approach for those sorts of calculations. Our
experience with algebraic approaches to QM such as infinite component
wavefunctions leads us to the expectation that the issue will be resolved
satisfactorily in due course. The harmonic oscillator, for example, can be
described as well via a purely algebraic formulation as it can via a purely
wave mechanical approach, and this makes a QDN description of that system
relatively easy to develop \cite{J2005D,JRT-2007}.

On the positive side, apart from giving a novel perspective on physical
reality, QDN gives a useful computational methodology readily applicable to
certain branches of quantum optics. In particular, the formalism is closely
allied to quantum computation, which gives it a modern flavour. It should be
possible to encode QDN in computer algebra packages, thereby opening the
door to the efficient calculation of quantum amplitudes for the outcomes of
quantum optics experiments of arbitrary complexity.

\section{Why CM appears to work}

Before $1900$, there were very few indications that there was anything wrong
with CM, the Rayleigh-Jeans ultraviolet catastrophe and the classically
unaccountable stability of matter being perhaps the most important of these.
CM works as well as it does because of a number of interlinked factors
working together. It is important to understand these factors, because they
have played important roles in the development of CM and QM.

First, there is the crucial role of technology. QM was discovered and
formulated only after certain advances in technology had been made,
particularly in spectroscopy. Without these advances, the flaws in CM would
have remained hidden, the relatively small size of Planck's constant being a
major contributor to this.

A second factor is that objects in the real world generally involve
extremely large numbers of degrees of freedom, which tend to behave
collectively as if the principles of CM were valid. QM experiments are
generally characterized by the careful way in which environmental factors
are excluded or controlled so as to allow focus on only a very few specially
selected degrees of freedom, such as electron spin. Good examples are
double-slit experiments in quantum optics, the Stern-Gerlach experiment, and
high energy particle scattering experiments. It is only under the most
carefully controlled conditions that quantum processes reveal their
spectacular non-classical properties clearly.

A third factor is the relative persistence in time of many structures or
patterns in the environment, compared with the timescales typical of quantum
experiments. This, coupled with the tendency of the human brain to objectify
complex phenomena, particularly when they re-occur with predictable and
well-defined characteristics, leads to a mental image of the universe as
divided into separate objects, such as observers, apparatus, and SUOs.

These images have limitations and can break down spectacularly in the
quantum domain. For example, electrons are generally regarded as point-like
objects with a well-defined mass, because in many experiments, that is a
good approximation. However, from the point of view of quantum field theory,
any charged particle is surrounded by a cloud of virtual photons which is
constantly interacting with its environment at long range. In consequence,
the full electron propagator has a cut rather than a simple pole\footnote{%
In relativistic quantum field theory, a simple pole in a propagator
corresponds to the possibility of detecting a particle in the conventional
sense of the word.}, which means that electron mass is contextual, i.e.,
depends on what is being measured and how. Another example is the simplest
atomic system, hydrogen, which is far from being just an electron bound to a
proton.

\section{The road to QM}

The first real crack in the classical world view of reality came in $1900$
with Planck's paper on the quantization of energy \cite{PLANCK-1900}. An
important fact about Planck's paper is that he did \textit{not} propose that
the electromagnetic radiation field itself contains quanta of energy. Planck
referred only to the behaviour of atomic oscillators absorbing and emitting
radiation, which they were postulated to do in a discrete way. If we take
the liberty of regarding atoms as detectors of radiation rather than being
SUOs themselves, then it seems not unreasonable to interpret Planck's
article as the first real paper on an instrumentalist approach to QM.
Planck's idea is at odds with CM because it is difficult if not impossible
to reconcile his vision of discrete energy levels in atomic oscillators with
the assumed existence of continuous Maxwellian electromagnetic fields
propagating between those atoms.

The crack opened wider when Bohr published his model of the hydrogen atom in
$1913$ \cite{BOHR-1913}. From the perspective of CM, this model is a mass of
inconsistencies and contradictions like Planck's idea; the classical
electron is assumed to be held in its atomic orbit by classical forces but
is not permitted to spiral into the nucleus under the effects of the
inevitable radiation damping predicted by Maxwellian electrodynamics. It is
simply impossible to understand how discrete energy levels could occur and
persist if the electromagnetic field is described in terms of the continuum
dynamics equations of Maxwell. However, as with Planck's idea discussed
above, we can rationalize Bohr's model to some extent if we interpret his
atoms as being part of the detecting apparatus and not SUOs.

By interpreting their work in this way, the \textquotedblleft old quantum
mechanics\textquotedblright\ (OQM) of Planck, Bohr and Sommerfeld may be
regarded not as a collection of ad hoc ideas swept aside by the sudden
discovery of wave mechanics by Schr\"{o}dinger in $1926$, but as important
steps in the development of a new approach to observation. The culmination
of that development was Heisenberg's seminal paper \cite{HEISENBERG-1925} in
$1925$ of what subsequently became known as matrix mechanics.

\section{Heisenberg's core philosophy}

Heisenberg had been a student of Sommerfeld's and contributed to OQM. He
came to the conclusion that the principles of CM were incorrect and
discovered how to remedy the situation. His vision about reality was
remarkably consistent and clear in the years $1925-27$. Above all, it was
radical, with extremely deep implications. What he wrote about electron
trajectories remains very disturbing, presenting a picture of a reality
which has no existence or meaning other than through the processes of
observation. In his ground-breaking $1927$ paper on the uncertainty
principle \cite{HEISENBERG-1927} he wrote: \textquotedblleft \emph{I believe
that one can fruitfully formulate the origin of the classical `orbit' in
this way: the `orbit' comes into being only when we observe it.}%
\textquotedblright\ This is a complete rejection of CM principles.

Although Heisenberg's matrix mechanics was accepted at the time, his core
philosophy did not take hold generally and his algebraic formalism was soon
overwhelmed by Schr\"{o}dinger's wave mechanical approach. Moreover, it had
to contend with Einstein's approach to physics, which in contrast is quite
classical. Einstein remained to the end of his days a leading supporter of
the classical world view. In $1905$ he published his famous papers on
special relativity (SR) and on the photo-electric effect. Although
conventional wisdom suggests that these papers overthrew classical
principles, we argue that they actually reinforced their core values,
because the imagery is entirely classical.

The fact is, SR is \textit{not} a theory that describes observer-SUO
dynamics, but a comparison of different observers' accounts of the same SUO,
given that each observer sees it classically. This assumes that extraction
of information about an SUO can come cost free to both observer and SUO. SR
in its traditional formulation simply does \emph{not} incorporate the
fundamental quantum principle that an act of observation (i.e., any process
which extracts information) necessarily changes the state of an SUO.

To illustrate the pitfalls when QM is mixed with relativity, consider a
single photon. Not only does SR allow us to talk of such a thing as an
object in its own right, but actually gives the Doppler shift between the
two frequencies associated with that photon as seen by two different,
relatively moving observers. The problem is that in reality, a single photon
can be observed by one observer only. What is meaningful is a comparison of
what each observer \emph{would have seen} if they had in fact been the one
who had observed the photon. This touches on the logical-philosophical
notion of \emph{counterfactuality}. Counterfactuality, or discussion of
might-have-beens and what-ifs, is a safe exercise in a world run on
classical principles, but a dangerous one when quantum processes are
involved.

A committed relativist's counter-argument to our concerns would be that a
proper SQM discussion of a single photon should really be a statistical one,
i.e., given in terms of ensembles, and that Doppler shifts and suchlike
should only be inferred from that form of discussion. This would be in fact
an argument in our favour, because it demonstrates that the conventional
formulation of SR, which makes no reference to ensembles or statistical
principles, is too simplistic and takes no account of what really goes on
during a quantum experiment.

As for the photoelectric effect, Einstein's vision of quanta residing in the
electromagnetic field is a clear attempt to maintain a classical world view.
It represents a move away from what Planck wrote about, i.e., the atomic
oscillators, towards a perceived SUO, the electromagnetic field, the
properties of which it is assumed are being studied. It should be admitted
that there is room for the SUO concept, \textit{when it works and does not
mislead}, and it is undeniable that Einstein's papers had enormous impact on
subsequent physics. However, that does not mean that the ideas in those
papers represent the actual quantum physics of the process of observation in
an adequate way.

An indication of how hard it was for Heisenberg's ideas to be accepted was
the speed with which his approach was abandoned once Schr\"{o}dinger
published his papers on wave mechanics in $1926$. This was partly due to the
much better and well-known computational technology associated with
wave-mechanical linear differential equations, in contrast to the generally
intractable nonlinear algebraic formalism introduced by Heisenberg. An
equally important factor was the ease with which waves can be visualized in
the mind's eye. Visualization of objects or waves in space is an important
feature in CM, so to people thoroughly conditioned to that way of thinking,
Heisenberg's abstract approach would inevitably appear intangible and
perhaps absurd.

It was Max Born, mentor and collaborator of Heisenberg's, who played the
principal role in destroying the classical interpretation of Schr\"{o}%
dinger's waves by interpreting them in terms of probability \cite{BORN-1926}%
. As with most historical matters, things are rarely as clear-cut as
tradition and conventional wisdom suggest. According to Born's Nobel lecture
\cite{BORN-1954}, it was Einstein himself who motivated him at a crucial
stage to develop the statistical interpretation of the wave-function.

As an aside, towards the end of his Nobel lecture, Born stated that he
believed in particles, but qualified this with the view that
\textquotedblleft \emph{Every object that we perceive appears in innumerable
aspects. The concept of the object is the invariant of all these aspects}%
.\textquotedblright\ This is equivalent to the position taken in QDN: the
particle concept is a resum\'{e} of certain consistent patterns of behaviour
in our observations.

Few theorists followed Heisenberg's specific philosophical path after $1926$%
. An important exception was Feynman, who whilst still a student, began to
think about photons as a manifestation of the properties of apparatus,
rather than as intrinsic \textquotedblleft things\textquotedblright\ in
their own right. His ideas are well expressed in the quotation given at the
start of this review \cite{FEYNMAN-THESIS}. The objective of his doctoral
thesis was to describe the interaction of particles, such as electrons,
without the intervening fields (the electromagnetic field in the case of
electrons). In this he was only partially successful and subsequently
recanted his views, once the QED calculation of vacuum polarization had
turned out to be so successful. Before that happened, however, he had
developed the path integral formulation of quantum mechanics, with deep and
lasting consequences for modern physics. We shall show that QDN is fully
consistent with Feynman's path integral formulation.

It is worth trying to identify the reason for Feynman's (albeit limited)
success in his instrumentalist programme. Although quantum field theory
treats all fields in a democratic fashion, it does not ignore their
individual properties. Electrons are fermions, whereas photons are bosons.
This is a difference that gives electrons more of a permanence, or identity,
than photons\footnote{%
It is hard to avoid using the conventional language of particles at this
point.}. There is no theorem which requires photon number to be conserved,
whereas total electric charge is conserved in any interaction. Provided
electron-positron pair production thresholds are not exceeded, then it can
be meaningful to think of interacting electrons as having an identity during
that interaction\footnote{%
In the case of purely electron-electron scattering, this viewpoint is
undermined somewhat by the indistingushability of these objects.}. Under
those circumstances, they can be regarded as part of the detecting equipment
involved in an experiment. For example, in any Feynman diagram involving in
and out electron propagator lines which do not run back in time, such as in
Compton scattering, we can think of each fermion line as the worldline of a
detector which responds to electromagnetic exchanges with other such
detectors. Feynman's aim in his thesis of eliminating the intervening
photons can be seen in this light as an instrumentalist approach to QM.

\section{The role of physical space}

The idea that physical space is a three-dimensional manifold with a metric
is such a central concept in CM and SQM that it is necessary to comment on
it here, because QDN regards it quite differently. In QDN, physical space is
regarded as contextual, i.e., manifests itself differently according to the
details of an experiment. There are some experiments, such as those in
quantum optics described later on in this review, where physical space is
frequently a secondary, or even irrelevant, factor. The idea that physical
space is a pre-existing manifold with a metric over which physical phenomena
act out their dynamics is neither correct nor incorrect, but useful insofar
as the experimental context justifies it. Certainly, we find it hard to
describe the structure of real \emph{apparatus} without invoking physical
space, but we have to question the need to believe in SUOs moving around it,
particularly in the case of photons and other elementary particles.
Wave-particle duality in SQM is a manifestation of something not quite
correct with such a belief.

A particular difficulty in explaining the QDN perspective on space is not
that it is wrong, but that all the evidence from the world around us seems
to contradict it. A typical objection, for example, would be that when
astronomers observe stars, these appear to have all the characteristics of
objects at great distances from the observers. Distant stellar images are
reduced in angular size and in intensity exactly as predicted by the
traditional model of space as a three-dimensional continuum.

There are several arguments against this sort of objection. First,
astronomers do not observe stars just like that, but are themselves embedded
in a vast amount of contextuality, preserved and carried forwards in time by
the phenomenon of persistence (mentioned above), and much of this is well
modelled using the physical space concept. Astronomers know they live on a
planet which is in a solar system, which is itself part of a vast galaxy,
and so on. All of this contextuality is an essential ingredient in the
interpretation of stellar observations. Because of the absence of such
contextual information, ancient astronomers made numerous stellar
observations but could not deduce our conventional spacetime model of the
universe. In consequence, their models of the universe were often quite
bizarre by modern standards.

A second argument concerns the observations themselves. Astronomers build up
pictures based on physical space only after large numbers of photons have
been captured from stellar sources. A single photon captured by a detector
gives no information by itself as to the nature of the source of that
photon, not even as to whether there has been a Doppler shift, or at what
distance the source is situated. The space concept is of limited value in
this case. At best, some information may be acquired about the approximate
direction of the source, but this requires some specific knowledge about the
apparatus, such as where it is pointing. This supports the QDN view that
contextual knowledge about apparatus is an essential ingredient in the
interpretation of observations.

A third argument concerns the Born interpretation of wave-functions and
wave-particle duality. It seems impossible to reconcile the classical notion
of an electromagnetic wave radiating from a star with single photon capture
a long way away from that star. Electromagnetic waves are best regarded as
probability amplitude waves, not as swarms of photons. Any attempt to view
quantum wave-functions in objective terms, such as in Bohmian mechanics \cite%
{BOHM-1952}, encounters great difficulties in reconciling spherically
symmetric wave propagation with the wave-function collapse associated with
single photon capture.

A fourth argument is that the numerous non-local correlation effects which
have been empirically confirmed to date demonstrate that distance seems to
have no significance as far as certain forms of quantum information are
concerned.

The QDN view of physical space lies at the core of its instrumentalist
philosophy. As with Heisenberg's matrix mechanics, this does not make it
easy to accept on an intuitive level, but that does not invalidate it. A
useful way to think about quantum experiments is to imagine the observer
from the position of a blind and deaf person receiving occasional discrete
tactile signals from their immediate surroundings, rather than from the
perspective of a viewer swamped by a continuum of audio-visual signals
appearing to come from all around them. If the QDN perspective on physical
space is correct, then one implication is that conventional approaches to
\textquotedblleft quantum gravity\textquotedblright\ should fail in the long
run, because they generally assume space to be some sort of continuous SUO
which needs to be quantized. The programme of quantum gravity might be as
futile as attempts to quantize the classical continuum equations of fluid
mechanics.

\

This concludes our historically flavoured motivation for QDN. In the next
part, we discuss the mathematics of our approach. It will be seen that
Heisenberg's original ideas can be encoded into a general mathematical
formalism virtually identical to that used in quantum computation.

\

\

\begin{center}
\textbf{\Large PART\ II: Formalism}
\end{center}

\section{Elementary signal detectors and signal bits}

\emph{Elementary signal detectors} (ESDs) are central to QDN. An ESD is any
physical device or procedure permitting an observer to extract classical
elementary yes/no information from it at a given time. This information is
an answer to the basic question: \textquotedblleft \emph{is there a signal
in this detector or not?}\textquotedblright .

An ESD need not be located in physical space at a specific place. In
practice, some degree of localization will be involved, because physics
laboratories tend to be localized in space and time. The information carried
by a signal from an ESD need not represent a localized quantity such as
position, either. ESDs are used in QDN to model quantum outcomes of
experiments, and are analogous to projection operators in SQM, with some
important differences.

QDN assumes that all measurements in physics can be described in terms of
amplifications of quantum signals from collections of ESDs, to such levels
that observers in laboratories can interrogate their apparatus in a
classical way. What this means is that an observer can always determine a
yes/no answer unambiguously from an ESD, \emph{if they choose to look}.
Underneath this classical veneer, however, there exists the quantum domain
where ESDs operate according to the basic quantum rules described in this
review. How signal amplification occurs is of course important, but beyond
the scope of this particular review.

An objection to these ideas, coming from experimentalists, would be that
they do not do physics in such a way. In fact, careful examination
invariably shows that they do. Ultimately, everything experimentalists see,
record and interpret is expressible in terms of vast but finite numbers of
yes/no answers to basic questions. For example, when an experimentalist
gives the $x-y$ coordinates of a black spot on an otherwise white screen,
those two numbers represent a really economical way of summarizing a vast
amount of discrete information, most of it contextual and arising from the
initial set-up of the apparatus. These two numbers are a concise way of
saying that the answer to the basic question \emph{is this spot black?} is
\emph{no} for every one of the countless spots on the screen, except for
just one of them, for which the answer is \emph{yes}.

Experimentalist may report their results in terms of real numbers, but in
reality all they are dealing with are good discrete approximations to an
imagined continuous reality. QDN cannot concede any argument on this point,
precisely because its goal is to model reality as it is experienced, not as
it is imagined to be. On this account, experimentalists should try to avoid
expressions such as \textquotedblleft \emph{We found the position of the
particle to be here\textquotedblright }, because the associated imagery can
be misleading. Ideally, they should simply say what sort of signals they
have observed in their apparatus, and leave any interpretation as to what
those signals mean to theorists.

To illustrate the difference between how SQM and QDN model quantum outcomes,
consider an idealized Stern-Gerlach experiment, illustrated in Figure $1$.
In such an experiment, it would be found that when an electron\footnote{%
In the original experiment, the electron was carried by an ion.} had passed
through the strong inhomogeneous magnetic field in the middle of the device,
there would be two distinct regions or spots on the detector screen where it
could land and be detected. Each electron passing through the apparatus
would land on only one of these sites each time. Which site was landed on
could not be predicted in advance in general (this depends on the way the
electron was prepared), but a consistent probability distribution would be
built up after sufficient runs of the experiment had taken place.

The SQM formalism assigns a ket vector $|$\emph{up}$\rangle $ to the state
of those electrons which had landed on one particular spot, and a ket vector
$|$\emph{down}$\rangle $ to the state of those which had landed on the other
spot. These two vectors are then assumed to form an orthonormal basis set
for a two-dimensional Hilbert space called a quantum bit, or qubit.

\begin{figure}[t!]
\centerline{\includegraphics[width=4.0in]{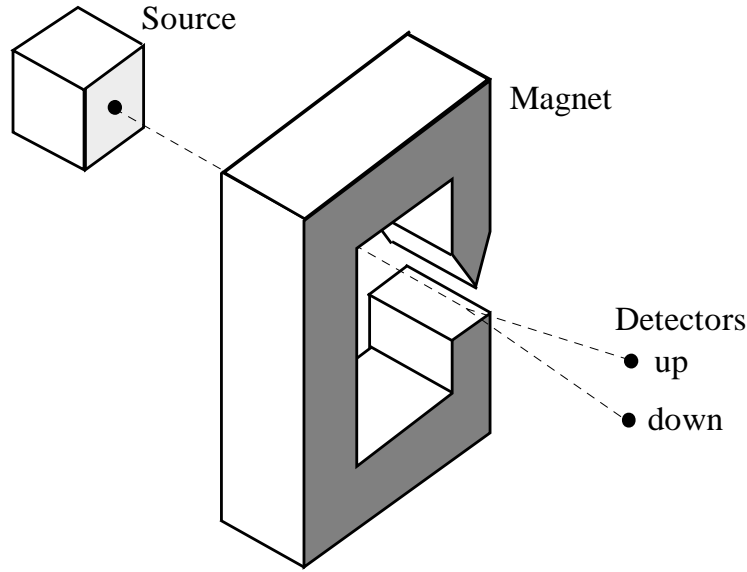}} \vspace*{8pt}
\caption{The Stern-Gerlach experiment.}
\end{figure}
SQM assumes that each time a single electron passes through the apparatus,
only one outcome signal is generated, either at the \emph{up }spot or else
at the \emph{down }spot\emph{. }The physical justification for this
assumption stems from the law of electric charge conservation, which has
never been observed to be violated, and from the fact that electrons cannot
be split into fractionally charged objects.

Whilst SQM takes a minimalist approach to the modelling of outcomes, QDN
appears to go the other way and allows for other logical possibilities to
exist \emph{in principle. }For example,\emph{\ }suppose the observer knew
that a single electron had been sent into the device. Without knowing in
advance the details of the apparatus, it is logically possible to imagine
that when the \emph{up} and \emph{down} sites were looked at separately, no
electron signal would be seen at either, indicating a net loss of electric
charge. Another logical possibility is that an electron signal would be
found at each of them, suggesting a total of two electrons had emerged.

The use of electric charge conservation to rule out each of these exotic
possibilities in SQM is specific to the details of the experiment; it could
not be applied for example in the case of electrically neutral particles
such as neutrons. It is in fact possible to do analogous non-linear optics
experiments with photons where such exotic outcomes could occur. In
spontaneous parametric down-conversion for example, a single incoming photon
could stimulate a crystal in a device to emit two photons, each of which
could be detected at a different site, giving rise to a total of two signals.

Once such logical possibilities are taken into account, it becomes clear
that the important factor determining which signals could actually be
observed is the physics of the apparatus. It is the apparatus which
determines the \emph{dynamical} possibilities of signal outcomes; if the
apparatus is changed then the dynamics changes, and that in turn generates
different signal possibilities\footnote{%
When we use the term \emph{apparatus}, we include the preparation devices,
the outcome detectors, and by implication, the observer's knowledge of their
equipment.}.

According to QDN principles, the basic signal question should be asked of
each ESD available to the observer at the same time, independently and
irrespective of any answer obtained from any other ESD at that time. This is
not what happens in the SQM approach to the Stern-Gerlach experiment, where
an examination of either one of the two outcome spots is assumed to imply
the signal state at the other. QDN represents the two possible outcomes of
the Stern-Gerlach experiment by two separate qubits $\mathcal{Q}^{1}$ and $%
\mathcal{Q}^{2}$, as shown in figure $2$, rather than the one qubit used in
SQM.

\begin{figure}[t!]
\centerline{\includegraphics[width=4.0in]{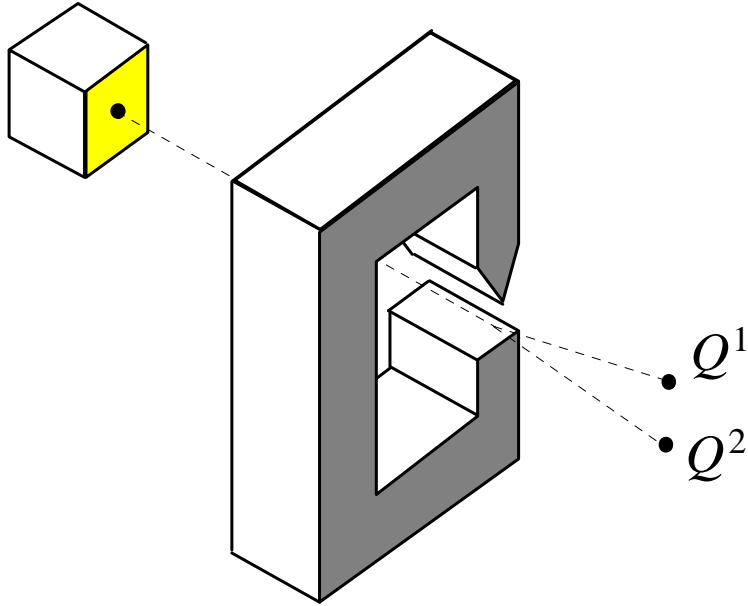}} 
\caption{Qubit assignment for the Stern-Gerlach experiment.}
\end{figure}
More generally, a quantum experiment with $k$ outcomes is described in QDN
by $k$ ESD qubits. This includes all situations in SQM\ where the projection
valued measure (PVM) formalism applies and extends naturally to those where
the more general positive operator-valued measure (POVM) formalism is
appropriate.

This rule has the consequence that the Hilbert spaces dealt with in QDN tend
to have much larger dimensions than their corresponding SQM\ analogues.
However, this cannot be avoided and generally has physical significance. For
example, QDN can deal with a many-signal scenario as easily as it deals with
a single-signal scenario, something which SQM does with some difficulty
through the use of Fock space, or with greater difficulty through the
apparatus of quantum field theory. QDN can also deal readily with multiple
observers either acting independently or interacting dynamically.

Over the next few sections, we shall mostly suppress any explicit reference
to time. How dynamically evolving networks are dealt with is quite involved
and is discussed in detail from section \ref{Dynamics} onwards. Before we
can do that, we have to establish what we mean by a quantum register and the
concept of \emph{Heisenberg net}, and in order to do that, we need to
discuss the properties of individual ESD qubits further.

\section{\label{Bits}Single ESDs}

Associated with each ESD qubit $\mathcal{Q}^{i\text{ }}$ is a \emph{%
preferred orthonormal basis}, denoted by $\mathsf{B}^{i}$. In QDN its
elements are denoted by $|0)_{i}$ and $|1)_{i}$\footnote{%
We use round bracket notation $|\Psi )$ in QDN to denote a \emph{labstate},
or quantum state of the observer's apparatus, reserving the more traditional
Dirac notation $|\Psi \rangle $ (with angular brackets) to denote a state of
an SUO when we are using SQM.}, and have the following interpretation. If
the observer knew \emph{with certainty before they looked} that the $i^{th}$
ESD would show nothing, i.e., be in its \textquotedblleft
void\textquotedblright\ (i.e., no-signal) state, then the \emph{anticipated
state}\textbf{\ }of that ESD would be represented by $|0)_{i}$. Conversely,
if the observer knew for sure that the ESD would be in its fired, or signal
state, if it were looked at, then that anticipated state of that ESD would
be represented by $|1)_{i}.$

Turning now to the calculation of signal outcome probabilities, there is a
fundamental difference between how an ESD would be treated in CM compared to
how it would be treated in QDN. In CM, an observer would assign Bayesian
(conditional) probabilities\emph{\ }$P($no-signal$|\xi )$ and $P($signal$%
|\xi )$ to the two possible mutually exclusive anticipated states of a
classical ESD, where $\xi $ represents the a priori information (i.e., the
context) held by the observer, such that%
\begin{equation}
P\left( \text{no-signal}|\xi \right) +P(\text{signal}|\xi )=1.
\end{equation}

In QDN, in contrast, the observer assigns a \emph{labstate}$|\Psi )$ to the
anticipated state of an ESD (represented by qubit $\mathcal{Q})$, of the form%
\begin{equation}
|\Psi )\equiv \alpha |0)+\beta |1),  \label{101}
\end{equation}%
where $\alpha $ and $\beta $ are complex numbers such that $|\alpha
|^{2}+|\beta |^{2}=1$. At this point, the objects $|0)$ and $|1)$ are
assumed to be vectors in the qubit Hilbert space $\mathcal{Q}$, so that
vector addition is mathematically defined.

The signal states $|0)$, $|1)$ represent mutually exclusive outcomes and the
interpretation of (\ref{101}) is given by the Born probability rule \cite%
{BORN-1926}. Using linearity and the inner product rule%
\begin{equation}
(i|j)=\delta _{ij},\ \ \ 0\leqslant i,j\leqslant 1,
\end{equation}%
the Born rule gives the two conditional outcome probabilities $P(0|\Psi )$, $%
P(1|\Psi )$, corresponding to $P($no-signal$|\xi )$ and $P($signal$|\xi)$
respectively, to be
\begin{equation}
P\left( 0|\Psi \right) \equiv |(0|\Psi )|^{2}=|\alpha |^{2},\ \ \ P\left(
1|\Psi \right) \equiv |(1|\Psi )|^{2}=|\beta |^{2}.
\end{equation}%
At the one-ESD level there is no obvious advantage in using a quantum
description rather than a classical one. The fundamental difference comes in
only when we deal with networks of qubits representing two or more ESDs.

Still on the one-ESD level, there are a number of qubit operators which play
important roles in the construction of operators essential to the many-qubit
discussion in the next section. For each qubit $\mathcal{Q}^{i}$, these are
the projection operators
\begin{equation}
P_{i}^{0}\equiv |0)_{i}(0|,\ \ \ P_{i}^{1}\equiv |1)_{i}(1|
\end{equation}%
and the signal creation and destruction operators%
\begin{equation}
A_{i}^{+}\equiv |1)_{i}(0|,\ \ \ A_{i}\equiv |0)_{i}(1|.
\end{equation}%
The signal operators $A_{i}^{+}$ and $A_{i}$ play a particularly significant
role in the theory, their mathematical properties being intimately related
to the physics of ESDs. These operators satisfy the \emph{signal bit algebra}%
\begin{equation}
A_{i}A_{i}=A_{i}^{+}A_{i}^{+}=0,\ \ \ \{A_{i},A_{i}^{+}\}=I_{i}\text{,}
\end{equation}%
where $I_{i}\equiv P_{i}^{0}+P_{i}^{1}$ is the identity operator for $%
\mathcal{Q}^{i}$ and no sum is implied over the index $i.$ The nilpotency
rule $A_{i}^{+}A_{i}^{+}=0$ encodes the physical fact that a given ESD
cannot be used to generate two or more signals simultaneously, i.e., an ESD
obeys the two-valued logic that it can be observed only in its void state or
else in its signal state.

An important point in QDN that a single signal from an ESD could represent
what is interpreted as a many-particle state in SQM. What matters here is
the context in which the signal is received.

\section{Quantum registers and Heisenberg nets}

We turn now to the more complicated but typical situation where an
experiment involves two or more ESDs. Before we give further details,
however, we need to clarify what QDN assumes about the evolution of
apparatus in time, because this affects the modelling. QDN is designed to
reflect the behaviour of apparatus in the real world and so it cannot be
assumed in general that a given observer's apparatus is constant in time,
even during a given run of an experiment. Although many experiments are
performed with apparatus that appears to be constant in time, that is by no
means a universal rule. Individual runs of certain experiments can involve
enormous intervals of time between state preparation and observation, as
always happens in the case of astrophysical observations of stars and
galaxies. In such cases, light from a distant star may be received by
astronomers long after that star had ceased to exist as a star.

In the PVM formulation of SQM \cite{VON-NEUMANN:1955,PERES:1993}, it is
generally assumed that state vectors of SUOs evolve in Hilbert spaces of
fixed dimension. Any time dependence of the apparatus itself, such as
externally imposed time-dependent electromagnetic fields, is generally
encoded into an explicit time dependence in the Hamiltonian. This approach
is consistent with the idea that an experiment extracts information from an
SUO, and whilst its states may change in time, its essential character
remains constant. In SQM, this approach was eventually recognized as too
limited, and so the POVM formalism was developed to deal with the
possibility that the number of outcome possibilities is different to the
dimension of the Hilbert space used \cite{PERES:1993}.

In contrast, QDN assumes from the outset that the Hilbert space representing
outcome possibilities is always different from one time step to the next,
even if the dimensionality remains constant, and even if the ESDs involved
in the experiment appear to persist over several time steps.

To be specific, at any given instant $n$ of the observer's time, their
apparatus, \emph{as regarded at that time by that observer, }will be denoted
by $\mathcal{A}_{n}$. This will consist of a countable number $r_{n}$ of
ESDs, $\mathcal{D}_{n}^{i}$, where $i$ runs from $1$ to $r_{n}$. In the
description of real experiments, $r_{n}$ will always be finite, an important
point at odds with standard assumptions in SQM. The harmonic oscillator is
an example where SQM assumes that there is an infinite number of possible
states of the system. In reality, there are no harmonic oscillators, just
various approximations to them.

The question as to whether $r_{n}$ is finite or not is central to many if
not all of the technical difficulties encountered in the refinement of SQM
known as quantum field theory. The harmonic oscillator appears to be
intimately involved in all of these problems in one way or another. Although
the mathematical properties of the quantized harmonic oscillator play an
essential role in accounting for the particle concept in free quantum field
theory, those same properties generate fundamental problems in interacting
field theories. For instance, the ultraviolet divergences encountered in
most Feynman loop integrals are linked to the unbounded energy spectrum of
the SQM oscillator, whilst infrared divergences are linked to the assumed
continuity of spacetime and the zero-point energy of the SQM oscillator.

Given $\mathcal{A}_{n}$, its associated ESDs are represented by a set of
signal qubits $\{\mathcal{Q}_{n}^{i}:i=1,2,\ldots ,r_{n}\}$, where qubit $%
\mathcal{Q}_{n}^{i}$ is identified with ESD $\mathcal{D}_{n}^{i}$. Together,
all the signal qubits associated with $\mathcal{A}_{n}$, plus the
information held by the observer about the physical significance of those
qubits constitute what we call a \emph{quantized detector network }(QDN), or
\emph{Heisenberg net}, denoted by $H_{n}$. The word \emph{net} here comes
from an analogy with a fisherman's net, which is spread out over space at a
particular time in an attempt to catch fish, the difference being that in
quantum experiments, the intention is to catch information.

The number of qubits in a Heisenberg net will be called the \emph{rank} $%
r_{n}$ of the net. These qubits when tensored together form an $2^{r_{n}}$%
-dimensional Hilbert space $\mathcal{R}^{r_{n}}$ known as a quantum
register. A fundamental property of any quantum register of rank greater
than one is that it contains entangled states as well as separable states.
Entanglement in QDN is regarded as an attribute of the observer's
information about their apparatus, and not as an intrinsic property of SUOs,
as is often implied in SQM terminology. QDN tries to avoid terms such as
\textquotedblleft entangled photon\textquotedblright \thinspace\ etc., but
we reserve the right to use such terminology occasionally, provided it does
not mislead. The concept of an entangled labstate is perfectly acceptable in
QDN.

\section{The signal basis}

In the following, we discuss a collection of qubits at a single instant of
the observer's time, so for convenience we shall suppress any reference to
time in this section. In the general theory given later there will be a
temporal subscript $n$ associated with every dynamical variable, including
the rank of the Heisenberg net.

Given a rank-$r$ quantum register $\mathcal{R}^{r}\equiv \mathcal{Q}%
^{1}\otimes \mathcal{Q}^{2}\otimes \ldots \otimes \mathcal{Q}^{r}$, then the
preferred basis $\mathsf{B}_{r}$ consists of all possible \emph{signal basis
states}, each of which is a tensor product of the form $|\varepsilon
_{1})_{1}\otimes |\varepsilon _{2})_{2}\otimes \ldots \otimes |\varepsilon
_{r})_{r}.$ Here, the \emph{occupancies} $\varepsilon _{i}$ are all either
zero or unity and $\{|0)_{i},|1)_{i}\}$ is the preferred basis for $\mathcal{%
Q}^{i}$. $\mathsf{B}_{r}$ will be referred to as the \emph{signal basis}.

There are $d\equiv 2^{r}$ distinct elements in $\mathsf{B}_{r}$, and
together they constitute an orthonormal basis for the Hilbert space $%
\mathcal{R}^{r}.$ For example,
\begin{equation}
\mathsf{B}_{2}=\{|0)_{1}\otimes |0)_{2},|1)_{1}\otimes
|0)_{2},|0)_{1}\otimes |1)_{2},|1)_{1}\otimes |1)_{2},\}  \label{111}
\end{equation}%
is a signal basis for the four dimensional vector space $\mathcal{R}^{2}$.
Labels are used in QDN to identify individual signal qubits, such as the
subscripts on the RHS of (\ref{111}). Therefore, the left-right ordering in
tensor products is not significant, provided the qubit identifier labels are
shown. We employ the convention that the quantum registers $\mathcal{Q}%
^{1}\otimes \mathcal{Q}^{2}$ and $\mathcal{Q}^{2}\otimes \mathcal{Q}^{1}$
are regarded as the same thing. For example, $|1)_{1}\otimes
|0)_{2}=|0)_{2}\otimes |1)_{1}$.

It is generally more useful to use the simplified notation%
\begin{equation}
|\varepsilon _{1}\varepsilon _{2}\ldots \varepsilon _{r})\equiv |\varepsilon
_{1})_{1}\otimes |\varepsilon _{2})_{2}\otimes \ldots \otimes |\varepsilon
_{r})_{r},
\end{equation}%
where the different elements in $\mathsf{B}_{r}$ are identified with the
different possible finite binary sequences $\{\varepsilon _{1},\varepsilon
_{2},\ldots \varepsilon _{r}\}$ of length $r$. For example,
\begin{equation}
\mathsf{B}_{2}=\{|00),|10),|01),|11)\}.  \label{222}
\end{equation}%
This notation no longer has individual qubit labelling, and therefore, the
left-right ordering is now significant. Generally, the $i^{th}$ element $%
\varepsilon _{i}$ of the sequence $\{\varepsilon _{1},\varepsilon
_{2},\ldots ,\varepsilon _{r}\}$ will represent the signal status of ESD $%
\mathcal{D}^{i}$.

The physical significance of the signal basis is best illustrated by
examples. The state $|000\ldots 0)$ is called the \emph{void state}. If the
observer examined the apparatus when it was in that labstate, every ESD
would be found in its void or no-signal state. Another example involves a
rank-$4$ Heisenberg net. The interpretation of a signal state such as $%
|0110) $ is that, if the apparatus were described by that particular
labstate at a given time, then provided nothing had changed, the observer
would find ESD $1 $ in its void state, ESD $2$ in its signal state, ESD $3$
in its signal state, and ESD $4$ in its void state.

An important feature of a signal basis is that each of its elements $%
|\varepsilon_{1}\varepsilon_{2}\ldots\varepsilon_{r})$ is a maximally
separable element of the quantum register $\mathcal{R}^{r}$, i.e., has no
degree of entanglement whatsoever, relative to the observer's information
about the current apparatus. Signal basis elements are identifiable with
classical signals because of this separability, and this provides a bridge
between the quantum processes being investigated and the classical
information extracted from them.

The assumption that a QDN observer can look at two or more different ESDs
\textquotedblleft simultaneously\textquotedblright\ implies that such an
observer has to be a non-local concept, simply because ESDs are invariably
separated in physical space. An observer in ESD is more like an collective
of local observers in relativity, each of which sits at a particular place
in a given frame of reference and observes what happens locally.

This goes some way towards accounting for the source of non-locality
problems in SQM. The QDN interpretation of wave-particle duality is not that
it arises from any bizarre property of an SUO, but originates from the
context of observation. Any observation which involves looking at two or
more ESDs simultaneously requires a great deal of pre-arrangement, which
comes at considerable cost in various ways. One cost is the need for the
\emph{space} concept itself, which is synonymous with the idea that
different objects exist at different places. Such thinking is reminiscent of
Mach's ideas concerning the origin of inertia \cite{MACH-1912}.

Associated with any signal basis $\mathsf{B}_{r}$ is its dual signal basis
\newline
$\mathsf{B}_{r}^{\ast }\equiv \left\{ (\varepsilon _{1}\varepsilon
_{2}\ldots \varepsilon _{r}|:\varepsilon _{i}=0,1,\text{ for }i=1,2,\ldots
,r\right\} $, which is the preferred basis for the dual quantum register $%
\mathcal{R}_{r}^{\ast }$. Elements of these two bases satisfy the $2^{2r}$
relations
\begin{equation}
(\varepsilon _{1}\varepsilon _{2}\ldots \varepsilon _{r}|\varepsilon
_{1}^{\prime }\varepsilon _{2}^{\prime }\ldots \varepsilon _{r}^{\prime
})=\delta _{\varepsilon _{1}\varepsilon _{1}^{\prime }}\delta _{\varepsilon
_{2}\varepsilon _{2}^{\prime }}\ldots \delta _{\varepsilon _{r}\varepsilon
_{r}^{\prime }},  \label{189}
\end{equation}%
where $\delta _{ij}$ is the Kronecker delta. The interpretation of the dual
basis is that its elements represent all possible \emph{maximal questions}
about the current signal status of the apparatus as a whole, i.e., an
examination of all of the ESDs in the net at a given time. For example, $%
(011|$ represents the simultaneous asking of three elementary questions:
\textquotedblleft \emph{is ESD }$1$\emph{\ in its void state}, \textbf{and}
\emph{is ESD }$2$\emph{\ in its fired state}, \textbf{and }\emph{is ESD }$3$%
\emph{\ in its fired state}?\textquotedblright , asked of an apparatus with
three ESDs. If in fact ESD $1$ were in its void state \textbf{and} ESD $2$
were in its fired state \textbf{and }ESD $3$ were in its fired state, then
the answer would be \emph{one} , which means \textquotedblleft
yes\textquotedblright . Otherwise, the answer would be \emph{zero}, which
means \textquotedblleft no\textquotedblright .

A useful feature of the formalism is that for a given maximal question, only
one basis signal state out of all possible basis signal states can return
the answer \textquotedblleft yes\textquotedblright , corresponding to a
probability amplitude of one. Therefore, for apparatus of any rank or
complexity, all but one of the basis signal states will return an answer
\textquotedblleft no\textquotedblright\ to any maximal question, which
greatly simplifies many calculations.

Another important point about QDN which goes to the heart of the
classical-quantum debate is that QDN permits linear superpositions of signal
basis vectors (i.e., a labstate can in principle be any element of the
quantum register spanned by the signal basis), but superpositions of maximal
questions are not allowed, by fiat. Only individual maximal questions are
classically meaningful. Essentially, the dual signal basis \emph{defines}
what is meant by a semi-classical observer in the theory. This gives QDN a
clear advantage over SQM, particularly over those variants such as Everett's
relative state theory which suffer from a lack of a preferred basis \cite%
{EVERETT-1957}. The contextual difference between the signal basis and its
dual means that the mathematical relationship $\langle \phi |\psi \rangle
=\langle \psi |\phi \rangle ^{\ast }$, associated with time reversal in SQM,
has to be treated cautiously in QDN. Time reversal experiments do not
actually reverse time, but examine the evolution of an initial state in one
experiment to the possible outcome states of another. In all discussions of
time reversal experiments in QDN, the labstate space $\mathcal{R}^{r}$
always plays a different physical role compared with its dual, $\mathcal{R}%
_{r}^{\ast }$.

\section{The computation basis}

The signal basis notation $|\varepsilon _{1}\varepsilon _{2}\ldots
\varepsilon _{r})$ is useful in some respects but less so in others. A
frequently more useful but equivalent notation for the elements of the
preferred basis $\mathsf{B}_{r}$ is given by writing
\begin{equation}
\mathsf{B}_{r}=\{|0),|1),|2),\ldots ,|2^{r}-1)\},
\end{equation}%
where
\begin{equation}
|\sum_{i=1}^{r}\varepsilon _{i}2^{i-1})\equiv |\varepsilon _{1}\varepsilon
_{2}\ldots \varepsilon _{r}).
\end{equation}%
For example, $\mathsf{B}_{2}=\left\{ |0),|1),|2),|3)\right\} ,$ where%
\begin{equation}
|0)\equiv |00),\ \ \ |1)\equiv |10),\ \ \ |2)\equiv |01),\ \ \ |3)=|11)\text{%
.}
\end{equation}%
Context will generally make the meaning clear whenever there is an
accidental numerical ambiguity, such as with the state $|11)$. This could
mean either the rank-$2$ element $|1)_{1}\otimes |1)_{2}$ described in the
occupation notation, or else an element such as $|1)_{1}\otimes
|1)_{2}\otimes |0)_{3}\otimes |1)_{4}\otimes |0)_{5}\otimes |0)_{6}\ldots $
in a rank-$4$ or greater register, expressed in the computation notation.

The computation notation generally has the advantage of being more compact
and is suited to many but not all calculations. The dual preferred basis $%
\mathsf{B}_{r}^{\ast }$ can also be expressed in computation terms, and then
the inner product relations (\ref{189}) take the form%
\begin{equation}
(i|j)=\delta _{ij}\text{,}\ \ \ 0\leqslant i,j<2^{r},
\end{equation}%
which is very useful.

A disadvantage of the computation notation is that it masks the signal
properties of a given state. For example, the state $|3)$ could represent
the labstate $|11)$ of a rank-$2$ apparatus, or the labstate $|11000)$ of a
rank-$5$ apparatus. However, as in the case of accidental numerical
ambiguity mentioned above, context will generally make it clear what is
meant by a given expression.

The computation basis is useful for expressing operators over the register,
which are generally be denoted in blackboard font in QDN. For example, the
register identity operator $\mathbb{I}_{r}$ can be expressed in the form%
\begin{equation}
\mathbb{I}_{r}=\sum_{k=0}^{2^{r}-1}|k)(k|.  \label{333}
\end{equation}

\section{Signal operators}

Associated with a rank-$r$ quantum register $\mathcal{R}^{r}$ is a number of
important operators connected to the physics of observation, and these will
appear frequently throughout the formalism. The most important of these are
the $r$ signal operators $\mathbb{A}_{i}^{+}$, $i=1,2,\ldots ,r$ and their
adjoints $\mathbb{A}_{i}$. These operators are defined in terms of the
one-qubit operators discussed in section $\ref{Bits}$, viz.,%
\begin{equation}
\mathbb{A}_{i}^{+}\equiv I_{1}I_{2}\ldots I_{i-1}A_{i}^{+}I_{i+1}\ldots
I_{r},\ \ \ i=1,2,\ldots ,r
\end{equation}%
where the subscripts on the right-hand side label individual signal qubits
and the tensor product symbol $\otimes $ has been suppressed for notational
economy. These operators satisfy the following relations, which we shall
refer to as the \emph{signal algebra}:

\noindent for $i=1,2,\ldots ,r$ we have%
\begin{equation}
\mathbb{A}_{i}\mathbb{A}_{i}=\mathbb{A}_{i}^{+}\mathbb{A}_{i}^{+}=0,\ \ \ \
\ \left\{ \mathbb{A}_{i},\mathbb{A}_{i}^{+}\right\} =\mathbb{I}_{r},
\label{Sig-1}
\end{equation}%
\noindent whilst for $i\neq j$, we have%
\begin{equation}
\lbrack \mathbb{A}_{i},\mathbb{A}_{j}]=[\mathbb{A}_{i}^{+},\mathbb{A}%
_{j}^{+}]=[\mathbb{A}_{i},\mathbb{A}_{j}^{+}]=0.  \label{Sig-2}
\end{equation}%
The signal algebra gives QDN a particular \textquotedblleft
flavour\textquotedblright ; sometimes it looks like a theory with fermions
and sometimes like a theory with bosons. At the signal level however, we are
dealing with neither concept specifically; the signal algebra is determined
by the physics of observation as it relates to apparatus and has its own
logic which is distinct to that of conventional particle physics.

It is remarkable that not long after the discovery of QM by Heisenberg and
Schr\"{o}dinger, Jordan and Wigner showed how to describe fermions in
quantum register terms \cite{JORDAN+WIGNER-1928,BJORKEN+DRELL:1965}. Their
construction of local fermionic quantum field operators requires tensor
product contributions from all of the qubits in a quantum register. In a QDN
approach to fermionic quantum fields \cite{J2005A}, their techniques were
used to describe fermionic fields using an infinite rank quantum register
associated with a net of ESDs distributed throughout all of physical space.
Because the Jordan-Wigner construction requires non-trivial contributions
from all qubits in the register, fermionic fields are manifestly and
inherently non-local in QDN.

\section{Signal classes.}

The preferred basis $\mathsf{B}_{r}$ for a rank-$r$ Heisenberg net has $%
2^{r} $ elements. These can be classified into subsets referred to as \emph{%
signal classes}. The zero-signal class $\mathcal{C}_{r}^{0}$ consists of
just one element, the void state, denoted by $|0)$ in the computation basis.
The one-signal class $\mathcal{C}_{r}^{1}$ consists of all elements in $%
\mathsf{B}_{r}$ of the form $\mathbb{A}_{i}^{+}|0)=|2^{i-1})$, $i=1,2,\ldots
,r$, and there are exactly $r$ of these. Likewise, the two-signal class $%
\mathcal{C}_{r}^{2}$ consists of all elements in $\mathsf{B}_{r}$ of the
form $\mathbb{A}_{i}^{+}\mathbb{A}_{j}^{+}|0)=|2^{i-1}+2^{j-1})$ for $i\neq
j $. The nilpotency of the signal operators eliminates states such as $%
\mathbb{A}_{i}^{+}\mathbb{A}_{i}^{+}|0)$ from further consideration.

More generally, the $k$-signal class $\mathcal{C}_{r}^{k}$ consists of basis
states of the form $\mathbb{A}_{i_{1}}^{+}\mathbb{A}_{i_{2}}^{+}\ldots
\mathbb{A}_{i_{k}}^{+}|0)$, for all integers $1\leqslant i_{1}<i_{2}<\ldots
<i_{k}\leqslant r$, and there are precisely $\tbinom{r}{k}\equiv r!/k!(r-k)!$
such states. If $c_{k}$ represents the number of elements in the $k$-signal
class, then $\sum_{k=0}^{r}c_{k}$ $=2^{r}$ as expected.

Some experience with QDN soon confirms the following rules:

\

\noindent $i)\qquad $the void state $|0)$ represents a state of the
apparatus with no signal anywhere, and this is the analogue of the vacuum%
\textbf{\ }state in quantum field theory;

\

\noindent $ii)\qquad $the one-signal states correspond often but not always
to what are called one-particle\textbf{\ }states in SQM, and so on;

\

\noindent $iii)\qquad $there is no universal rule which forbids labstates
which are superpositions of elements of different signal classes. Another
way of saying this is that signal number is not generally a conserved
quantity. However, a number of important examples can be discussed which
behave in such a way that it looks as if particle number was conserved. This
depends on the dynamics of the apparatus.

\section{Computation basis representation of signal operators.}

The signal operators $\mathbb{A}_{i}^{+}$ may be represented in the
computation basis as follows. First, consider any finite non-negative
integer $k$. There is always a unique representation of $k$ in the form of
its \emph{binary decomposition}, defined by%
\begin{equation}
k=k_{[1]}+k_{[2]}2+k_{[3]}2^{2}+\ldots +k_{[i]}2^{i-1}+\ldots
k_{[p_{k}]}2^{p_{k}-1}.
\end{equation}%
Here the binary coefficients $k_{[i]}$ are each either zero or unity and $%
p_{k}$ is some finite integer called the minimum rank of\emph{\ }$k$.\textbf{%
\ }This is the rank of the smallest quantum register with a preferred basis
containing $|k)$. For example, $9=1.2^{0}+0.2^{1}+2^{2}+1.2^{3}$, so $%
9_{[1]}=1$, $9_{[2]}=0$, $9_{[3]}=0$, $9_{[4]}=1$ and the minimum rank of $9$
is $4$.

The binary decomposition of an integer permits a description of a typical
computation basis element $|k)$ in terms of signal operators acting on the
void state, i.e.,
\begin{equation}
|k)=\mathbb{A}_{i_{1}}^{+}\mathbb{A}_{i_{2}}^{+}\ldots \mathbb{A}_{i_{\max
}}^{+}|0)\text{,}
\end{equation}%
where $i_{1}$, $i_{2},\ldots ,i_{\max }$ are all the non-zero binary
coefficients of $k$. For example,%
\begin{equation}
|9)=\mathbb{A}_{1}^{+}\mathbb{A}_{4}^{+}|0).  \label{139}
\end{equation}%
The value of this representation of a basis element is that the right-hand
side (\ref{139}) is independent of the rank of the register involved, apart
from the requirement that it has rank $4$ or more.

We can invert the process and describe the signal operators in terms of the
computation basis. Given the binary decomposition of $k$, then if $|k)$ is
one of the elements of a preferred basis $\mathsf{B}_{r}\equiv \left\{
|k):0\leqslant k\leqslant 2^{r}-1\right\} $ we may write
\begin{equation}
\mathbb{A}_{i}^{+}|k)=\bar{k}_{[i]}|k+2^{i-1}),
\end{equation}%
for $i=1,2,\ldots ,r$. Here $\bar{k}_{[i]}\equiv 1-k_{[i]}$ and we adopt the
rule that $\bar{k}_{[i]}|k+2^{i-1})=0$ if $\bar{k}_{[i]}=0$, even if there
is no actual element $|k+2^{i-1})$ in the given preferred basis\footnote{%
This is analogous to definitions such as $0!\equiv 1$, which greatly
enhances notation.}. We note the rule%
\begin{equation}
\mathbb{A}_{i}^{+}|0)=|2^{i-1}),
\end{equation}%
which is useful in applications to quantum optics involving one-signal
labstates..

Since the $|k)$ elements form a complete orthonormal set, we may use the
resolution of the identity (\ref{333}) to deduce that%
\begin{equation}
\mathbb{A}_{i}^{+}=\sum_{k=0}^{2^{r}-1}\bar{k}_{[i]}|k+2^{i-1})(k|.
\label{555}
\end{equation}%
For example, for a rank-$3$ QDN, we find the computation basis
representation (CBR)%
\begin{align}
\mathbb{A}_{1}^{+}& =|1)(0|+|3)(2|+|5)(4|+|7)(6|,  \notag \\
\mathbb{A}_{2}^{+}& =|2)(0|+|3)(1|+|6)(4|+|7)(5|, \\
\mathbb{A}_{3}^{+}& =|4)(0|+|5)(1|+|6)(2|+|7)(3|.  \notag
\end{align}%
It is easy to verify that such specific representations of the signal
operators satisfy the signal algebra (\ref{Sig-1}-\ref{Sig-2}). Note that a
CBR depends on rank; $A_{1}^{+}=|1)(0|$ for a rank-$1$ register but $\mathbb{%
A}_{1}^{+}=|1)(0|+|3)(2|$ for a rank -$2$ register, $\mathbb{A}%
_{1}^{+}=|1)(0|+|3)(2|+|5)(4|+|7)(6|$ for a rank-$3$ register, and so on.

In general, the CBR for any signal operator in a rank-$r$ register consists
of a sum of $2^{r-1}$ transition operators, all of which annihilate each
other including themselves. Likewise, a product $\mathbb{A}_{i}^{+}\mathbb{A}%
_{j}^{+}$ of two different signal operators can be expressed as a sum of $%
2^{r-2}$ transition operators which mutually annihilate, and so on. This
process of representation can be continued until we arrive at the \emph{%
saturation operator }$\mathbb{A}_{1}^{+}\mathbb{A}_{2}^{+}\ldots \mathbb{A}%
_{r}^{+}\equiv |2^{r}-1)(0|$, which creates the antithesis of the void
state, the fully saturated signal state $|2^{r}-1)\equiv |111\ldots 1)$,
when applied to the void state.

A particularly useful expression for the signal operators is obtained by
writing (\ref{555}) in the form%
\begin{equation}
\mathbb{A}_{i}^{+}=|2^{i-1})(0|+\mathbb{X}_{i}^{+},  \label{econ}
\end{equation}%
where the operator $\mathbb{X}_{i}^{+}\equiv \sum_{k=1}^{2^{r}-1}\bar{k}%
_{[i]}|k+2^{i-1})(k|$ annihilates the void state. This expression can be
used to greatly simplify calculations for those experiments involving
one-signal outcomes, such as single-photon quantum optics experiments.

\section{Persistence}

A conventional assumption in SQM is that pure states of an SUO may be
represented by time-dependent elements of a fixed Hilbert space. The chosen
Hilbert space is usually assumed fixed for two reasons. First, there is the
conditioned belief that an SUO \textquotedblleft exists\textquotedblright\
in time as a separate entity, at least long enough for the observer to study
it. Another contributory factor is the \emph{persistence of the apparatus},
or the tendency of actual apparatus to exist in its original form and
functionality in a laboratory before and after its useful role has ended.

Most physics experiments deal with persistent apparatus. That is generally
arranged by the observer as a matter of economy: experimentalists generally
do not have the resources to reconstruct their apparatus for each run.

There are situations however where persistence cannot be assumed. For
example, astronomers can catch light from a supernova only during an
extremely limited time, and that particular observation cannot be repeated.
What helps them is the vast numbers of photon signals that they can detect
during that limited time.

A similar issue arises in quantum cosmology. The universe is believed to be
expanding, and on that account, any approach to quantum cosmology should
take the attendant irreversibility into consideration and not treat the
evolution of the universe in traditional SQM\ terms, as if it were an SUO
being studied in a typical laboratory with persistent apparatus. The
expansion of the universe means there is no true persistence.

In QDN, individual ESDs are never persistent. Each ESD is assigned a
particular time at which it operates as an ESD, and outside of that time,
has no role in the formalism. This is the QDN analogue of the concept of an
event in relativity. Early versions of QDN work did make some use of
persistence \cite{J2004C}, but this simply increases the number of qubits
used in the formalism in a harmless way. Some of the examples discussed in
this review will assume a form of persistence when it is economical to do
so. In particular, our discussion of particle decay experiments involves a
description of how information from a given ESD is propagated forwards in
time, and this requires a careful discussion of what is meant by persistence.

\section{Observers and time}

Observers generally come equipped with their own sense of time, and quantum
experiments are carried out relative to that time. Relativity teaches that
there are two time concepts with different properties; coordinate (or
manifold) time and proper (or process) time. In both SR and GR (general
relativity), the former time concept is used to label events in spacetime
and is generally locally integrable. This means that spacetime can be
discussed in terms of coordinate patches \cite{SCHUTZ-1980}, such that
within a given coordinate patch, events can be labelled by spacetime
coordinates in a path-independent way. On the other hand, proper time is
non-integrable, which is to say that it depends on the particular dynamical
path taken between initial and final events.

In QDN, the time parameter associated with an experiment can normally be
identified with the proper time of an idealized inertial observer moving
along a timelike worldline, and for whom their laboratory appear to be at
rest at all times. However, it is just as easy to discuss inter-frame
physics, which is a discussion of experiments which start in one inertial
frame and end up in another. What is important in such situations is the
identification of what in SR and GR are known as spacelike hypersurfaces;
these are the analogue of instants of the observer's time in QDN.

In the real world, observers have finite existence: they come and go.
Observers and their apparatus are created at certain times and disappear at
later times, as seen by other observers in the wider universe. QDN as
formulated here allows for a discussion of different observers, each with
their individual time parameters and lifetimes. The use of quantum registers
also raises the possibility of accounting for the origin of various
temporally related concepts such as light cones, time dilation and other
metric-based phenomena in terms of Heisenberg net dynamics. A useful way to
discuss what is going on is in terms of the causal sets, the structures of
which arise naturally within quantum register dynamics \cite{J2005A}.

During their operational lifetimes, observers quantify their time in terms
of real numbers, usually read off from clocks. Most clocks give only a crude
estimate of the passage of time, and as a result, the ordinary human
perception of time as a one-dimensional continuum is just a convenient
approximation. The classical view of time is that it is a continuum at all
scales and for all phenomena. Certainly, things appear consistent with that
view in the ordinary world.

In quantum mechanics however, the situation is quite different. What matters
in a quantum experiment is information acquisition from the observer's
apparatus and this can only ever be done in a discrete way, regardless of
any theoretical assumption to the contrary \cite{MISRA+SUDARSHAN-1977}.
Whilst an observer's effective sense of time can be modelled accurately as
continuous, it is certainly the case that an observer can look at an ESD and
determine its status in a discrete way only. There are no truly
continuous-in-time observations. It is important here to distinguish between
what happens actually in experiments and what theorists would like to assume
happens in experiments.

The discreteness of the information extraction process forms the basis of
the time concept in QDN. In general, a given observer will represent the
state of their apparatus (the labstate) at a finite sequence of their own
(observer) times, denoted by the integer $n$. In QDN, a labstate at time $n$
will be denoted by $|\Psi ,n)$.

In QDN, a time $n+1$ is always regarded as definitely \emph{later}\textbf{\ }%
than time $n$. There is no scope in QDN for the concept of closed timelike
curve (CTC) found in some GR spacetimes, such as the G\"{o}del model \cite%
{GODEL-1949}. There is no need either to assume that the temporal interval $%
[t_{n},t_{n+1}]$ represents the same amount of physical duration as any
other interval $[t_{m},t_{m+1}]$.

\section{The Born probability rule}

One of the most significant attributes of quantum processes is the
randomness of quantum outcomes. Given identical state preparation, different
runs of a given experiment generally demonstrate controlled
unpredictability; the observer knows all about the range of possible
outcomes before observation, but cannot in general say beforehand which one
will occur for any particular run.

In practice the SQM approach to probability works well and we use it in QDN.
The Born probability rule \cite{BORN-1926} in SQM states that if a final
state $|\Psi \rangle $ is represented by a superposition of the form%
\begin{equation}
|\Psi \rangle =\sum_{i=1}^{d}\Psi ^{i}|i\rangle \text{,}
\end{equation}%
where the possible outcomes are represented by orthonormal vectors $%
|i\rangle $, $i=1,2,\ldots ,d$, then the conditional probability $P_{i}$ of
outcome $|i\rangle $ is given by $P_{i}=|\langle i|\Psi \rangle |^{2}$, if
the final state is normalized to unity.

This rule is used in much the same way in QDN, as follows. Consider a pure
labstate $|\Psi ,n)$ at time $n$. This can always be expanded in terms of
the computational basis $\mathsf{B}_{n}$ at that time in the form
\begin{equation}
|\Psi ,n)=\sum_{i=0}^{2^{r_{n}}-1}\Psi ^{i}|i,n)\text{,}
\end{equation}%
where $\sum_{i=0}^{2^{r_{n}}-1}|\Psi ^{i}|^{2}=1$. Labstates are always
normalized to unity, and because the signal basis states form a complete
orthonormal basis set, we may immediately read off the various signal state
conditional probabilities $P_{i}$, which are given by the rule%
\begin{equation}
P_{i}\equiv |(i,n|\Psi ,n)|^{2}=|\Psi ^{i}|^{2},\ \ \ 0\leqslant i<2^{r_{n}}.
\end{equation}%
$P_{i}$ is the conditional (Bayesian) probability for the observer to find
the apparatus in signal state $|i,n)$ at time $n$, \emph{if the observer
looked at their apparatus at that time}. These probabilities are conditional
on the observer being sure, just before they look, that the labstate at time
$n$ is $|\Psi ,n)$.

There is no natural restriction in QDN to labstates which are eigenstates of
signal number, i.e., superpositions of basis states from different signal
classes are permitted in principle. QDN is analogous in this respect to the
Fock space extension of Schr\"{o}dinger wave mechanics and to quantum field
theory.

\section{\label{Dynamics}Principles of QDN dynamics}

We are now in a position to discuss the principles of labstate dynamics from
the perspective of a single observer. At time $n$, this observer will hold
in their memory current information about their apparatus $\mathcal{A}_{n},$
the associated Heisenberg net $H_{n}$, and the labstate $|\Psi ,n)$, all at
that time. An analogous statement will hold for each time in a finite
sequence of times $n$ running from some integer $M$ to some other integer
such that $N>M$. QDN does not assume observers exist over unbounded
intervals of time, so its formalism is valid only over restricted ranges of
time.

We restrict attention to pure labstates throughout this and subsequent
sections for reasons of economy. A mixed-state, density matrix approach to
QDN dynamics should be straightforward to develop and is left for future
articles.

For the most basic sort of experiment, labstate preparation will be assumed
to have taken place by initial time $M$ and outcome detection is to take
place at final time $N$. For each integer $n$ such that $M\leqslant
n\leqslant N$, the observer associates with their apparatus $\mathcal{A}_{n}$
at that time a Heisenberg net $H_{n}$. This net consists of a finite number $%
r_{n}$ of qubits, $\mathcal{Q}_{n}^{1}$, $\mathcal{Q}_{n}^{2},\ldots ,%
\mathcal{Q}_{n}^{r_{n}}$, each qubit $\mathcal{Q}_{n}^{i}$ representing the $%
i^{th}$ signal detector $\mathcal{D}_{n}^{i}$ in $\mathcal{A}_{n}$. The
tensor product of all of these qubits is the quantum register $\mathcal{R}%
_{n}$, with preferred basis $B_{n}$ consisting of the $2^{r_{n}}$ basis
signal states.

There is no requirement in QDN or implication in our notation for the ESD
represented by $\mathcal{Q}_{n+1}^{i}$ to be related in any obvious way to
the ESD represented by $\mathcal{Q}_{n}^{i}$, i.e., we do not assume
persistence. In other words, successive quantum registers are completely
different Hilbert spaces, even if $r_{n+1}=r_{n}$. This is one of the
factors which makes QDN more general in its scope than SQM.

At time $n$, the observer describes the quantum state of their apparatus at
that time by a labstate $|\Psi ,n)$, which is some normalized vector in $%
\mathcal{R}_{n}$. Using the computational basis notation, this state can be
written in the form%
\begin{equation}
|\Psi ,n)=\sum_{k=0}^{2^{r_{n}}-1}\Psi _{n}^{k}|k,n),
\end{equation}%
where the signal basis $B_{n}\equiv \left\{ |k,n):0\leqslant
k<2^{r_{n}}\right\} $ satisfies the inner product rule $(k,n|l,n)=\delta
_{kl}$ and $\sum_{k=0}^{2^{r_{n}}-1}|\Psi _{n}^{k}|^{2}=1.$

A given run of an experiment will be described by the observer in terms of a
sequence $\left\{ |\Psi ,n):M\leqslant n\leqslant N\right\} $ of normalized
labstates, each element of which is associated with a particular Heisenberg
net $H_{n}$, followed by state outcome at time $N$. The question now is how
successive labstates relate to each between times $M$ and $N$.

Provided each run is prepared in the same way, and provided the apparatus
during each run is controlled in the same way, we can discuss a typical
labstate $|\Psi ,n)$ as a representative for an ensemble of runs. QDN
follows SQM in this respect. It is only at time $N$, when the observer
actually looks at all the detectors in $H_{n}$, do we encounter any run
dependence, on account of the inherent quantum randomness of the outcome of
any given run. We shall discuss this part of a run further on.

The dynamical transition from labstate $|\Psi ,n)$ to labstate $|\Psi ,n+1)$
involves a mapping from one quantum register $\mathcal{R}_{n}$ to another, $%
\mathcal{R}_{n+1}$. This leads us to give the following definitions and
theorems, which have proved useful in QDN.

\subsection{Born maps and semi-unitarity}

\noindent \textbf{Definition }$\mathbf{1}$:\qquad A \emph{Born map} is a
norm-preserving map from one Hilbert space $\mathcal{H}$ to some other
Hilbert space $\mathcal{H}^{\prime }$; if $\Psi $ in $\mathcal{H}$ is mapped
into $\mathfrak{B}(\Psi )\equiv \Psi ^{\prime }$ in $\mathcal{H}^{\prime }$
by a Born map $\mathfrak{B}$, then $(\Psi ^{\prime },\Psi ^{\prime })=(\Psi
,\Psi )$.

\

Born maps are used in QDN in order to preserve total probabilities (hence
the terminology), but unfortunately, their properties are insufficient to
model all quantum processes. Born maps are not necessarily linear, as can be
seen from the elementary example $\mathfrak{B}(\Psi )=|\Psi |\Phi ^{\prime }$
for all $\Psi $ in $\mathcal{H}$, where $\Phi ^{\prime }$ is a fixed element
of $\mathcal{H}^{\prime }$ normalized to unity and $|\Psi |$ is the norm of $%
\Psi $ in $\mathcal{H}$. To go further, it is necessary to impose linearity.

\

\noindent \textbf{Definition }$\mathbf{2}$:\qquad A \emph{semi-unitary
operator}\textbf{\ }is a linear Born map. If $U$ is such a map then for any
elements $\psi $, $\phi $ in $\mathcal{H}$ and complex $\alpha ,\beta $, we
may write $|\alpha \psi +\beta \phi |\ =|\alpha U\left( \psi \right) +\beta
U\left( \phi \right) |$.

\

The following theorems are relatively easy to prove and left to the reader:

\

\noindent \textbf{Theorem }$\mathbf{1}$: \qquad A semi-unitary operator from
$\mathcal{H}$ to $\mathcal{H}^{\prime }$ exists if and only if $\dim
\mathcal{H\leqslant }\dim \mathcal{H}^{\prime }$.

\

\noindent \textbf{Theorem }$\mathbf{2}$:\qquad If $U$ is a semi-unitary
operator from $\mathcal{H}$ to $\mathcal{H}^{\prime }$, then $U^{+}U=I$,
where $I$ is the identity operator over $\mathcal{H}$.

\

\noindent \textbf{Corollary }$\mathbf{1}$:\qquad A semi-unitary operator
preserves inner products and not just norms.

\

\noindent \textbf{Theorem }$\mathbf{3}$:\textbf{\qquad }If $U$ is a
semi-unitary operator from $\mathcal{H}$ to $\mathcal{H}^{\prime }$ and $%
\dim \mathcal{H}=\dim \mathcal{H}^{\prime }$, then $U^{+}$ is also a
semi-unitary operator from $\mathcal{H}^{\prime }$ to $\mathcal{H}$. For
such an operator, $U^{+}U=I$ and $UU^{+}=I^{\prime }$.

\

\noindent \textbf{Definition }$3$:\qquad An operator $U$ satisfying the
conditions of Theorem $3$ will be called \emph{unitary}.

\subsection{Application to dynamics}

It is normally assumed in QDN that a labstate $|\Psi ,n)$ in $\mathcal{R}%
^{n} $ at time $n$ is mapped into a labstate $|\Psi ,n+1)$ in $\mathcal{R}%
^{n+1}$ by some Born map $\mathfrak{B}_{n}$\footnote{%
Our convention is $\mathfrak{B}_{n}(|\Psi ,n))=|\Psi ,n+1)$ rather than $%
\mathfrak{B}_{n}(|\Psi ,n))=|\Psi ^{\prime },n+1)$, analogous to $%
U(t^{\prime },t)|\Psi ,t\rangle =|\Psi ,t^{\prime }\rangle $ in SQM.}.
Because $(\Psi ,n+1|\Psi ,n+1)=(\Psi ,n|\Psi ,n)$ under such a map, the Born
rule used in conjunction with the signal bases $\mathsf{B}_{n}$ and $\mathsf{%
B}_{n+1}$ means that total probability is conserved. This is not the same
thing as conservation of signal, charge, particle number, or any other
quantum variable.

Three scenarios are possible:

\

\noindent $i)\qquad \mathfrak{B}_{n}$ is non-linear:

By Theorem $1$, non-linearity is necessary if the rank $r_{n}$ of $\mathcal{R%
}^{n}$ is greater than the rank $r_{n+1}$ of $\mathcal{R}^{n+1}$, but can
arise even if this is not the case. For example, switching off any apparatus
at time $n+1$ would be modelled by the Born map $\mathfrak{B}_{n}(|\Psi
,n))=|0,n+1)$ for any state $|\Psi ,n)$ in $\mathcal{R}^{n}$, where $|0,n+1)$
is the void state of the apparatus at time $n+1$.

Another example is \emph{state reduction }due to observation, i.e., if at
time $n+1$ the observer actually looks at the apparatus and determines its
signal status, then this would be modelled by the non-linear Born map $%
\mathfrak{B}_{n}(|\Psi ,n))=|k,n+1)$, where now $|k,n+1)$ was some element
of the signal basis $\mathsf{B}_{n+1}$, chosen randomly with a probability
weighting given by the Born rule. In this particular case, however, there
are actually \emph{two} labstates associated with time $n+1$: $|\Psi ,n+1)$
represents the state of the apparatus immediately prior to state reduction
whilst $|k,n+1)$ represents the actual observed outcome immediately after.

\

\noindent $ii)\qquad \mathfrak{B}_{n}$ is linear and $r_{n}=r_{n+1}$:

This scenario corresponds to unitary evolution in SQM, and to reflect this,
we use the notation $\mathfrak{B}_{n}(|\Psi ,n))\equiv \mathbb{U}%
_{n+1,n}|\Psi ,n)=|\Psi ,n+1)$. From Theorem $3$, $\mathbb{U}_{n+1,n}$ in
this case satisfies the rules%
\begin{equation}
\mathbb{U}_{n+1,n}^{+}\mathbb{U}_{n+1,n}=\mathbb{I}_{n},\ \ \ \ \ \mathbb{U}%
_{n+1,n}\mathbb{U}_{n+1,n}^{+}=\mathbb{I}_{n+1}  \label{345}
\end{equation}%
and is called unitary, being the formal analogue of a unitary operator in
SQM.

\

\noindent $iii)\qquad \mathfrak{B}_{n}$ is linear and $r_{n}<r_{n+1}$:

In this case we use the same notation as in case $ii)$ above, i.e., $%
\mathfrak{B}_{n}(|\Psi ,n))\equiv \mathbb{U}_{n+1,n}|\Psi ,n)=|\Psi ,n+1)$,
but now $\mathbb{U}_{n+1,n}$ is properly semi-unitary and only the first
relation $\mathbb{U}_{n+1,n}^{+}\mathbb{U}_{n+1,n}=\mathbb{I}_{n}$ in (\ref%
{345}) is true. Such a scenario arises in particle decay experiments, for
example. These are discussed in section $\ref{Decays}$.

\

We cannot in general expect the rank $r_{n}$ of the quantum register $%
\mathcal{R}^{n}$ to be constant with $n$, so if we wish to preserve
probability and restrict the dynamical evolution to be linear in the
labstate, then we have to assume%
\begin{equation}
r_{M}\leqslant r_{M+1}\leqslant \ldots \leqslant r_{n}\leqslant \ldots
\leqslant r_{N},  \label{semi}
\end{equation}%
where $M$ is the initial time of the experiment and $N>M$ is the final time.
From this, we can appreciate that unless experimentalists are extremely
careful, their Heisenberg nets will grow irreversibly in rank. On the other
hand, the particle decay experiments discussed in section $\ref{Decays}$
specifically require the rank to increase at each time step.

The use of Born maps means total probability is always conserved, even if
linearity is absent. In principle, therefore, QDN allows for a discussion of
non-linear quantum mechanics, still based on most of the familiar Hilbert
space concepts used in SQM. As we have mentioned in case $i)$, necessarily
non-linear processes such as state preparation, state reduction, the
switching on and off of apparatus, etc., which are outside the scope of
unitary (Schr\"{o}dinger) evolution in SQM, can all be discussed in QDN in
terms of non-linear Born maps. We shall not focus further on this aspect of
the theory in this review, save to comment here that teleportation
experiments will involve such maps during intermediate times.

Henceforth, our interest will generally be in experiments based on linear
quantum processes, so (\ref{semi}) will be taken to be true. For such an
experiment running from time $M$ to time $N>M$, and knowing $r_{n}\leqslant
r_{n+1}$, then the labstate $|\Psi ,n)$ will change according to the rule%
\begin{equation}
|\Psi ,n)\rightarrow |\Psi ,n+1)\equiv \mathbb{U}_{n+1,n}|\Psi ,n),\ \ \
M\leqslant n<N\text{,}
\end{equation}%
where $\mathbb{U}_{n+1,n}$ is a semi-unitary operator (this terminology will
be used from now on even in the case where $r_{n}=r_{n+1}).$

The computational bases at times $n$ and $n+1$ can be used to represent $%
\mathbb{U}_{n+1,n}$. Specifically, we may write%
\begin{equation}
\mathbb{U}_{n+1,n}|i,n)=\sum_{j=0}^{\bar{d}_{n+1}}U_{n+1,n}^{j,i}|j,n+1),
\end{equation}%
where $\bar{d}_{n+1}\equiv 2^{r_{n+1}}-1$ and the coefficients $\left\{
U_{n+1,n}^{j,i}\right\} $ are complex numbers satisfying the semi-unitary
matrix conditions%
\begin{equation}
\sum_{j=0}^{\bar{d}_{n+1}}\left[ U_{n+1,n}^{j,i}\right] ^{\ast
}U_{n+1,n}^{j,k}=\delta _{ik},\ \ \ 0\leqslant i,k\leqslant \bar{d}_{n+1}.
\label{prob}
\end{equation}%
Using completeness, we arrive at the representation%
\begin{equation}
\mathbb{U}_{n+1,n}=\sum_{j=0}^{\bar{d}_{n+1}}|j,n+1)U_{n+1,n}^{j,i}(i,n|.
\label{dyn}
\end{equation}%
From this, we deduce that the adjoint operator $\mathbb{U}_{n+1,n}^{+}$ is
given by%
\begin{equation}
\mathbb{U}_{n+1,n}^{+}=\sum_{j=0}^{\bar{d}_{n+1}}|i,n)\left[ U_{n+1,n}^{j,i}%
\right] ^{\ast }(j,n+1|.
\end{equation}%
This is an operator from $\mathcal{R}^{n+1}$ to $\mathcal{R}^{n}$ which is
not semi-unitary, or even a Born map in general, if $r_{n}<r_{n+1}$.

A useful way of thinking about and constructing semi-unitary operators is in
terms of complex vectors. To each element $|i,n)$ of a given signal basis $%
\mathsf{B}_{n}\equiv \{|i,n):i=0,2,\ldots ,\bar{d}_{n}\}$, we associate a
complex vector $\mathbf{a}_{i}$ with $\bar{d}_{n+1}$ components,
corresponding to the image of $|i,n)$ under the action of $\mathbb{U}%
_{n+1,n} $. Specifically, we define the components ($\mathbf{a}_{i})^{j}$ of
$\mathbf{a}_{i}$ by the rule%
\begin{equation}
(\mathbf{a}_{i})^{j}\equiv U_{n+1,n}^{j,i}.
\end{equation}%
Then the set of complex vectors $\left\{ \mathbf{a}_{i}:i=0,1,\ldots ,\bar{d}%
_{n}\right\} $ satisfy the orthonormality relations%
\begin{equation}
\mathbf{a}_{i}^{\ast }\cdot \mathbf{a}_{j}\equiv \sum_{k=0}^{\bar{d}_{n+1}}%
\left[ (\mathbf{a}_{i})^{k}\right] ^{\ast }(\mathbf{a}_{j})^{k}=\delta
_{ij},\ \ \ 0\leqslant i,j\leqslant \bar{d}_{n}.
\end{equation}%
It is now obvious from these orthonormality relations why semi-unitarity
operators cannot exist if $\bar{d}_{n}>\bar{d}_{n+1}$. For example, it is
not possible to find a set of three or more mutually orthogonal non-zero
complex vectors in a two-dimensional complex space.

\section{The signal theorem}

The mathematical properties of semi-unitary operators and their relationship
to signal bases have an important bearing on the permitted physics of QDN
dynamics. Consider an experiment at times $n$ and $n+1$ and assume
semi-unitarity. At time $n$ the labstate $|\Psi ,n)$ is given by a
superposition of signal states from signal basis $\mathsf{B}_{n}\equiv
\left\{ |i,n):0\leqslant i<d_{n}\right\} $ whilst the labstate $|\Psi ,n+1)$
is given as a superposition of signal states from signal basis $\mathsf{B}%
_{n+1}\equiv \left\{ |i,n+1):0\leqslant i<d_{n+1}\right\} $. Because of
linearity, the crucial question as far as the dynamics is concerned is how
individual signal states evolve. Semi-unitarity imposes the following
constraint, which we call the \emph{signal theorem}:

\

\noindent \textbf{Theorem }$\mathbf{4}$:\qquad Two different signal basis
states $|i,n)$ and $|j,n)$ in a signal basis $\mathsf{B}_{n}$ cannot evolve
by semi-unitary dynamics into labstates which have only one signal basis
state in common.

\

\noindent \textbf{Proof}:\qquad Take $0\leqslant i<j<2^{r_{n}}$. Suppose $%
|i,n)$ evolves by semi-unitarity dynamics into a labstate according to the
rule
\begin{equation}
|i,n)\rightarrow \mathbb{U}_{n+1,n}|i,n)=\alpha |k,n+1)+|\phi ,n+1),
\end{equation}%
whilst $|j,n)$ evolves according to the rule%
\begin{equation}
|j,n)\rightarrow \mathbb{U}_{n+1,n}|j,n)=\beta |k,n+1)+|\psi ,n+1).
\end{equation}%
Here $k$ is some integer in the semi-open interval $[0,2^{r_{n+1}})$, $%
\alpha $ and $\beta $ are non-zero complex numbers, and $|\phi ,n+1)$ and $%
|\psi ,n+1)$ are elements in $\mathcal{R}_{n+1}$ sharing no signal states in
common either with each other or with $|k,n+1)$ in their computational basis
expansions. From Corollary $1$,\ semi-unitarity preserves inner products and
not just norms, so we must have%
\begin{equation}
0=(i,n|j,n)=(i,n|\mathbb{U}_{n+1,n}^{+}\mathbb{U}_{n+1,n}|j,n)=\alpha ^{\ast
}\beta ,
\end{equation}%
because $|k,n+1)$, $|\phi ,n+1)$ and $|\psi ,n+1)$ share no signal states in
common and are therefore mutually orthogonal. This establishes the theorem.

\

The signal theorem leads to the following important result for conventional
physics. Suppose an observer constructs an apparatus which, if prepared at
time $n$ to be in its void state, would remain in that state at time $n+1$.
If the dynamics is semi-unitary, then we may write%
\begin{equation}
|0,n)\rightarrow \mathbb{U}_{n+1,n}|0,n)=|0,n+1).  \label{1239}
\end{equation}%
This condition models an important physical property expected of most
laboratory apparatus; we would not expect equipment which had been switched
off to spontaneously generate outcome signals subsequently, unless it was
interfered with by some external agency. An apparatus which satisfies (\ref%
{1239}) will be called \emph{isolated} (between times $n$ and $n+1)$ on that
account. The analogue of such a situation in Schwinger's source theoretic
approach to quantum field theory \cite{SCHWINGER:1969} would be one where
the external sources were switched off during some interval of time, so that
the vacuum (empty space) remained unchanged during that time\footnote{%
We use the term \emph{void state} in QDN rather than \emph{vacuum} in order
to avoid unwarranted imagery associated with the space concept. Likewise, we
avoid the term \emph{ground state} to avoid unwarranted associations with
Hamiltonians and energy.}.

Suppose now that, given such an isolated apparatus, the observer had instead
prepared at time $n$ some labstate $|\Psi ,n)$ of the form%
\begin{equation}
|\Psi ,n)=\sum_{i=1}^{\bar{d}_{n}}\Psi ^{i}|i,n)\text{,}
\end{equation}%
i.e., a labstate with no void component (note the summation runs from unity,
not zero). Then for isolated apparatus under semi-unitary evolution, the
signal theorem tells us that there can be no void component in the labstate
at time $n+1$, and so we may write%
\begin{equation}
|\Psi ,n)\rightarrow \mathbb{U}_{n+1,n}|\Psi ,n)=\sum_{j=1}^{\bar{d}%
_{n+1}}\Phi ^{j}|j,n+1)\text{,}
\end{equation}%
where%
\begin{equation}
\Phi ^{j}=\sum_{i=1}^{\bar{d}_{n}}U_{n+1,n}^{j,i}\Psi ^{i}.
\end{equation}%
This is an important result, because it tells us that under normal
circumstances, apparatus does not normally fall into its void state during
an experiment, unless forced to do so by an external agency, such as the
observer switching it off.

\

\noindent \textbf{Example }$\mathbf{1}$:\qquad Consider an isolated rank-$1$
apparatus evolving into a rank-$1$ apparatus between times $n$ and $n+1$
under semi-unitary evolution. Then by a suitable choice of phase of basis
elements, we may always write%
\begin{align}
|0,n)& \rightarrow \mathbb{U}_{n+1,n}|0,n)=|0,n+1),  \notag \\
|1,n)& \rightarrow \mathbb{U}_{n+1,n}|1,n)=|1,n+1),
\end{align}%
from which we conclude the dynamics is essentially trivial.

\ \

The following example is important, as it models what happens in various
quantum optics modules such as beam-splitters and Wollaston prisms.

\

\noindent \textbf{Example }$\mathbf{2}$:\qquad Consider an isolated rank-$2$
apparatus evolving into a rank-$2$ apparatus between times $n$ and $n+1$
under semi-unitary evolution. Then isolation means that we must have $%
\mathbb{U}_{n+1,n}|0,n)=|0,n+1)$. Suppose further that it is known that any
one-signal state always evolves into a one-signal state. Then we may write%
\begin{align}
\mathbb{A}_{1,n}^{+}|0,n)& \equiv |1,0)\rightarrow \mathbb{U}%
_{n+1,n}|1,n)=\alpha |1,n+1)+\beta |2,n+1), \\
\mathbb{A}_{2,n}^{+}|0,n)& \equiv |2,n)\rightarrow \mathbb{U}%
_{n+1,n}|2,n)=\gamma |1,n+1)+\delta |2,n+1),
\end{align}%
where the coefficients satisfy the constraints%
\begin{equation}
|\alpha |^{2}+|\beta |^{2}=|\gamma |^{2}+|\delta |^{2}=1,\ \ \ \alpha ^{\ast
}\gamma +\beta ^{\ast }\delta =0\text{.}
\end{equation}%
Then by the signal theorem, the two signal state $\mathbb{A}_{1,n}^{+}%
\mathbb{A}_{2,n}^{+}|0,n)\equiv |3,n)$ must necessarily evolve into a
two-signal state, i.e., according to the rule
\begin{equation}
|3,n)\rightarrow \mathbb{U}_{n+1,n}|3,n)=|3,n+1),
\end{equation}%
modulo phase.

\

The following application of the signal theorem is surprising and somewhat
counterintuitive, because what appears to be a trivial mathematical result
rules out an entire class of physics experiment:

\

\noindent \textbf{Example }$\mathbf{3}$:\qquad Suppose an experimentalist
prepares a rank-two, one-signal labstate of the form
\begin{equation}
|\Psi ,n)=\left( \alpha \mathbb{A}_{1,n}^{+}+\beta \mathbb{A}%
_{2,n}^{+}\right) |0,n),
\end{equation}%
where $|\alpha |^{2}+|\beta |^{2}=1$. Suppose further that the dynamics is
semi-unitary and that the apparatus at time $n+1$ is of rank three. Then by
the signal theorem, semi-unitary evolution such that
\begin{align}
\mathbb{A}_{1,n}^{+}|0,n)& \rightarrow \mathbb{U}_{n+1,n}\mathbb{A}%
_{1,n}^{+}|0,n)=(a\mathbb{A}_{1,n+1}^{+}+b\mathbb{A}_{2,n+1}^{+})|0,n+1),\ \
\ |a|^{2}+|b|^{2}=1,  \notag \\
\mathbb{A}_{2,n}^{+}|0,n)& \rightarrow \mathbb{U}_{n+1,n}\mathbb{A}%
_{2,n}^{+}|0,n)=(c\mathbb{A}_{2,n+1}^{+}+d\mathbb{A}_{3,n+1}^{+})|0,n+1),\ \
\ |c|^{2}+|d|^{2}=1,
\end{align}%
is not possible.

\

This result tells us that a double-slit type of experiment where each slit
has only one quantum outcome site in common with the other cannot be
physically constructed. Experiments where two or more quantum outcome sites
are in common are possible, and then inevitable quantum interference terms
will occur in final state amplitudes. For example, in a standard double-slit
experiment, every site on the detector screen is affected by the presence of
each of the two slits.

This result reinforces an important lesson in QM: we cannot simply add
pieces of apparatus together and expect the result to conform to an addition
of classical expectations. A double-slit experiment where both slits are
open is not equivalent to two single-slit experiments run coincidentally and
simultaneously.

\section{Path summations}

The QDN formulation of dynamics has all the hallmarks of the Feynman path
integral formulation of SQM \cite{FEYNMAN+HIBBS:1965}, with several
important differences: time is not continuous, the Hilbert space changes at
each intermediate time-step and is assumed finite dimensional, and there is
no need to introduce a Lagrangian or Hamiltonian.

A typical run or repetition of a basic experiment will be assumed to start
at time $M$ and finish at a later time $N>M$. Given labstate preparation at
time $M$, there will be semi-unitary evolution through a sequence of
apparatus stages $\left\{ \mathcal{A}_{n}:M<n<N\right\} $, at which times
the observer does not look at their ESDs, and then an outcome detection
phase at the final time $N$. At the final time $N$, the observer looks at
all of their ESDs and finds out which element of $\mathsf{B}_{N}$
corresponds to the observed set of signal and void outcomes, for that given
run. The objective in practice is to compare the statistical distribution of
observed outcomes with the theoretically derived conditional probability $%
P\left( k,N|\Psi ,M\right) $ for each of the possible final state signal
basis outcomes $|k,N)$, \thinspace $0\leqslant k<2^{r_{N}}$.

Semi-unitary evolution will be assumed to hold between times $M$ and $N$,
i.e., condition (\ref{semi}) is valid. Given an initial labstate $|\Psi ,M$)$%
\equiv \sum_{i=0}^{\bar{d}_{M}}\Psi ^{i}|i,M)$, the next labstate is given
by $|\Psi ,M+1)=\mathbb{U}_{M+1,M}|\Psi ,M)$, where $\mathbb{U}_{M+1,M}$ is
semi-unitary, and so on, until finally we may write%
\begin{equation}
|\Psi ,N)=\mathbb{U}_{N,N-1}\mathbb{U}_{N-1,N-2}\ldots \mathbb{U}%
_{M+1,M}|\Psi ,M),\ \ \ \ \ N>M\text{.}
\end{equation}%
Inserting a resolution of each evolution operator of the form (\ref{dyn}),
the final state can be expressed in the form%
\begin{equation}
|\Psi ,N)=\sum_{j_{N}=0}^{\bar{d}_{N}}\sum_{j_{N-1}=0}^{\bar{d}_{N-1}}\ldots
\sum_{j_{M}=0}^{\bar{d}%
_{M}}|j_{N},N)U_{N,N-1}^{j_{N},j_{N-1}}U_{N-1,N-2}^{j_{N-1},j_{N-2}}\ldots
U_{M+1,M}^{j_{M+1},j_{M}}\Psi ^{j_{M}}.
\end{equation}%
We may immediately read off from this expression the coefficient of the
signal basis vector $|i,N).$ This gives the QDN analogue of the SQM\ Feynman
amplitude $\langle \Phi _{\text{final}}^{i}|\Psi _{\text{initial}}\rangle $
for the initial state $|\Psi _{\text{initial}}\rangle $ to go to a
particular final outcome state $|\Phi _{\text{final}}^{i}\rangle $. In our
case, what we are actually reading off is $\mathcal{A}\left( i,N|\Psi
,M\right) $, the amplitude for the labstate to propagate from its initial
state $|\Psi ,M)$ and then be found in signal basis state $|i,N)$ at time $N$%
. We find%
\begin{equation}
\mathcal{A}\left( i,N|\Psi ,M\right) =\sum_{j_{N-1}=0}^{\bar{d}_{N-1}}\ldots
\sum_{j_{M}=0}^{\bar{d}%
_{M}}U_{N,N-1}^{i,j_{N-1}}U_{N-1,N-2}^{j_{N-1},j_{N-2}}\ldots
U_{M+1,M}^{j_{M+1},j_{M}}\Psi ^{j_{M}}.  \label{amp}
\end{equation}

The required conditional probabilities are obtained from the Born rule as
discussed above, and so we conclude%
\begin{equation}
P\left( i,N|\Psi,M\right) =|\sum_{j_{N-1}=0}^{\bar{d}_{N-1}}\ldots
\sum_{j_{M}=0}^{\bar{d}%
_{M}}U_{N,N-1}^{i,j_{N-1}}U_{N-1,N-2}^{j_{N-1},j_{N-2}}\ldots
U_{M+1,M}^{j_{M+1},j_{M}}\Psi^{j_{M}}|^{2}.
\end{equation}
By writing the amplitude (\ref{amp}) in the form
\begin{equation}
\mathcal{A}\left( i,N|\Psi,M\right) =\sum_{j=0}^{\bar{d}%
_{N-1}}U_{N,N-1}^{i,j}\mathcal{A}\left( j,N-1|\Psi,M\right) ,
\end{equation}
it is easy to use the semi-unitary matrix conditions (\ref{prob}) to prove
that%
\begin{equation}
\sum_{i=0}^{\bar{d}_{N}}P\left( i,N|\Psi,M\right) =1,
\end{equation}
which means total probability is conserved, as expected.

Feynman derived his path integral for continuous time QM by discretizing
time and then taking the limit of the discrete time interval going to zero.
Technical problems occur in the taking of this limit and because of these,
the path integral in its original formulation \cite{FEYNMAN+HIBBS:1965} is
generally regarded as ill-defined. However, it is an invaluable heuristic
tool which provides the best way to discuss the quantization of certain
classical theories for which other approaches prove inadequate. In QDN, time
is discrete and in that sense we follow Feynman's lead, whilst avoiding the
pitfalls associated with the continuum limit, which we do not take in QDN.

This completes our introduction to the QDN formalism. An obvious extension
would be a description of mixed labstates, but there is no room to discuss
this here and this topic is left to future articles.

\newpage

\begin{center}
\textbf{\Large PART\ III: Applications}
\end{center}

\section{Preparation switches and outcome detectors}

For a typical experiment, we represent each ESD $\mathcal{D}_{n}^{i}$ in the
apparatus at time $n$ by a separate qubit $\mathcal{Q}_{n}^{i}$. For typical
experiments which start at time zero, there will be a single source
preparing the initial state, called a \emph{preparation switch}. Such an ESD
should more properly be referred to as an ESS (elementary signal source),
because that is the role that it is playing. In general, it will be clear
from context when an ESD is acting as a signal source or as a signal
detector, so we shall not differentiate between the two concepts further,
except in our discussion of QDN and relativity in section $31$. It is
possible to have two or more preparation switches simultaneously, as in the
case of multi-source photon interference experiments.

A preparation switch is a qubit which represents one of the possible sources
of a labstate, and this may be the result of non-linear evolution. In such a
situation, the evolution leading up to state preparation may be modelled
using a Born map, rather than a semi-unitary operator.

Between state preparation and detection, qubits will normally be involved in
semi-unitary evolution, corresponding to Schr\"{o}dinger unitary evolution
in SQM. Such intermediate qubits will not be involved in signal information
extraction directly, but play a role in the evolution of the labstate.
Essentially, they act as detectors of signals generated earlier in the
experiment, and act as preparation switches for signals detected later
elsewhere. The superposition of labstates originating from numbers of such
preparation switches is part and parcel of the quantum properties of QDNs.

\section{The double-slit experiment}

We are now in a position to discuss the application of QDN to a real
experiments. In this section we show how easily and generally we can discuss
the double-slit experiment, which demonstrates quantum interference.

A typical double-slit experiment involves three distinct pieces of
equipment, shown in Figure $3$. These are the source $A$ of some collimated
beam of particles such as photons or electrons, a pair of slits $B$ and $C$,
and a detecting screen $DE$.

\begin{figure}[th]
\centerline{\includegraphics[width=3.0in]{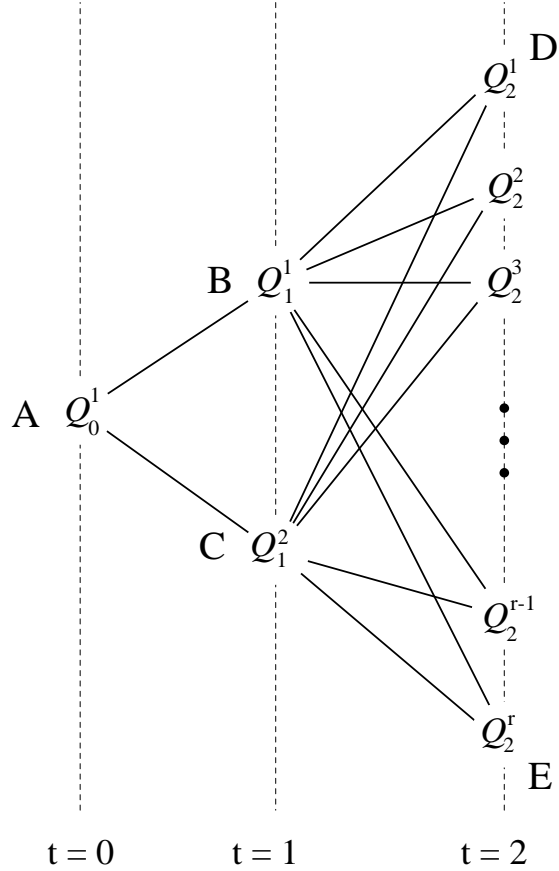}} \vspace*{8pt}
\caption{A double-slit experiment.}
\end{figure}

The source $A$ is represented by a single qubit labelled $\mathcal{Q}%
_{0}^{1} $ at time $t=0$ in the figure, which acts as a preparation switch.
Given successful preparation, the labstate at time zero is given by $|\Psi
,0)=\mathbb{A}_{1,0}^{+}|0,0)$. Subsequently, a signal could be detected at
either of the two slits $B$, $C$. In the formalism, these are represented by
qubits $\mathcal{Q}_{1}^{1}$and $\mathcal{Q}_{1}^{2}$ respectively, so the
labstate is assumed to evolve according to the rule%
\begin{equation}
|\Psi ,0)\equiv \mathbb{A}_{1,0}^{+}|0,0)\rightarrow \mathbb{U}_{1,0}\mathbb{%
A}_{1,0}^{+}|0,0)=\left\{ \alpha \mathbb{A}_{1,1}^{+}+\beta \mathbb{A}%
_{2,1}^{+}\right\} |0,1),
\end{equation}%
where $|\alpha |^{2}+|\beta |^{2}=1$ and the apparatus assumed isolated.

Now if either slit $B$ or $C$ were blocked off, the other slit would act as
a preparation switch at time $1$ with $r_{2}$\ possible one-signal outcomes
subsequently at time $2$. Each one of these potential outcomes would be
detected at an ESD $\mathcal{D}_{2}^{i}$ on the screen $DE$, and hence
identified with a qubit $\mathcal{Q}_{2}^{i}$ in $H_{2}$, the Heisenberg net
at time $2$. Here $i$ runs from $1$ to $r_{2}$, the number of distinct sites
on the screen where a signal could be seen. Therefore we may write%
\begin{equation}
\mathbb{A}_{i,1}^{+}|0,1)\rightarrow \mathbb{U}_{2,1}\mathbb{A}%
_{i,1}^{+}|0,1)=\sum_{j=1}^{r_{2}}\Psi ^{j,i}\mathbb{A}_{j,2}^{+}|0,2),
\end{equation}%
where $i=1,2$ and $\Psi ^{j,i}\equiv U_{2,1}^{2^{j-1},2^{i-1}}$. This
assumes that one-signal states evolve into one-signal states only, an
assumption which will undoubtedly depend on the physics of the situation.
The double slit experiment in quantum optics is normally conducted in
regimes where single-photon dynamics holds, so this is assumed here.

Semi-unitarity of evolution between times $1$ and $2$ imposes the rules%
\begin{equation}
\sum_{j=1}^{r_{2}}[\Psi ^{j,i}]^{\ast }\Psi ^{j,k}=\delta _{ik},\ \ \ \ \
i,k=1,2.
\end{equation}%
With this information and assuming both slits open, we conclude that the
labstate evolves from the initial labstate via the rule%
\begin{equation}
|\Psi ,0)\rightarrow \mathbb{U}_{2,1}\mathbb{U}_{1,0}|\Psi
,0)=\sum_{j=1}^{r_{2}}\{\alpha \Psi ^{j,1}+\beta \Psi ^{j,2}\}\mathbb{A}%
_{j,2}^{+}|0,2).
\end{equation}%
We may immediately read off from this the final state amplitudes and hence
determine the conditional probability of a signal being seen at any given
site. We find%
\begin{equation}
P\left( j,2|\Psi ,0\right) =|\alpha |^{2}|\Psi ^{j,1}|^{2}+|\beta |^{2}|\Psi
^{j,2}|^{2}+\alpha ^{\ast }\beta \lbrack \Psi ^{j,1}]^{\ast }\Psi
_{j,2}+\alpha \beta ^{\ast }\Psi ^{j,1}[\Psi ^{j,2}]^{\ast },
\end{equation}%
for an outcome such that detector $j$ registers a signal and all the other
remain void. The first two terms on the right hand side correspond to
classical expectations whilst the remaining terms correspond to quantum
interference terms. It is easy to prove that total probability is conserved,
i.e., $\sum_{j=1}^{r_{2}}P(j,2|\Psi ,0)=1.$

If the labstate arrived at the two slits such that the coefficients $\alpha $
and $\beta $ were randomly correlated in phase, then a classical ensemble
averaging procedure would eliminate the interference terms in the above.
Then we would have$\ P\left( j,2|\Psi ,0\right) =|\alpha |^{2}|\Psi
^{j,1}|^{2}+|\beta |^{2}|\Psi ^{j,2}|^{2}$, which corresponds to classical
probability expectations. In such a case, the experiment would look like one
with two independent sources. From this sort of discussion, we conclude that
it is the experimental context which is the source of wave-particle duality,
and not the properties of an SUO alone.

\section{Beam splitters}

Quantum optics experiments are remarkable in consisting generally of modular
components, such as beam-splitters, Wollaston prisms, mirrors,
phase-shifters and other devices, connected by photonic pathways. In this
section we discuss beam-splitters, which are mechanisms for superposing
quantum amplitudes. In general such a module will form part of a greater
network, as in the Mach-Zehnder interferometer discussed below, but in order
to understand its structure better we shall first discuss the beam-splitter
as if it were a complete piece of apparatus.

A typical beam-splitter $BS$ consists of two input ports labelled $a$ and $b$
in Figure $4$, and two output ports labelled $c$ and $d$. Considering a
beam-splitter as a separate piece of apparatus in its own right, an
experiment involving it would be described in QDN by a rank-$2$ net $%
\mathcal{Q}_{n}^{a}\otimes \mathcal{Q}_{n}^{b}$ evolving to a rank-$2$ net $%
\mathcal{Q}_{n+1}^{c}\otimes \mathcal{Q}_{n+1}^{d}$.

\

\begin{figure}[th]
\centerline{\includegraphics[width=2.0in]{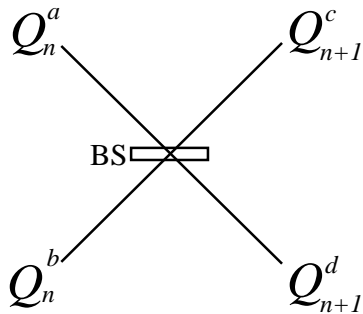}} \vspace*{8pt}
\caption{A beam-splitter.}
\end{figure}

At this point we simplify our notation further, in a form which has proved
convenient for rapid calculations. Using the computation basis, signal
states at time $n$ are written $i\equiv |i,n)$ whilst those at time $n+1$
are written $\bar{j}\equiv |j,n+1)$. We write $i\cdot j\equiv |i,n)(j,n|$, $%
\bar{\imath}\cdot j\equiv |i,n+1)(j,n|$, $A_{a}^{+}\equiv \mathbb{A}%
_{a,n}^{+}$, $\bar{A}_{c}^{+}\equiv \mathbb{A}_{c,n+1}^{+}$, $U\equiv
\mathbb{U}_{n+1,n}$, and so on. In this notation, the signal operators at
time $n$ are given by%
\begin{equation}
A_{a}^{+}=1\cdot 0+3\cdot 2,\ \ \ A_{b}^{+}=2\cdot 0+3\cdot 1
\end{equation}%
and similarly for $\bar{A}_{c}^{+}$, $\bar{A}_{d}^{+}$.

Next, we specify the dynamics in terms of the computation basis vectors. The
properties of a beam-splitter are encapsulated by the rules%
\begin{equation}
\begin{array}{lll}
0\rightarrow U0=\bar{0}, & \ \ \ \ \  & 1\rightarrow U1=\alpha\bar{1}+\beta%
\bar{2}, \\
2\rightarrow U2=\gamma\bar{1}+\delta\bar{2}, &  & 3\rightarrow U3=\bar{3},%
\end{array}
\label{beam}
\end{equation}
where the complex coefficients $\alpha,\beta,\gamma,\delta$ satisfy the
semi-unitary relations%
\begin{equation}
|\alpha|^{2}+|\beta|^{2}=|\gamma|^{2}+|\delta|^{2}=1,\ \ \
\alpha^{\ast}\gamma+\beta^{\ast}\delta=0.
\end{equation}
This assumes the beam-splitter is lossless. It is not necessary at this
stage to assume the symmetrical case $|\alpha|^{2}\ =|\beta|^{2}=\tfrac{1}{2}
$. In general, our formalism will be applied in the ideal, lossless and
non-symmetric case.

The relations (\ref{beam}) have a wider applicability than just to
beam-splitters. They can be used also to describe the Stern-Gerlach
experiment, which involves electrons rather than photons. In that
experiment, it is conventional to use only one incoming port, and then the
experiment looks formally like a rank-$1$ net evolving into a rank-$2$ net.
However, we could in principle reverse one of the out-beams back into the
same inhomogeneous magnetic field and in that case, we would expect that the
single reversed beam would split into two outgoing components, not one. One
of these would be identifiable with the original in-port beam, whilst the
existence of the other one would demonstrate that in fact the Stern-Gerlach
experiment has essentially the same architecture as a beam-splitter.

Both beam-splitter and Stern-Gerlach devices represent specific realizations
of a $2-2$ process, involving a rank-$2$ register evolving unitarily into
another rank-$2$ register. The only significant difference between these
realizations are the specific values of the coefficients $\alpha ,\beta
,\gamma $ and $\delta $, which are context-dependent, i.e., depend on the
information held by the observer as to the physical interpretation of the
signals being generated. The fact that the same structure of mathematics can
be used for beam-splitters and Stern-Gerlach experiments reflects the fact
that QDN can be regarded as a general formalism for the description of the
logical architecture of any quantum experiment.

From the above, we use completeness to deduce%
\begin{equation}
U=\bar{0}\cdot 0+\left\{ \alpha \bar{1}+\beta \bar{2}\right\} \cdot
1+\left\{ \gamma \bar{1}+\delta \bar{2}\right\} \cdot 2+\bar{3}\cdot 3.
\end{equation}%
Because there is no change of rank, this operator satisfies the unitarity
rules $U^{+}U=I$, $UU^{+}=\bar{I}$, where $I$ and $\bar{I}$ are the identity
operators for the registers $\mathcal{Q}_{n}^{a}\otimes \mathcal{Q}_{n}^{b}$
and $\mathcal{Q}_{n+1}^{c}\otimes \mathcal{Q}_{n+1}^{d}$ respectively. In
order to prove the relationship $UU^{+}=\bar{I}$, we note that $U^{+}$ is in
this case also a semi-unitary operator, and therefore the coefficients $%
\alpha ,\beta ,\gamma $ and $\delta $ satisfy the conjugate relations%
\begin{equation}
|\alpha |^{2}+|\gamma |^{2}=|\beta |^{2}+|\delta |^{2}=1,\ \ \ \alpha ^{\ast
}\beta +\gamma ^{\ast }\delta =0.
\end{equation}%
Hence we conclude $|\alpha |=|\delta |$ and $|\beta |=|\gamma |$. With this
information it is easy to relate our matrices to those given by Zeilinger in
his matrix formulation of the beam-splitter \cite{ZEILINGER-1981}.

With all this information, we can work out the rules for the evolution of
the signal operators. We find%
\begin{align}
A_{a}^{+} & \rightarrow UA_{a}^{+}U^{+}=\left\{ \alpha\bar{1}+\beta\bar {2}%
\right\} \cdot\bar{0}+\bar{3}\cdot\left\{ \gamma^{\ast}\bar{1}+\delta^{\ast}%
\bar{2}\right\} ,  \notag \\
A_{b}^{+} & \rightarrow UA_{b}^{+}U^{+}=\left\{ \gamma\bar{1}+\delta\bar {2}%
\right\} \cdot\bar{0}+\bar{3}\cdot\left\{ \alpha^{\ast}\bar{1}+\beta^{\ast}%
\bar{2}\right\} .
\end{align}
It can be verified that the signal algebra (\ref{Sig-1}-\ref{Sig-2}) is
invariant to this transformation.

The rules for the evolution of the signal operators can be given in the more
useful form%
\begin{align}
A_{a}^{+} & \rightarrow UA_{a}^{+}U^{+}=\alpha\bar{A}_{c}^{+}+\beta\bar {A}%
_{d}^{+}+\bar{X}_{a},  \notag \\
A_{b}^{+} & \rightarrow UA_{b}^{+}U^{+}=\gamma\bar{A}_{c}^{+}+\delta\bar {A}%
_{c}^{+}+\bar{X}_{b},
\end{align}
where%
\begin{equation}
\bar{X}_{a}\equiv\bar{3}\cdot\lbrack(\gamma^{\ast}-\beta)\bar{1}+(\delta
^{\ast}-\alpha)\bar{2}],\ \ \ \bar{X}_{b}\equiv\bar{3}\cdot\left[
(\alpha^{\ast}-\gamma)\bar{1}+(\beta^{\ast}-\delta)\bar{2}\right]
\end{equation}
are operators which annihilate the void state. The significance of this
representation of the dynamics is that for single-signal processes, such as
one-photon processes commonly studied in quantum optics, the $\bar{X}$
operators can be ignored. Therefore, in the general situation shown in
Figure $4$, the dynamics of the in-ports associated with a beam-splitter can
be written in the effective form%
\begin{align}
\mathbb{A}_{a,n}^{+} & \rightarrow\mathbb{U}_{n+1,n}\mathbb{A}_{a,n}^{+}%
\mathbb{U}_{n+1,n}^{+}\sim\alpha\mathbb{A}_{c,n+1}^{+}+\beta \mathbb{A}%
_{d,n+1}^{+},  \notag \\
\mathbb{A}_{b,n}^{+} & \rightarrow\mathbb{U}_{n+1,n}\mathbb{A}_{b,n}^{+}%
\mathbb{U}_{n+1,n}^{+}\sim\gamma\mathbb{A}_{c,n+1}^{+}+\delta \mathbb{A}%
_{d,n+1}^{+},
\end{align}
provided we restrict attention to one-photon processes. This justifies our
usage of such approximations in our earlier work, which were derived using
intuition \cite{J2004C}.

\section{The Mach-Zehnder interferometer}

In the following, we shall drop explicit reference to time. The reasons for
this are related to persistence and the relationship between photons and
wave-trains, and are discussed further in later sections.

\

\begin{figure}[h!]
\centerline{\includegraphics[width=4.0in]{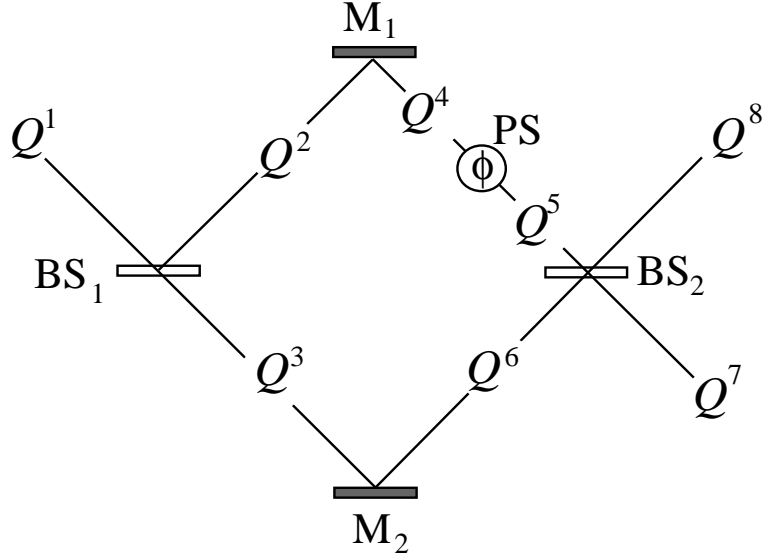}} \vspace*{8pt}
\caption{The Mach-Zehnder interferometer}
\end{figure}

The well-known Mach-Zehnder (MZ) interferometer, shown in Figure $5$,
provides a useful testbed for the application of our formalism. A
monochromatic beam of light is first passed through a beam-splitter $BS_{1}$%
. Each emergent component is then passed onto a separate mirror, denoted $%
M_{1}$ or $M_{2}$, which changes its phase by an amount $e^{i\mu_{1}}$ or $%
e^{i\mu_{2}}$ respectively. The component reflected from $M_{1}$ is then
passed through a sample or device which changes its phase by an amount $%
e^{i\phi}$, and then both components are passed through another
beam-splitter $BS_{2}$ onto two photon detectors. In Figure $5$ we show
qubits which might be placed at various places to detect photon signals,
\emph{if that were desired or planned by the experimentalist}.

The strategy here is to invoke just as many qubits as needed to describe the
essential logical structure of the device. If we wished to model the physics
in a much more complete but inefficient and mostly redundant way, we might
consider a much greater number of qubits, placed virtually everywhere along
the optical paths that the light might take. This would then begin to look
more like a quantum field theoretic description of the experiment. It is
actually rare for quantum field theory to be applied in full detail to a
quantum experiment performed over finite regions of time and space, and QDN
is no different in this respect. Our experience with the formalism suggests
that the qubits shown, labelled from $1$ to $8,$ represent a reasonable
basis to describe the experiment.

In real experiments with such devices, the various parts of the apparatus
persist during any given run, and we find it useful to encode that feature
here. Hence we shall work with a rank-$8$ quantum register, with the various
qubits labelled $1$ to $8$ as shown. Qubit $1$ acts as a preparation switch
whilst qubits $7$ and $8$ act as outcome detectors.

Consider a pulse of light starting at the preparation switch $\mathcal{Q}%
^{1} $ at time $0$. The initial state is denoted by
\begin{equation}
|\Psi ,0)=\mathbb{A}_{1,0}^{+}|0,0).
\end{equation}%
After passage through the first beam-splitter, $BS_{1}$, the discussion in
the preceding section justifies us in writing%
\begin{equation}
\mathbb{A}_{1,0}^{+}\rightarrow \mathbb{U}_{1,0}\mathbb{A}_{1,0}^{+}\mathbb{U%
}_{1,0}^{+}\sim \alpha _{1}\mathbb{A}_{2,1}^{+}+\beta _{1}\mathbb{A}%
_{3,1}^{+},
\end{equation}%
because single photon signal dynamics is assumed. Here $|\alpha
_{1}|^{2}+|\beta _{1}|^{2}=1$ and only one in-port of $BS_{1}$ is being
used. Reflection at the mirrors $M_{1}$, $M_{2}$ gives phase changes of the
form%
\begin{equation}
\mathbb{A}_{2,1}^{+}\rightarrow e^{i\mu _{1}}\mathbb{A}_{4,2}^{+},\ \ \
\mathbb{A}_{3,1}^{+}\rightarrow e^{i\mu _{2}}\mathbb{A}_{6,2}^{+}.
\end{equation}%
The phase-shift device $PS$ gives%
\begin{equation}
\mathbb{A}_{4,2}^{+}\rightarrow e^{i\phi }\mathbb{A}_{5,2}^{+,}
\end{equation}%
where we assume there is essentially zero time delay. Finally, passage
through the second beam-splitter $BS_{2}$ gives%
\begin{equation}
\mathbb{A}_{5,2}^{+}\rightarrow \alpha _{2}\mathbb{A}_{8,3}^{+}+\beta _{2}%
\mathbb{A}_{7,3}^{+},\ \ \ \mathbb{A}_{6,2}^{+}\rightarrow \gamma _{2}%
\mathbb{A}_{8,3}^{+}+\delta _{2}\mathbb{A}_{7,3}^{+}.
\end{equation}%
Taking all changes together, we may write%
\begin{align}
|\Psi ,0)=\mathbb{A}_{1,0}^{+}|0,0)\rightarrow & \left[ \alpha _{1}\alpha
_{2}e^{i(\mu _{1}+\phi )}+\beta _{1}\gamma _{2}e^{i\mu _{2}}\right] \mathbb{A%
}_{8,3}^{+}|0,3)+  \notag \\
& [\alpha _{1}\beta _{2}e^{i(\mu _{1}+\phi )}+\beta _{1}\delta _{2}e^{i\mu
_{2}}]\mathbb{A}_{7,3}^{+}|0,3).  \label{444}
\end{align}%
From this we may immediately read off the two detection amplitudes:%
\begin{align}
\mathcal{A}\left( 8,3|\Psi ,0\right) & =\alpha _{1}\alpha _{2}e^{i(\mu
_{1}+\phi )}+\beta _{1}\gamma _{2}e^{i\mu _{2}},  \notag \\
\mathcal{A}\left( 7,3|\Psi ,0\right) & =\alpha _{1}\beta _{2}e^{i(\mu
_{1}+\phi )}+\beta _{1}\delta _{2}e^{i\mu _{2}}.
\end{align}%
Final state probabilities are then given by the rule%
\begin{equation}
P\left( 8|\Psi \right) =|\mathcal{A}\left( 8,3|\Psi ,0\right) |^{2},\ \ \
P\left( 7|\Psi \right) =|\mathcal{A}\left( 7,3|\Psi ,0\right) |^{2},
\end{equation}%
and it may be verified that total probability is conserved. By tuning the
various constants $\alpha _{1},\beta _{2},\ldots ,\phi ,$ the usual standard
constructive and destructive quantum interference phenomena can be
demonstrated readily, so we shall not discuss those effects further here,
save for the following subsection.

\subsection{The identity interferometer}

As a check on the formalism, we consider the special case when the mirrors
and the phase-shifter have no effect (i.e., $\mu_{1}=\mu_{2}=\phi=0)$ and
the second beam-splitter, $BS_{1}$, effectively undoes the action of the
first one, $BS_{2}$ This is achieved by setting%
\begin{equation}
\alpha_{2}=\alpha_{1}^{\ast},\beta_{2}=\gamma_{1}^{\ast},\gamma_{2}=\beta
_{1}^{\ast}\text{, }\delta_{2}=\delta_{1}^{\ast},
\end{equation}
which is equivalent in SQM\ of representing the effect of $BS_{2}$ by the
adjoint of the operator representing $BS_{1}.$ This gives%
\begin{equation}
\mathcal{A}\left( 8,3|\Psi,0\right) =|\alpha_{1}|^{2}+|\beta_{1}|^{2}=1,\ \
\ \mathcal{A}\left( 7,3|\Psi,0\right) =\alpha_{1}\gamma
_{1}^{\ast}+\beta_{1}\delta_{1}^{\ast}=0,
\end{equation}
as expected.

\section{Nested and serial networks}

One of the advantages of the QDN formalism is that it allows an efficient
approach to the calculation of amplitudes for extended networks, such as
those consisting of two or more M-Z interferometers coupled together in
series or parallel, with a consequent increase in the number of out-ports.
We can imagine \textquotedblleft plugging in\textquotedblright\ as many M-Z
modules into networks of any desired topology as we wish, and the formalism
should be able to handle them relatively straightforwardly. Currently, we
are investigating the viability of encoding the formalism into a
computer-algebra package, which should give an efficient method for
calculating outcome probabilities for networks of great complexity.

\begin{figure}[h!]
\centerline{\includegraphics[width=6.0in]{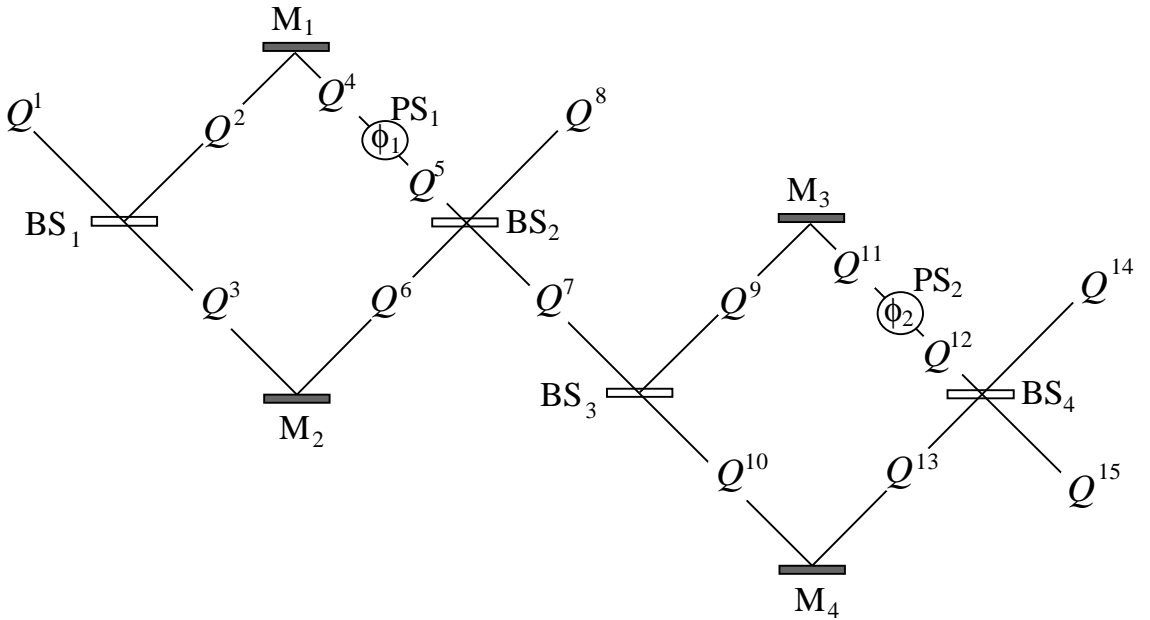}} \vspace*{8pt}
\caption{Nested Mach-Zehnder interferometers}
\end{figure}

As an example of a serial network, consider the network shown in Figure $6$,
where one of the out-ports (number $7)$ of one M-Z interferometer is
channelled into the in-port of a duplicate interferometer. The total number
of detector sites is now increased to three, which are represented by qubits
$8$, $14$ and $15.$

We may use the above calculation to determine the further evolution of
output port $7$, which gives%
\begin{align}
\mathbb{A}_{7}^{+}\sim & \left[ \alpha_{3}\alpha_{4}e^{i(\mu_{3}+\phi_{2})}+%
\beta_{3}\gamma_{4}e^{i\mu_{4}}\right] \mathbb{A}_{14}^{+}+  \notag \\
&
[\alpha_{3}\beta_{4}e^{i(\mu_{2}+\phi_{2})}+\beta_{3}\delta_{4}e^{i\mu_{4}}]%
\mathbb{A}_{15}^{+},
\end{align}
where we do not show the dependence on time. Inserting this into (\ref{444})
gives%
\begin{align}
|\Psi)=\mathbb{A}_{1}^{+}|0)\rightarrow & \left[ \alpha_{1}\alpha
_{2}e^{i(\mu_{1}+\phi)}+\beta_{1}\gamma_{2}e^{i\mu_{2}}\right] \mathbb{A}%
_{8}^{+}|0)+  \notag \\
& [\alpha_{1}\beta_{2}e^{i(\mu_{1}+\phi)}+\beta_{1}\delta_{2}e^{i\mu_{2}}]
\left[ \alpha_{3}\alpha_{4}e^{i(\mu_{3}+\phi_{2})}+\beta_{3}\gamma
_{4}e^{i\mu_{4}}\right] \mathbb{A}_{14}^{+}|0)+  \notag \\
&
[\alpha_{1}\beta_{2}e^{i(\mu_{1}+\phi)}+\beta_{1}\delta_{2}e^{i\mu_{2}}][%
\alpha_{3}\beta_{4}e^{i(\mu_{2}+\phi_{2})}+\beta_{3}\delta_{4}e^{i\mu_{4}}]%
\mathbb{A}_{15}^{+}|0),
\end{align}
from which the outcome probability amplitudes can be read off immediately.

The potential complexity of such networks does not mean more than one photon
would be detected in any given run. That depends on the dynamics of
individual modules. If for example, one of the above output ports was fed
into a module that produced a two-signal outcome, corresponding to a
two-photon state, then we would find a final state with two-signal
components. This is easily encoded into our formalism.

The potential value of QDN is that an increase in the complexity of a
network can be handled relatively easily. In SQM, an increase in the number
of out-ports requires special attention. It was this sort of issue that led
to the development of the POVM formalism, because in the more conventional
PVM\ approach, the dimensions of the Hilbert space involved is considered
fixed \cite{VON-NEUMANN:1955}. Once an arbitrary number of output channels
is involved, the simple PVM strategy fails.

We shall illustrate in the next section the difference between the POVM
approach and our approach by analyzing a network studied by Brandt using the
conventional POVM formalism. In fact, it is known that the POVM formalism
can be turned into a PVM formalism by embedding everything into a
sufficiently big Hilbert space. This may involve the introduction of an
auxiliary space known as an ancilla. Generally, we do not find that approach
so convincing, as it is not obvious to us how the auxiliary spaces relate to
the apparatus.

It is at this point that a criticism of QDN could be made: it looks like
another way of embedding the conventional formalism into a sufficiently big
Hilbert space so as to allow the operation of a PVM formalism. Neumark's
theorem says that this is always possible. However, that would be to miss
the point. The QDN strategy is based on the physics of observation and not
by any desire to have a PVM formalism \emph{per se}, although that is always
desirable. A qubit is not introduced into our approach unless there is a
physical reason for it and every aspect of the formalism is justifiable in
terms of the experimental procedures carried out by the observer. The
orthonormality of the signal basis comes from the classical nature of
observation as it is done in real experiments.

A more serious criticism of our formalism, already mentioned, is that it
leads to very large Hilbert spaces. In the example shown in Figure $6$, the
fifteen qubits form a quantum register of dimension $2^{15}=32768$, which
seems excessive considering there are just three out-ports. However, the
dimensions of the quantum registers we employ seem excessive simply because
persistence is assumed and there is so much more we can do with the
formalism. At each point where we place a qubit, we can insert additional
modules such as beam-splitters and so on, leading to new experimental
architecture. There is no reason to restrict this to diagrams which are
planar, either. We can consider time-dependent apparatus (which we did not
do in our discussion of the M-Z interferometer above), and we can consider
situations where several runs coexist at the same time. What this means is
that parts of a network may be showing signals which would conventionally be
identified with one in-state, whilst other parts would behave is if they
were carrying signals from an earlier or later run. Something like this
happens in high-energy particle physics experiments, when separate pulses of
charged particles are kept circulating in a collider ring all at the same
time until they are diverted and smashed into a target at separate times.

In the sort of calculations we have encountered, the relatively high
dimensionality of the quantum registers involved have not been a problem,
because as we have shown, the calculations can be handled usually in terms
of the signal operators $\mathbb{A}_{i}^{+}$, which leads to a great deal of
economy.

Another point in defence of QDN is that its objective of modelling apparatus
rather than SUOs would create difficult problems for SQM. In particular,
quantum field theory requires relatively sophisticated mathematical
machinery, and any attempt to model apparatus in terms analogous to those of
QDN would require heavy computation. Schwinger's source theory approach is
perhaps the closest quantum field theoretic analogue to QDN \cite%
{SCHWINGER:1969}. There are also parallels between S-matrix theory and QDN,
particularly in the way on-shell amplitudes are iterated \cite{ELOP:1966}.

\section{Time dependence and Bohmian mechanics}

In the above calculation, the time dependence was dropped for notational
convenience. The reason why this is permissible in a number of quantum
optics scenarios is that \textquotedblleft monochromatic\textquotedblright\
photons are associated with wave-trains of approximately fixed wavelength $%
\lambda$, and the physical length of such a wave-train can be very long
compared to $\lambda$ itself. When such a long wave-train passes through an
apparatus, the result is that over an extended period of time, it looks as
if the electromagnetic field is persistent throughout the apparatus, and so
time-dependence can be ignored.

If a relatively short wave-train is passed through a Mach-Zehnder network,
however, it is possible that no interference would take place, because waves
components travelling along different optical paths would arrive at the
second beam-splitter at different times and so miss each other. In such a
scenario, our approach would have to take the time dependence of the
labstate more carefully into account, most probably by introducing
additional qubits to incorporate the additional optical path lengths, but
this would not be an issue which the formalism could not handle
straightforwardly. Give it enough qubits and QDN can model the universe \cite%
{J2005A}.

Wave-trains do not \textquotedblleft contain\textquotedblright\ photons as
such. Paul \cite{PAUL-2004} gives examples of experiments where the average
number of photons per wave-train is less than one. Such wave-trains are
properly regarded as probability amplitudes, not structural components of
SUOs in the way Schr\"{o}dinger believed.

Bohmian mechanics \cite{BOHM-1952} is the outcome of the type of thinking
that motivated first de Broglie and then Schr\"{o}dinger towards the
development of wave-mechanics. It retains the interest of a proportion of
theorists who adhere to the conditioning of the classical world view, and
can be thought of as the opposite side of the spectrum of quantum theories
to QDN. Rather than abandon SUOs, Bohmian mechanics turns the wave-function
itself into a type of SUO, with an assumed classical existence in one form
or another. Such theories have little or no regard for the issues involved
with information extraction, and are regarded by us as severely flawed in
that respect.

An issue in Bohmian mechanics, where it is imagined that particles are
guided by \textquotedblleft pilot waves\textquotedblright , is the existence
of \textquotedblleft empty\textquotedblright\ waves containing no particle.
In the double slit experiment, for example, Bohmian theorists would have to
accept that according to their principles, a particle was guided to go
through one slit and not the other, which means that an empty wave has
passed through one of the slits. It is not clear in that approach what
happens to these empty waves once the particle has landed on a screen.
Presumably, there has to be some mechanism in Bohmian mechanics for signals
to be sent from the impact site to kill off any empty waves still
propagating elsewhere in the universe. No such requirement occurs in QDN.

\section{A POVM network calculation}

We turn now to a variant of the Mach-Zehnder interferometer studied by
Brandt using the POVM formalism in SQM \cite{BRANDT-1999,BRANDT-2002}. The
network is shown in Figure $7$.

\begin{figure}[t]
\centerline{\includegraphics[width=4.0in]{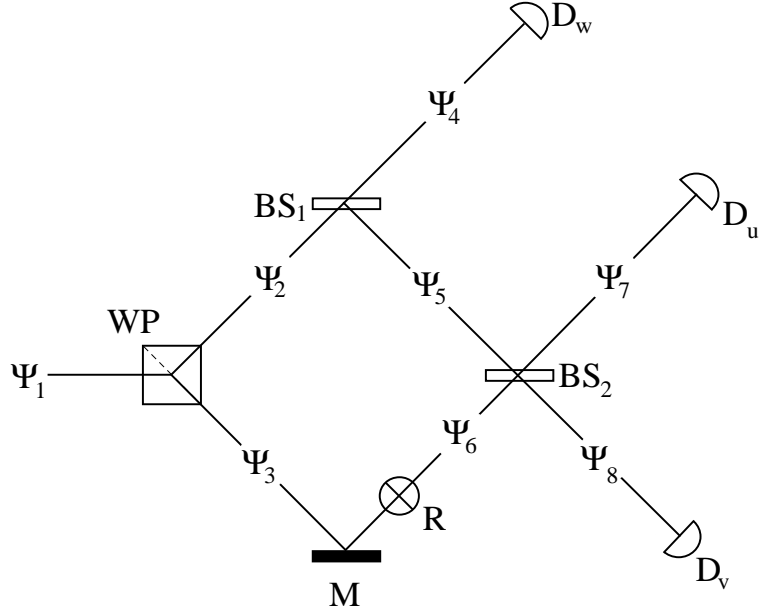}} \vspace*{8pt}
\caption{Brandt's POVM network.}
\end{figure}

In standard terminology, a beam $\Psi_{1}$ passes through a Wollaston prism $%
WP$, and its output channels $\Psi_{2}$, $\Psi_{3}$ are passed to a
beam-splitter $BS_{1}$ and mirror $M$ respectively. Beam-splitter $BS_{1}$
has transmission and reflection coefficients characterized by an angle $%
\theta$ in a specific way. Its transmitted wave $\Psi_{4}$ is detected at
detector $D_{w}$ whilst the reflected wave $\Psi_{5}$ is passed into
beam-splitter $BS_{2}.$ After wave $\Psi_{3}$ has reflected off mirror $M$,
its polarization is rotated by $90^{\circ}$ at $R$ and then it is passed on
to beam-splitter $BS_{2}$ in order to interfere with $\Psi_{5}.$ Detectors $%
D_{u}$ and $D_{v}$ act as out-ports for beam-splitter $BS_{2}.$

In his discussion, Brandt takes the initial state $|\Psi_{1}\rangle$ to be a
linear combination of non-orthogonal normalized states $|u\rangle$, $%
|v\rangle$ such that $|\Psi_{1}\rangle=\alpha|u\rangle+\beta|v\rangle$, with
$\langle u|v\rangle=\cos\theta\neq-1$ and%
\begin{equation}
|\alpha|^{2}+|\beta|^{2}+(\alpha^{\ast}\beta+\alpha\beta^{\ast})\cos
\theta=1.  \label{BRANDT}
\end{equation}
The outcomes are discussed in terms of POVM operators $E_{u\text{ }},E_{v}$
and $E_{w}$, which satisfy the relations%
\begin{equation}
E_{u}=\frac{I_{\mathcal{H}}-|v\rangle\langle v|}{1+\cos\theta},\ E_{v}=\frac{%
I_{\mathcal{H}}-|u\rangle\langle u|}{1+\cos\theta},\ \ \ E_{w}=I_{\mathcal{H}%
}-E_{u}-E_{v}.
\end{equation}
Here $I_{\mathcal{H}}\equiv E_{u}+E_{v}+E_{w}$ is the identity operator for
the two-dimensional Hilbert space with non-orthogonal normalized basis $%
\{|u\rangle,|v\rangle\}$. The three outcome probabilities are given by
\begin{equation}
\begin{array}{ccccc}
P\left( u|\Psi_{1}\right) & = & \langle\Psi_{1}|E_{u}|\Psi_{1}\rangle & = &
|\alpha|^{2}(1-\cos\theta), \\
P\left( v|\Psi_{1}\right) & = & \langle\Psi_{1}|E_{v}|\Psi_{1}\rangle & = &
|\beta|^{2}(1-\cos\theta), \\
P\left( w|\Psi_{1}\right) & = & \langle\Psi_{1}|E_{w}|\Psi_{1}\rangle & = &
|\alpha+\beta|^{2}\cos\theta,%
\end{array}%
\end{equation}
which add up to unity.

The quantum register discussion assigns qubits as shown in Figure $8$, so we
need at least a rank-$8$ register.

\begin{figure}[t]
\centerline{\includegraphics[width=4.0in]{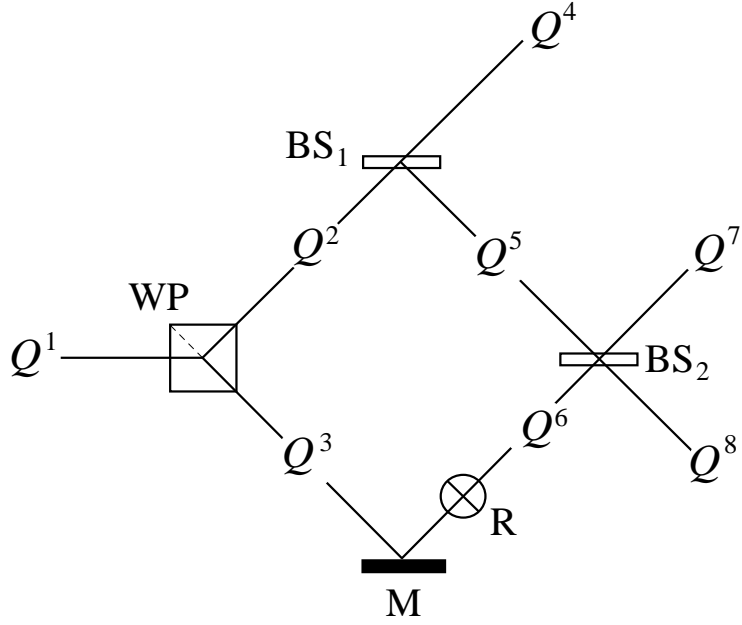}} \vspace*{8pt}
\caption{Qubit assignment for Brandt's network.}
\end{figure}

The QDN calculation is based on Brandt's parametrization of the various
modules and goes as follows.

\noindent$i)$ The rules for a Wollaston prism are virtually the same as for
a beam-splitter, except for having a different physical interpretation. The
two output ports, for example, carry different photon polarizations.
Wollaston prism $WP$ gives

\begin{equation}
\mathbb{A}_{1}^{+}\rightarrow\left( \alpha+\beta\right) \cos(\frac{_{1}}{^{2}%
}\theta)\mathbb{A}_{2}^{+}+\left( \alpha-\beta\right) \sin(\frac{_{1}}{^{2}}%
\theta)\mathbb{A}_{3}^{+},
\end{equation}
where the angle $\theta$ is related to $\alpha$ and $\beta$ by relation (\ref%
{BRANDT});

\noindent$ii)$ the beam-splitter $BS_{1}$ gives

\begin{equation}
\mathbb{A}_{2}^{+}\rightarrow\sqrt{1-\tan^{2}(\frac{_{1}}{^{2}}\theta )}%
\mathbb{A}_{4}^{+}+i\tan(\frac{_{1}}{^{2}}\theta)\mathbb{A}_{5}^{+}.
\end{equation}
Here the beam-splitter is finely tuned to the effects of the Wollaston prism
$WP$ in a particular way, it being assumed that this is possible in a real
experiment;

\noindent$iii)$ the mirror $M$ and the $90^{\circ}$ polarization rotation $R$
give

\begin{equation}
\mathbb{A}_{3}^{+}\rightarrow-\mathbb{A}_{6}^{+};
\end{equation}

\noindent$iv)$ the beam splitter $BS_{2}$ gives

\begin{equation}
\mathbb{A}_{5}^{+}\rightarrow\frac{i}{\sqrt{2}}\mathbb{A}_{7}^{+}+\frac {1}{%
\sqrt{2}}\mathbb{A}_{8}^{+},\ \ \ \mathbb{A}_{6}^{+}\rightarrow\frac {1}{%
\sqrt{2}}\mathbb{A}_{7}^{+}+\frac{i}{\sqrt{2}}\mathbb{A}_{8}^{+}.
\end{equation}
Here, beam-splitter $BS_{2}$ is assumed symmetric.

The net result is the evolution rule%
\begin{equation}
\mathbb{A}_{1}^{+}\rightarrow\left( \alpha+\beta\right) \sqrt{\cos\theta }%
\mathbb{A}_{4}^{+}-\alpha\sqrt{1-\cos\theta}\mathbb{A}_{7}^{+}+i\beta \sqrt{%
1-\cos\theta}\mathbb{A}_{8}^{+},
\end{equation}
or
\begin{equation*}
|\Psi_{out})=\left( \alpha+\beta\right) \sqrt{\cos\theta}|2^{3})-\alpha
\sqrt{1-\cos\theta}|2^{6})+i\beta\sqrt{1-\cos\theta}|2^{7}).
\end{equation*}
From this we readily work out the conditional probabilities%
\begin{equation}
\begin{array}{ccccc}
P\left( w|\Psi_{0}\right) & \equiv & |(2^{3}|\Psi_{out})|^{2} & = & |\alpha+
\beta|^{2}\cos\theta, \\
P\left( u|\Psi_{0}\right) & \equiv & |(2^{6}|\Psi_{out})|^{2} & = &
|\alpha|^{2}\left( 1-\cos\theta\right) , \\
P\left( v|\Psi_{0}\right) & \equiv & |(2^{7}|\Psi_{out})|^{2} & = &
|\beta|^{2}\left( 1-\cos\theta\right) ,%
\end{array}%
\end{equation}
in complete agreement with Brandt's calculation.

A few comments about the respective merits of the two approaches are in
order here. First, the POVM approach attempts to keep the discussion to
within the Hilbert space of the original state, i.e., in two dimensions in
this particular example. The need to accommodate three possible outcomes
then forces a break with the strict PVM formalism of von Neumann \cite%
{VON-NEUMANN:1955}, which would require all outcome states to be orthogonal.
The quantum register calculation works naturally within a PVM setting
because of the physics of observation. QDN was not motivated by any desire
to use the Neumark theorem, which says that a POVM description can always be
replaced by an equivalent PVM setting by introducing extra Hilbert space
dimensions.

Another issue is the physical significance of the non-orthogonality of the
vectors $|u\rangle $ and $|v\rangle $ in the POVM discussed here. Our
approach avoids non-orthogonality because we adhere to the principle that
the dual register basis consists of elements representing classically
meaningful questions, and these are mutually exclusive. It is not clear to
us what the physical significance of the non-zero inner product $\langle
u|v\rangle =\cos \theta $ is, other than as formal device, within an SQM\
setting, designed to reflect the physical properties of the preparation
apparatus, not those of the detectors $D_{u}$ and $D_{v}$.

A final point about POVMs is that they are usually applied in situations
involving mixed states, and the density matrix formulation becomes
necessary. The QDN formalism should be able to deal with such situations
readily. A particularly interesting possibility arises here: the idea that
the apparatus itself could become mixed (i.e., uncertain). In such a
scenario, the observer might have to assign a probability distribution to
more than one Heisenberg net at a given time, and these nets could have
different ranks. Such a scenario would require an enhanced version of
density matrix theory, perhaps representing a step in the development of a
more comprehensive theory of the physics of observation than is currently
available. This is one of the lines of further enquiry which we hope to
report on presently.

\section{Higher signal-rank experiments\hfill\hfill}

The examples studied above correspond to signal-rank one dynamics, such as
one-photon experiments in quantum optics, but these are a subset of all
possible experiments. We can just as easily discuss scenarios involving two
or more photons, and these raise the possibility of entanglement, in
addition to the superposition involved in the one-photon case.

\subsection{Independent experiments}

In the real world, different laboratories conduct quantum optics experiments
independently of each other, and any comprehensive theory should be able to
describe that scenario naturally, and allow for the possibility of dynamical
interaction between laboratories. QDN describes such scenarios naturally by
tensoring separate quantum registers together into larger super-registers. A
super-labstate in such a super-register representing independent
laboratories would remain factorizable as long as there was no information
exchange between those laboratories in any form. Each factor would be
associated with a separate laboratory conducting its own experiment.
Separability would endure until such a time as there was an interaction
between laboratories, at which point the super-labstate would become
entangled.

To illustrate the point, consider two independent Stern-Gerlach experiments
carried out simultaneously, according to some superobserver in contact with
each independent laboratory. Figure $9$ show the basic set-up. Experiment $1$
has a preparation switch $\mathcal{Q}^{1}$ and output qubits $\mathcal{Q}%
^{2},\mathcal{Q}^{3}$ whilst Experiment $2$ has preparation switch $\mathcal{%
Q}^{4}$ and output qubits $\mathcal{Q}^{5}$ and $\mathcal{Q}^{6}$.

\begin{figure}[t]
\centerline{\includegraphics[width=2.5in]{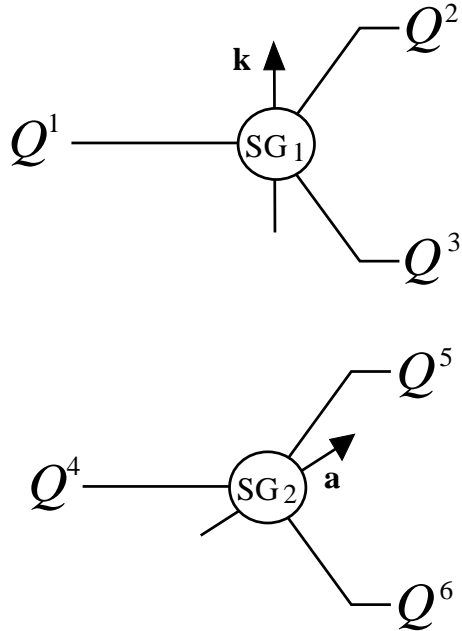}} \vspace*{8pt}
\caption{Super-network containing two subnetworks.}
\end{figure}

The super-observer describes the two experiments in terms of the rank-$6$
quantum register $\mathcal{R}^{6}\equiv\mathcal{Q}^{1}\otimes\mathcal{Q}%
^{2}\otimes\mathcal{Q}^{3}\otimes\mathcal{Q}^{4}\otimes\mathcal{Q}^{5}\otimes%
\mathcal{Q}^{6}$. Assuming each laboratory starts a run at the same time,
then the initial labstate is given by%
\begin{equation}
|\Psi_{in})=\mathbb{A}_{1}^{+}\mathbb{A}%
_{4}^{+}|0)=|100100)=|2^{0}+2^{3})=|9)=|100)_{123}\otimes|100)_{456},
\end{equation}
where the last expression on the right-hand side shows the separability of
the labstate. This means that according to the super-observer, their
labstate is a two-signal state, but each independent laboratory would think
it was dealing with a one-signal state.

If each experiment is truly independent and carries out a run simultaneously
with the other laboratory according to the time of the super-observer, then
evolution is given by%
\begin{align}
\mathbb{A}_{1}^{+} & \rightarrow\alpha\mathbb{A}_{2}^{+}+\beta\mathbb{A}%
_{3}^{+},\;\;\;\;\;|\alpha|^{2}+|\beta|^{2}=1,  \notag \\
\mathbb{A}_{4}^{+} & \rightarrow\gamma\mathbb{A}_{5}^{+}+\delta \mathbb{A}%
_{6}^{+},\;\;\;\;\;|\gamma|^{2}+|\delta|^{2}=1,
\end{align}
where independence of the laboratories means that there is no constraint
relating the coefficients $\{\alpha,\beta\}$ with the coefficient $\left\{
\gamma,\delta\right\} .$ The initial super-labstate then remains
factorizable, i.e.,%
\begin{equation}
|\Psi_{in})\rightarrow|\Psi_{out})=\left( \alpha\mathbb{A}_{2}^{+}+\beta%
\mathbb{A}_{3}^{+}\right) \left( \gamma\mathbb{A}_{5}^{+}+\delta\mathbb{A}%
_{6}^{+}\right) |0)=|\psi)_{123}\otimes|\phi)_{456},
\end{equation}
where
\begin{align}
|\psi)_{123} & \equiv\alpha|0)_{1}|1)_{2}|0)_{3}+\beta|0)_{1}|0)_{2}|1)_{3},
\notag \\
|\phi)_{456} & \equiv\gamma|0)_{4}|1)_{5}|0)_{6}+\delta|0)_{4}|0)_{5}|1)_{6}.
\end{align}

\subsection{Change of signal number experiments}

An interesting class of experiments involves changes of signal number
induced by entanglement, an important example being given by experiments
based on the EPR thought experiment \cite{EPR-1935}. It is most convenient
to discuss such an experiment in terms of electron spin or photon
polarization. We shall discuss it in terms of electron spin.

In the conventional terminology of SQM, such an experiment starts with the
preparation of an entangled state of an electron-positron pair, with total
spin zero. Subsequently, two observers, Alice and Bob measure spin
components using separate Stern-Gerlach machines. Alice filters only
electrons into her machine and observes their spin relative to the main
field of her apparatus, which is aligned along the $\mathbf{k}-$axis. The
other observer, Bob, filters only positrons into his machine, which has its
main magnetic field aligned along some other direction, $\mathbf{a}$.

\begin{figure}[t]
\centerline{\includegraphics[width=4.0in]{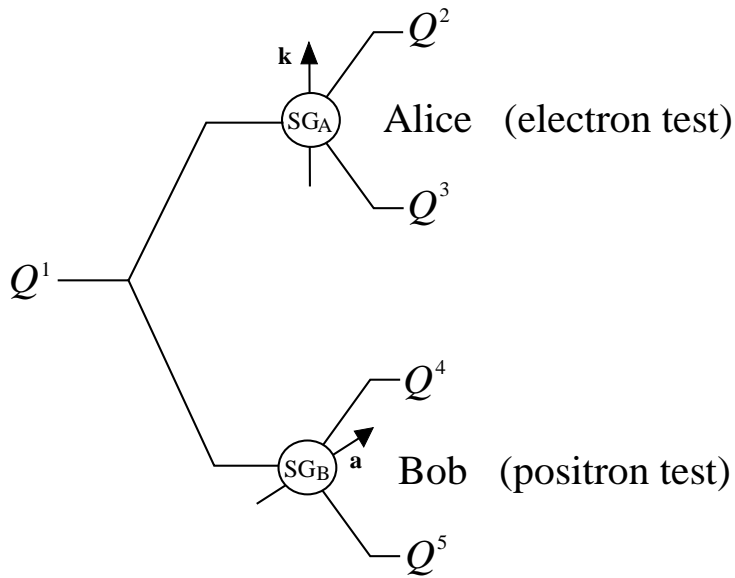}} \vspace*{8pt}
\caption{Change of signal rank experiment.}
\end{figure}

Assuming the initial state has total spin zero, the SQM representation of
the initial state is given by%
\begin{equation}
|\Psi\rangle=\frac{1}{\sqrt{2}}\left\{
\begin{array}{c}
\sin(\frac{_{1}}{^{2}}\theta)e^{-i\phi}|+\mathbf{k\rangle}_{-}\otimes |+%
\mathbf{a\rangle}_{+}+\cos(\frac{_{1}}{^{2}}\theta)e^{-i\phi}|+\mathbf{%
k\rangle}_{-}\otimes|-\mathbf{a\rangle}_{+} \\
-\cos(\frac{_{1}}{^{2}}\theta)|-\mathbf{k\rangle}_{-}\otimes|+\mathbf{%
a\rangle }_{+}+\sin(\frac{_{1}}{^{2}}\theta)|-\mathbf{k\rangle}_{-}\otimes |-%
\mathbf{a\rangle}_{+}%
\end{array}
\right\} ,
\end{equation}
where the subscript $(-)$ refers to the electron whilst $(+)$ refers to the
positron.

The QDN interpretation is inherently different to the SQM interpretation.
The initial labstate can be described as an entangled state of an electron
and a positron only if the apparatus after state preparation permits such
information to be extracted. According to our principles, positronium is
considered an unstable elementary particle if it is observed as such,
whereas if the apparatus can register electrons and positrons separately,
then positronium can be described as an entangled state of an electron and a
positron. This illustrates the point that entanglement is not an inherent
property of an SUO, but depends on the context of observation.

Figure $10$ shows the QDN qubit assignment for this experiment. In the QDN
description of the experiment, the preparation switch $\mathcal{Q}^{1}$
prepares an initial one-signal labstate $|\Psi ,0)\equiv \mathbb{A}%
_{1,0}^{+}|0,0)$. The subsequent dynamics is given by
\begin{align}
\mathbb{A}_{1,0}^{+}\rightarrow \mathbb{U}_{1,0}\mathbb{A}_{1,0}^{+}\mathbb{U%
}_{1,0}^{+}& \sim \frac{\sin (\frac{_{1}}{^{2}}\theta )e^{-i\phi }}{\sqrt{2}}%
\mathbb{A}_{2,1}^{+}\mathbb{A}_{4,1}^{+}+\frac{\cos (\frac{_{1}}{^{2}}\theta
)e^{-i\phi }}{\sqrt{2}}\mathbb{A}_{2,1}^{+}\mathbb{A}_{5,1}^{+}  \notag \\
& -\frac{\cos (\frac{_{1}}{^{2}}\theta )}{\sqrt{2}}\mathbb{A}_{3,1}^{+}%
\mathbb{A}_{4,1}^{+}+\frac{\sin (\frac{_{1}}{^{2}}\theta )}{\sqrt{2}}\mathbb{%
A}_{3,1}^{+}\mathbb{A}_{5,1}^{+},
\end{align}%
which means that the initial one-signal labstate evolves into an entangled
two-signal labstate.

\subsection{Two-particle interferometry}

\begin{figure}[h!]
\centerline{\includegraphics[width=4.0in]{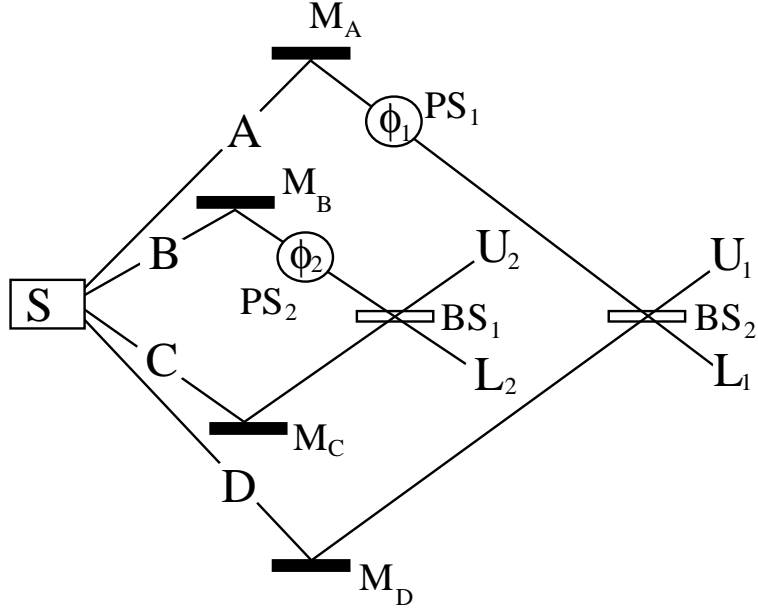}} \vspace*{8pt}
\caption{Experiment involving a two-photon source.}
\end{figure}

In $1989$, Horne, Shimony and Zeilinger discussed an experiment involving
two-photon interferometry \cite{HSZ-1989}. Their experimental is shown in
Figure $11$. A source $S$ prepares an entangled two-photon state. A set of
mirrors, $M_{A}$, $M_{B}$, $M_{C}$ and $M_{D}$ reflect various components of
the prepared state through phase-changers $PS_{1}$ and $PS_{2}$ and onto
beam-splitters $BS_{1}$ and $BS_{2}$. Finally, photon coincidence detectors $%
U_{1}$, $U_{2}$, $L_{1}$ and $L_{2}$ are placed as shown.

The initial state is described in terms of photon polarization vectors, and
is given in the form \cite{HSZ-1989}%
\begin{equation}
|\Psi_{in}\rangle=\frac{1}{\sqrt{2}}\left\{ |\mathbf{k}_{A}\rangle _{1}|%
\mathbf{k}_{C}\rangle_{2}+|\mathbf{k}_{D}\rangle_{1}|\mathbf{k}%
_{B}\rangle_{2}\right\} .
\end{equation}
A standard calculation then gives the coincidence probabilities%
\begin{equation}
\begin{array}{ccc}
P(U_{1},U_{2}|\phi_{1},\phi_{2})=P\left( L_{1},L_{2}|\phi_{1},\phi_{2}\right)
& = & \frac{1}{4}\left\{ 1+\cos(\phi_{2}-\phi_{1}+\theta)\right\} , \\
P(U_{1},L_{2}|\phi_{1},\phi_{2})=P\left( L_{1},U_{2}|\phi_{1},\phi_{2}\right)
& = & \frac{1}{4}\left\{ 1-\cos(\phi_{2}-\phi_{1}+\theta)\right\} ,%
\end{array}%
\end{equation}
where depends on the detailed relative placement of the mirrors, etc.

The QDN assignment of qubits is shown in Figure $12$. The preparation switch
prepares a labstate which evolves to the equivalent of an entangled
two-photon state, given by%
\begin{equation}
\mathbb{A}_{1}^{+}\rightarrow\frac{1}{\sqrt{2}}\left\{ \mathbb{A}_{2}^{+}%
\mathbb{A}_{4}^{+}+e^{i\theta}\mathbb{A}_{3}^{+}\mathbb{A}_{5}^{+}\right\} .
\end{equation}

\

\begin{figure}[h!]
\vspace*{10pt} \centerline{\includegraphics[width=4.0in]{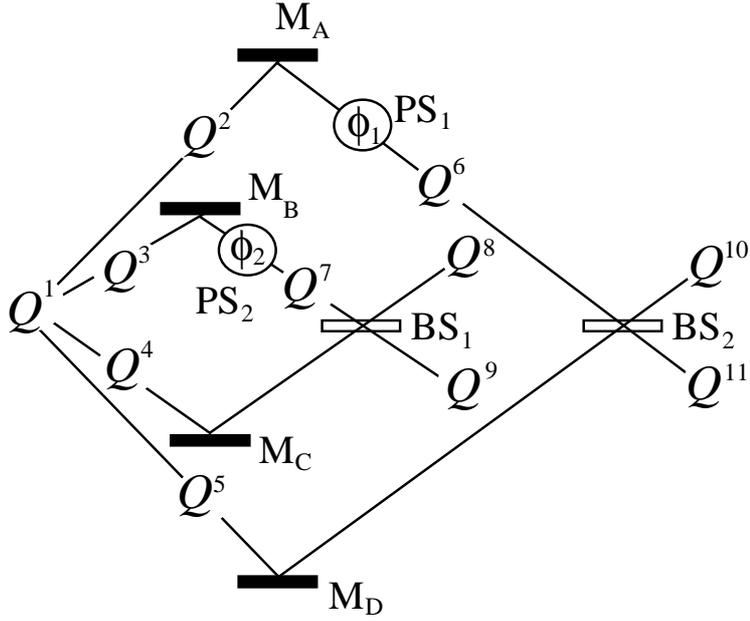}}
\caption{Qubit assignment for the experiment shown in Figure $11$.}
\end{figure}

Here the angle $\theta$ depends on the detailed placement of the various
pieces of equipment. Ignoring the phase changes at the mirrors, which are
the same and hence do not alter relative outcome probabilities, the various
modules have the following effects:

\noindent$i)$ The phase changers $PS_{1}$, $PS_{2}$ give%
\begin{equation}
\mathbb{A}_{2}^{+}\rightarrow e^{i\phi_{1}}\mathbb{A}_{6}^{+},\;\;\;\mathbb{A%
}_{3}^{+}\rightarrow e^{i\phi_{2}}\mathbb{A}_{7}^{+};
\end{equation}

\noindent$ii)$ the beam splitters $BS_{1}$, $BS_{2}$ are symmetric and give%
\begin{align}
\mathbb{A}_{7}^{+} & \rightarrow\frac{1}{\sqrt{2}}\left\{ \mathbb{A}%
_{9}^{+}+i\mathbb{A}_{8}^{+}\right\} ,\;\;\;\mathbb{A}_{4}^{+}\rightarrow
\frac{1}{\sqrt{2}}\left\{ \mathbb{A}_{8}^{+}+i\mathbb{A}_{9}^{+}\right\} ,
\notag \\
\mathbb{A}_{4}^{+} & \rightarrow\frac{1}{\sqrt{2}}\left\{ \mathbb{A}%
_{10}^{+}+i\mathbb{A}_{11}^{+}\right\} ,\;\;\;\mathbb{A}_{6}^{+}\rightarrow%
\frac{1}{\sqrt{2}}\left\{ \mathbb{A}_{11}^{+}+i\mathbb{A}_{10}^{+}\right\} .
\end{align}
Hence we arrive at the evolution rule%
\begin{align}
\mathbb{A}_{1}^{+} & \rightarrow\frac{1}{2\sqrt{2}}\left\{
e^{i\phi_{1}}-e^{i(\theta+\phi_{2})}\right\} \mathbb{A}_{8}^{+}\mathbb{A}%
_{11}^{+}+\frac{1}{2\sqrt{2}}\left\{
ie^{i\phi_{1}}+ie^{i(\theta+\phi_{2})}\right\} \mathbb{A}_{8}^{+}\mathbb{A}%
_{10}^{+}  \notag \\
& +\frac{1}{2\sqrt{2}}\left\{ ie^{i\phi_{1}}+ie^{i(\theta+\phi_{2})}\right\}
\mathbb{A}_{9}^{+}\mathbb{A}_{11}^{+}+\frac{1}{2\sqrt{2}}\left\{
-e^{i\phi_{1}}+e^{i(\theta+\phi_{2})}\right\} \mathbb{A}_{9}^{+}\mathbb{A}%
_{10}^{+}.
\end{align}

From this, we can immediately read off the various coincidence amplitudes,
and work out the coincidence probabilities, which are found to be%
\begin{equation}
\begin{array}{lcc}
P\left( 8\ \&\ 10|\Psi_{in}\right) =P\left( 9\ \&11|\Psi_{in}\right) & = &
\frac{1}{4}\left\{ 1+\cos\left( \phi_{2}-\phi_{1}+\theta\right) \right\} ,
\\
P\left( 8\ \&\ 11|\Psi_{in}\right) =P\left( 9\ \&\ 10|\Psi_{in}\right) & = &
\frac{1}{4}\left\{ 1-\cos\left( \phi_{2}-\phi_{1}+\theta\right) \right\} ,%
\end{array}%
\end{equation}
in agreement with the standard calculation.

We note that after passage through the mirrors, the labstate behaves as if
it were a super-labstate being analyzed by two separate laboratories, with
separate apparatus $BS_{1}$ and $BS_{2}$. The entanglement of the
super-labstate entanglement is then \textquotedblleft
undone\textquotedblright\ by the separate observations involved with these
beam-splitters, and these never show entanglement \emph{per se} in any
individual outcome.

\section{\label{Decays}Particle decays}

In this section we give a brief account of the QDN approach to particle
decays \cite{J2006M}. The approach extends naturally to describe the quantum
Zeno effect, the ammonium molecule and neutral Kaon decays.

We shall consider the quantum physics of what in SQM\ would be called an
unstable particle, the initial state $X$ of which can decay into some
multiparticle state $Y$. At all times total probability will be manifestly
conserved. The momenta of the particles will be ignored here, the discussion
being designed to illuminate the basic principles of the formalism only.

Typically, the sort of experiment of interest here can be repeated many
times, and the formalism gives the quantum description of an ensemble of
runs of a basic experiment. Clocks can always be reset, so a typical run of
the experiment may be taken to start at time $t=0$, at which time the
observer believes that they have prepared an $X$ state (to use the language
of SQM). In QDN, this is represented by the labstate $|\Psi ,0)\equiv
\mathbb{A}_{X,0}^{+}|0,0)$, which is automatically normalized to unity.

By time $1$, the labstate will have changed from $|\Psi ,0)$ to some new
labstate $|\Psi ,1)$ given by%
\begin{equation}
|\Psi ,1)=\alpha \mathbb{A}_{X,1}^{+}|0,1)+\beta \mathbb{A}%
_{Y_{1},1}^{+}|0,1),  \label{0111}
\end{equation}%
where the complex numbers $\alpha $ and $\beta $ satisfy the semi-unitarity
rule $|\alpha |^{2}+|\beta |^{2}=1$. The outcome possibilities of finding
the void state $|0,1)$ or the two-signal state $\mathbb{A}_{X,1}^{+}\mathbb{A%
}_{Y_{1},1}^{+}|0,1)$ are excluded on dynamical grounds: any run with either
of these outcomes would be discounted by the observer as contaminated by
external influences (as happens in real experiments). From (\ref{0111}), the
amplitude $\mathcal{A}(X,1|X,0)$ for the particle not to have decayed by
time $1$ is given by%
\begin{equation}
\mathcal{A}(X,1|X,0)\equiv (0,1|\mathbb{A}_{X,1}|\Psi ,1)=\alpha
\end{equation}%
whilst the amplitude $\mathcal{A}(Y,1|X,0)$ for the particle to have made
the transition to state $Y$ by time $1$ is given by%
\begin{equation}
\mathcal{A}(Y,1|X,0)\equiv (0,1|\mathbb{A}_{Y_{1},1}|\Psi ,1)=\beta .
\label{0666}
\end{equation}%
Total probability is therefore conserved. Note that on the right hand side
of (\ref{0666}), the label $Y$ is itself labeled by a subscript, in this
case the number $1$, which is the time at which the amplitude is calculated
for. The time at which a transition occurs is a crucial feature of the
analysis, being directly related to the measurement issues discussed by
Misra and Sudarshan \cite{MISRA+SUDARSHAN-1977}.

The above process conserves signal class, so the dynamics can be discussed
wholly in terms of the evolution of the signal operators rather than the
labstates. For instance, evolution from time $0$ to $1$ can be given in the
form%
\begin{equation}
\mathbb{A}_{X,0}^{+}\rightarrow \mathbb{U}_{1,0}\mathbb{A}_{X,0}^{+}\mathbb{U%
}_{1,0}^{+}=\alpha \mathbb{A}_{X,1}^{+}+\beta \mathbb{A}_{Y_{1},1}^{+},
\label{0888}
\end{equation}%
where $\mathbb{U}_{1,0}$ is a semi-unitary operator satisfying the rule $%
\mathbb{U}_{1,0}^{+}\mathbb{U}_{1,0}=\mathbb{I}_{0}$, with $\mathbb{I}_{0}$
being the identity for the initial register $\mathcal{R}_{0}\equiv \mathcal{Q%
}_{0}^{X}$. The above process involves a change in rank, since $\mathcal{R}%
_{1}\equiv \mathcal{Q}_{1}^{X}\mathcal{Q}_{1}^{Y_{1}}$. Because $r_{1}\equiv
\dim \mathcal{R}_{1}>r_{0}\equiv \dim \mathcal{R}_{0}$, semi-unitarity of
the evolution operator means that $\mathbb{U}_{1,0}\mathbb{U}_{1,0}^{+}\neq
\mathbb{I}_{1}$, which is equivalent to irreversibility in SQM.

The description of the next stage of the process, from time $1$ to time $2$,
is more subtle and involves the concept of \emph{null test} \cite{J2001A}.
In SQM, a null test is defined as any quantum test which extracts no
information from an initial state. Physically, this corresponds to passing
an outcome of a given apparatus through the same or equivalent apparatus,
the net effect being that the state remains unchanged. For example, an
electron emerging from a Stern-Gerlach apparatus $S_{0}$ in the spin-up
state would pass through another Stern-Gerlach apparatus $S_{1}$ unscathed
and still in its spin-up state, provided the magnetization axis of $S_{1}$
was in the same direction as that of $S_{0}$. In SQM, a null test is
modelled mathematically by the fact that an eigenstate of an operator is
also an eigenstate of the square of that operator.

In QDN, it is not the case that a null test involves no change whatsoever in
the observer's information of what is going on. The observer does have the
information that time has passed during the null test, and that fact is
registered in the observer's memory. Moreover, in QDN, a labstate \emph{%
always changes in time}, because the quantum register it is in changes with
time. What is relevant is the set of components of a labstate, relative to
the signal state basis at any given time. It is those components which are
related to outcome probabilities. If those components do not change, then
the observer may speak about the labstate as being constant in time, but the
observer will also have an awareness that the state is evolving in time as
well. In other words, the passage of time involves the observer as much as
it involves the labstate.

Considering the labstate of the above decay process at time $1$, there are
now two terms to consider. The first term in (\ref{0888}), $\alpha \mathbb{A}%
_{X,1}^{+}$, corresponding to a \emph{no decay} outcome by time $1$, can be
regarded as preparing at time $1$ an initial $X$ state which could
subsequently decay into a $Y$ state or not, with the same dynamical
characteristics as for the first stage of the run, i.e., between times $0$
and $1$. This assumes spatial and temporal homogeneity, a physically
reasonable assumption in the absence of gravitational fields and in the
presence of suitable apparatus. The second term, $\beta \mathbb{A}%
_{Y_{1},1}^{+}$, corresponds to \emph{decay having occurred during the first
time interval}. Such an outcome is regarded as irreversible in this example,
but this is not an inevitable assumption in general. Situations where the $Y$
state could revert back to the $X$ state are more complicated but of
empirical interest, such as in the ammonium maser and Kaon decay, discussed
elsewhere \cite{J2006M}.

Assuming homogeneity, the next stage of the evolution is given by%
\begin{eqnarray}
\mathbb{A}_{X,1}^{+} &\rightarrow &\mathbb{U}_{2,1}\mathbb{A}_{X,1}^{+}%
\mathbb{U}_{2,1}^{+}=\alpha \mathbb{A}_{X,2}^{+}+\beta \mathbb{A}%
_{Y_{2},2}^{+},  \notag \\
\mathbb{A}_{Y_{1},1}^{+} &\rightarrow &\mathbb{U}_{2,1}\mathbb{A}%
_{Y_{1},1}^{+}\mathbb{U}_{2,1}^{+}=\mathbb{A}_{Y_{1},2}^{+}.  \label{0222}
\end{eqnarray}%
The second equation is justified as follows. The decay term in (\ref{0888}),
proportional to $\mathbb{A}_{Y_{1},1}^{+}$ at time $1,$ corresponds to the
possibility of detecting a decay product state $Y$ at that time. Now there
is nothing which requires this information to be extracted precisely at that
time. The experimentalist could choose to delay information extraction until
some later time, effectively placing the decay product observation $``$on
hold \textquotedblright. As stated above, this may be represented in SQM\ by
passing a state through a null-test, which does not alter it. In QDN\ this
is represented by the second equation in (\ref{0222}). Essentially, quantum
information about a decay is passed forwards in time until it is physically
extracted.

The register $\mathcal{R}_{2}$ at time $2$ has rank three, being the tensor
product $\mathcal{R}_{2}=\mathcal{Q}_{2}^{X}\mathcal{Q}_{2}^{Y_{1}}\mathcal{Q%
}_{2}^{Y_{2}}$. Semi-unitary evolution from time zero to time $2$ therefore
gives%
\begin{equation}
\mathbb{A}_{X,0}^{+}\rightarrow \mathbb{U}_{2,1}\mathbb{U}_{1,0}\mathbb{A}%
_{X,0}^{+}\mathbb{U}_{1,0}^{+}\mathbb{U}_{2,1}^{+}=\alpha ^{2}\mathbb{A}%
_{X,2}^{+}+\alpha \beta \mathbb{A}_{Y_{2},2}^{+}+\beta \mathbb{A}%
_{Y_{1},2}^{+},  \label{123}
\end{equation}%
with the various probabilities being read off as the squared moduli of the
corresponding terms.

It will be apparent from a close inspection of (\ref{123}) that what appears
to look like a space-time description with a specific arrow of time is being
built up, with a memory of the change of rank of the QDN register at time $1$
being propagated forwards in time to time $2.$ This is represented by the
contribution involving $\mathbb{A}_{Y_{1},2}^{+},$ which is interpreted as a
potential decay process which may have occurred by time $1$, contributing to
the overall labstate amplitude at time $2$.

Subsequently the process continues in an analogous fashion, with the rank of
the register increasing by one at each timestep. By time $n$ the dynamics
gives%
\begin{equation}
\mathbb{A}_{X,0}^{+}\rightarrow \mathbb{U}_{n,0}\mathbb{A}_{X,0}^{+}\mathbb{U%
}_{n,0}^{+}=\alpha ^{n}\mathbb{A}_{X,n}^{+}+\beta \sum_{k=1}^{n}\alpha ^{k-1}%
\mathbb{A}_{Y_{k},n}^{+},
\end{equation}%
where $\mathbb{U}_{n,0}\equiv \mathbb{U}_{n,n-1}\mathbb{U}_{n-1,n-2}\ldots
\mathbb{U}_{1,0}$ is semi-unitary and satisfies the constraint $\mathbb{U}%
_{n,0}^{+}\mathbb{U}_{n,0}=\mathbb{I}_{0}$. From the above, the survival
probability $\Pr (X,n|X,0)$ that the original state has \emph{not} decayed
can be immediately read off and is found to be%
\begin{equation}
\Pr (X,n|X,0)=|\alpha |^{2n}.
\end{equation}
Provided $\beta \neq 0$, this probability falls monotonically with
increasing $n$, corresponding to particle decay.

The discussion at this point calls for some care with limits, because there
arises the theoretical possibility of encountering the quantum Zeno effect
\cite{MISRA+SUDARSHAN-1977,ITANO-1990}. In the following, it will be assumed
that $|\alpha |<1$, because $|\alpha |\ =1$ corresponds to a stable
particle, which is of no interest here.

Consider the physics of the situation. The calculated probabilities should
relate to the measured time $t$ as used by the observer in the laboratory.
The observer's time of observation $t$ has not been assumed to be a
continuous variable. The temporal label $n$ denoting the time of observation
corresponds to a physical time $t\equiv n\tau $, where $\tau $ is some
reasonably well-defined time characteristic of the apparatus. In the sort of
experiments relevant here, $\tau $ will be on a minute fraction of a second
scale, but certainly nowhere near Planck time scales. The smallest interval
that could be achieved in practice would be of the order $10^{-23}$ second,
which is on the shortest hadronic resonance scale, comparable with the time
light takes to cross a proton diameter. More realistic measurement scales,
involving electromagnetic processes, would be in the $10^{-9}-10^{-18}$
second range. Experimentalists would generally have a good understanding of
what $\tau $ was.

At first sight, we might have reason to believe that we can relate the
transition amplitude $\alpha $ to the characteristic time $\tau $ by the
rule
\begin{equation}
|\alpha |^{2}\equiv e^{-\Gamma \tau },  \label{0999}
\end{equation}%
where $\Gamma $ is a characteristic inverse time introduced to satisfy this
relation. Then the survival probability $P\left( t_{n}\right) $ would be
given by
\begin{equation}
P\left( t\right) \equiv \Pr (X,n|X,0)=e^{-\Gamma t},
\end{equation}%
which is the usual exponential decay formula. No imaginary term proportional
to $\Gamma $ in any supposed Hamiltonian or energy has been introduced in
order to obtain exponential decay.

A subtlety may arise here however. Expression (\ref{0999}) assumes that $%
|\alpha |^{2}$ is an analytic function of $\tau $ with a Taylor expansion of
the form%
\begin{equation}
|\alpha |^{2}=1-\Gamma \tau +O\left( \tau \right) ^{2},  \label{0552}
\end{equation}%
i.e., one with a non-zero linear term. Under such circumstances, the
standard result $\lim_{n\rightarrow \infty }(1-\frac{x}{n})^{n}=e^{-x}$
leads to the exponential decay law. The possibility remains, however, that
the dynamics is such that the linear term in (\ref{0552}) is zero, so that
the actual expansion is of the form%
\begin{equation}
|\alpha |^{2}=1-\gamma \tau ^{2}+O\left( \tau ^{3}\right) ,  \label{0551}
\end{equation}%
where $\gamma $ is a positive constant \cite{ITANO-1990}. Then in the limit $%
n\rightarrow \infty $, where $n\tau \equiv t$ is held fixed, the result is
given by%
\begin{equation}
\lim_{n\rightarrow \infty ,n\tau =t\ \text{fixed}}\left( 1-\gamma \tau
^{2}+O\left( \tau ^{3}\right) \right) ^{n}=1,
\end{equation}%
which gives rise to the quantum Zeno effect scenario. An expansion of the
amplitude of the form%
\begin{equation}
a=1+i\mu \tau +\nu \tau ^{2}+O(\tau ^{3})
\end{equation}%
is consistent with (\ref{0551}) for example, if $\mu $ is real and $\mu
^{2}+\nu +\nu ^{\ast }<0$.

To understand properly what is going on, it is necessary to appreciate that
there are two competing limits being considered: one where a system is being
repeatedly observed over an increasingly large macroscopic laboratory time
scale $t$, and another one where more and more observations are being taken
in succession, each separated on a time scale $\tau $ which is being brought
as close to zero as possible by the experimentalist. In each case, the limit
cannot be achieved in the laboratory. The result is that in such
experiments, the apparatus may play a decisive role in determining the
results. If the apparatus is such that (\ref{0552}) holds, then exponential
decay will be observed. On the other hand, if the apparatus behaves
according to the rule (\ref{0551}), or any reasonable variant of it, then
approximations to the quantum Zeno effect should be observed.

\section{QDN and Relativity}

Relativity and quantum mechanics are an explosive mix. Both of these
frameworks were developed around the start of the Twentieth Century and both
have been fully vindicated in their respective domains of applicability.
Currently, none of their core principles have been invalidated
experimentally. The problem is, they present radically different views of
physical reality.

On the one hand, relativity in both its special (SR) and general (GR) forms
is thoroughly based on a classical world view, in which well-defined SUOs
follow timelike trajectories in a four-dimensional spacetime continuum
endowed with a Lorentz-signature pseudo-Riemannian metric. GR goes so far as
to treat spacetime itself as a form of SUO, with its own dynamical rules and
evolution.

Observers appear to have more status in SR compared with GR. In SR, they are
generally associated with specific inertial frames, an idea which models
ordinary experience well. Physics experiments are almost always conducted in
laboratories which are tied to some local inertial frame (LIF). In such
frames, over limited intervals of time and space, all the laws of SR appear
to hold. Not all experiments involve just one LIF. For example, Doppler
shifts involve two such frames. We shall discuss the QDN approach to such
\emph{inter-frame} physics presently.

Although SR \emph{per se} takes no account of direct observer-SUO
interaction, it does impose some conditions on signal detection protocols,
which is of importance to us here. We have already mentioned that physical
signals from a source cannot be detected outside the forwards lightcone with
vertex at the source. Another condition is that whenever multiple signals
are received simultaneously (according to whatever consistent definition of
time the observer is using), then these are received over some space-like
hypersurface of Minkowski spacetime\footnote{%
We are justified in using the physical space concept here because we are
describing observers and their apparatus. In that context, it remains an
effective modelling tool.}. For those observers who regard themselves as at
rest in a given inertial frame, these space-like hypersurfaces are
space-like hyperplanes, in standard coordinates, labelled by the observer's
clock time (which also happens to be their coordinate time).

In GR, however, the status of observers is relegated somewhat, to the extent
that GR appears in places to eliminate the need for them completely. Indeed,
a core strategy in GR is to find a description of spacetime and SUOs which
is as independent of classical observers as possible. Traditionally, this
strategy has been deemed so important that it has been elevated to a
principle of physics and given the name \emph{principle of general covariance%
}, or coordinate frame independence of the laws of physics.

A particular problem in GR is that the two signalling protocols respected in
SR, i.e., that physical signals never propagate outside forwards lightcones
and that \textquotedblleft simultaneous\textquotedblright\ observations are
on spacelike hypersurfaces, can now run into difficulties. Some GR
spacetimes such as that of G\"{o}del \cite{GODEL-1949} contained closed
timelike curves (CTCs). For such spacetimes, not only is there no
possibility of a global foliation consisting of spacelike hypersurfaces
indexed by a time-like parameter (i.e., no universal observer), but there is
no consistent forwards direction for the irreversible acquisition of
information either. The Born probability interpretation of the wave-function
seems impossible to maintain in such cases\footnote{%
A way out of these problems is to assert that, for GR, physical observers
exist only over limited regions of spacetime which must never contain CTCs
and can have a local timelike foliation. This is consistent with QDN, which
views observers as transient structures.}.

On the other hand, QM cannot ignore observers or their apparatus. That has
been the central lesson taught by countless experiments, where the
principles of QM\ have been fully vindicated. Throughout this review, we
have put the case for the validity of Heisenberg's views about physical
reality. If these views are universally valid, then the inevitable
conclusion is that the principle of general covariance in the form given in
GR is too naive and should not be applied without a radical overhaul of its
meaning and the way it is applied to physics. We now discuss the QDN
approach to SR, which indicates in which direction such an overhaul might be
found.

\subsection{Lorentz transformations}

The principle of relativity states that the laws of physics are the same in
all standard inertial frames of reference, if gravitational effects are
excluded. This principle was used by Einstein to derive the Lorentz
transformation
\begin{equation}
t^{\prime }=\gamma (v)\left( t-vx/c^{2}\right) ,\ x^{\prime }=\gamma
(v)\left( x-vt\right) ,\ y^{\prime }=y,\ z^{\prime }=z,  \label{1118}
\end{equation}
$\gamma (v)=1/\sqrt{1-v^{2}/c^{2}}$, between two standard inertial frames $%
\mathcal{F}$, $\mathcal{F}^{\prime }$ in standard configuration, moving
apart with relative speed $v$ along the common $x$-direction.

A notable feature of the Lorentz transformation is the implied loss of
absolute simultaneity: a hyperplane of simultaneity $t^{\prime }=const$ in $%
\mathcal{F}^{\prime }$ is not a hyperplane of simultaneity in $\mathcal{F}$
and vice-versa. We now investigate the consequences of incorporating some
basic principles of quantum physics into the above. These principles were
not known before Heisenberg's formulation of quantum mechanics in $1925$
\cite{HEISENBERG-1925}, so it should not be surprising to find that they
modify the classical interpretation of (\ref{1118}).

In contrast to most approaches to relativity, we focus our attention not so
much on the frames of reference themselves (which can be thought of as
foliations of Minkowski spacetime with leaves consisting of hyperplanes of
simultaneity), but on two very specific hyperplanes of simultaneity, one in
each frame. We shall ignore the $y$ and $z$ coordinates because although
they cannot be ignored in the real world, they do not add any new insights
to the discussion here. Our discussion will be valid over limited regions of
spacetime where general relativistic effects can be ignored.

In classical relativity, it is generally assumed that signals can be sent
from one event to another, proved the latter is not outside the forwards
lightcone of the former. Although in reality any ESD involved in such
signalling has finite extent, i.e. is non-local, the scales involved in this
non-locality can be reduced to such levels relative to the distances
travelled by the signals that ESDs will be assumed to be localized, i.e.,
virtually pointlike.

The sort of experiments we discuss involve several such signals sent from
one physical apparatus consisting of two or more ESDs operating as sources,
and detected by another apparatus consisting of two or more ESDs operating
as detectors. Whenever the signalling ESDs and the detecting ESDs are not at
rest in a common inertial frame of reference (or not at rest in frames
related by a simple translation or spatial rotation), we shall refer to such
an experiment as an \emph{inter-frame experiment}.

Consider an inter-frame experiment conducted over a finite interval of time,
such that signals sent from ESDs $Q$ \& $P,$ at rest in frame $\mathcal{F}$
at time $t=0$, are observed by ESDs apparatus $Q^{\prime }P^{\prime }$ at
rest in frame $\mathcal{F}^{\prime }$ at time $T^{\prime }$, as measured in
that frame. Figure $13$ shows the essential details.

\begin{figure}[t]
\centerline{\includegraphics[width=5.0in]{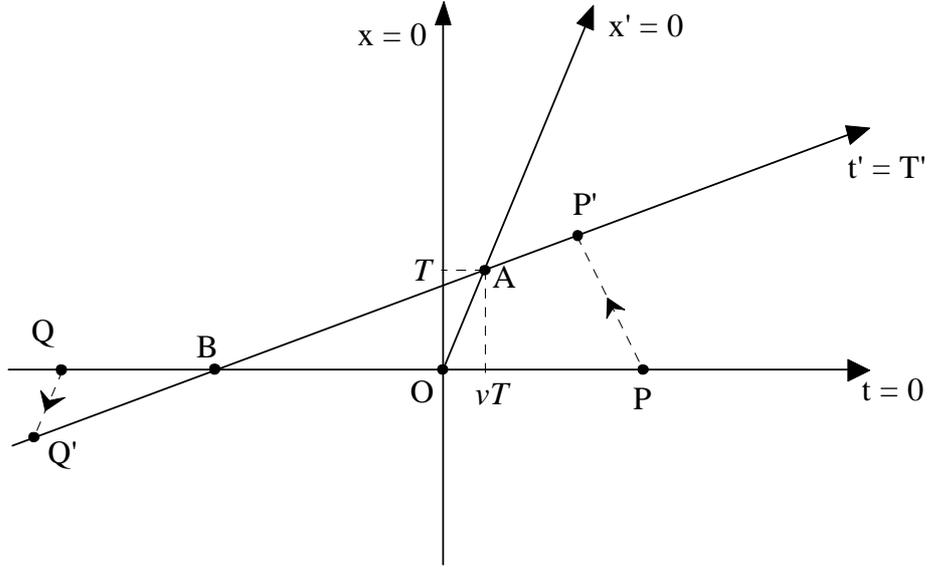}}
\caption{$B$ is the quantum horizon. }
\end{figure}

In Figure $13$, event $O$ is the common origin of spacetime coordinates
consistent with (\ref{1118}), events $P$ and $Q$ are simultaneous in $%
\mathcal{F}$ at time $t=0$ whilst events $P^{\prime }$and $Q^{\prime }$ are
simultaneous in $\mathcal{F}^{\prime }$ at time $t^{\prime }=T^{\prime }>0$.
We shall use the convention that an event $P$ has coordinates $(t_{P},x_{p})$
in $\mathcal{F}$, coordinates $\left[ t_{P}^{\prime },x_{P}^{\prime }\right]
$ in $\mathcal{F}^{\prime }$ and we write $P\sim (t_{P},x_{P})\sim \lbrack
t_{P}^{\prime },x_{p}^{\prime }]$. From Figure $13$, it will be seen that a
critical feature is event $B$, where the lines of simultaneity $t=0$ and $%
t^{\prime }=T^{\prime }$ intersect. Assuming the velocity component $v$ is
positive, then for $B$ we have
\begin{equation}
B\sim (0,-\frac{c^{2}T^{\prime }}{\gamma (v)v})\sim \lbrack T^{\prime },-%
\frac{c^{2}T^{\prime }}{v}].
\end{equation}

Event $B$ is the focus of attention in this article. A novel and interesting
interpretation of the significance of event $B$ can be found from basic
quantum mechanics. First, we recall that in de Broglie wave mechanics \cite%
{DEBROGLIE:1924}, the speed $w$ of a pilot wave associated with any material
physical particle moving with subluminal speed $v$ satisfies the relation $%
vw=c^{2}$. This suggests that such a pilot wave cannot be used to convey
physical signals, because it travels at superluminal speed, given $v<c$.
According to these ideas, event $B$ may be regarded by observers in $%
\mathcal{F}^{\prime }$ as the wave front at time $T^{\prime }$ of such a
pilot wave associated with a particle at rest in frame $\mathcal{F}$, if it
were sent out from $O$ in the same direction as $\mathcal{F}$ appears to
move. Note that this interpretation of event $B$ should be regarded as no
more than a mathematical curiosity, because any genuine de Broglie wave
would not be localized at a single point. Moreover, the significance of
event $O$ as the common origin of coordinates is an artefact do with the
choice of coordinates. Nevertheless, we shall argue below that $B$ does have
something critical to do with quantum processes.

We should ask why events such as $B$ do not appear in conventional physics.
There are several circumstances which normally conspire to mask the presence
of such events: $i)$ the speed of light is large on ordinary laboratory
scales, $ii)$ the time $T^{\prime }$ of observation is usually large and $%
iii)$ the relative speed $|v|$ is usually very small or even zero. In
consequence, $B$ is usually either at relatively large distance from the
origin of spatial coordinates $A$ in frame $\mathcal{F}^{\prime }$ or even
at spatial infinity.

In standard discussions of special relativity, therefore, event $B$ is
generally ignored, as it appears to be far removed from events $P$ and $%
P^{\prime }$ involved in the signalling experiment. In our case, matters are
different, because quantum mechanics is inherently non-local; even in the
case of single particle states, normalizable wave-functions have to have
spatial extent. This means that we must expect elements of non-locality in
both preparation and detection devices.

For this experiment, we imagine that a quantum state has been prepared by
apparatus $\mathcal{A}_{QP},$ at rest in frame $\mathcal{F}$, and a
contingent quantum outcome subsequently detected by apparatus $\mathcal{A}%
_{Q^{\prime }P^{\prime }}^{\prime }$, at rest in frame $\mathcal{F}^{\prime
} $. The critical word here is \textquotedblleft \emph{subsequently%
\textquotedblright }. Quantum physics, as it is performed in real
laboratories, can discuss only the possibility of quantum information
travelling forwards in time. Indeed, relativity itself insists that signals
cannot travel outside forwards lightcones. Therefore, according to both SR
and QM\ principles, both signal emitter and signal detector in any quantum
experiment must agree that the former acts before the latter. Otherwise, the
physical significance of the Born probability rule would be completely
undermined. In any quantum experiment, we cannot know the outcome of any run
before it is performed (except in the case of a null experiment, which
extracts no information). We shall call the requirement that $P$ is earlier
than $P^{\prime }$ in both frames of reference \emph{quantum causality.}

From Figure $13$, it is clear that there is no problem with quantum
causality as far as events $P$ and $P^{\prime }$ are concerned. But consider
events $Q$ and $Q^{\prime }$ on the other side of $B$. If quantum causality
is valid, then a signal prepared at $Q$ cannot be received by $Q^{\prime }$.
In essence, event $B$ acts a barrier to quantum causality, and on this
account we shall refer to $B$ as a \emph{quantum horizon. }

As we have just stated, such quantum horizons are ignored in conventional
physics, because under most circumstances, $B$ appears to be very far from
events such as $P$ and $P^{\prime }$. For instance, high energy particle
theory traditionally works with initial and final inertial frames which are
coincident, and initial and final scattering times are taken to be in the
remote past and remote future respectively. This means taking the scattering
limit $T^{\prime }$ $\rightarrow $ $\infty $, $v=0$ in the calculation of
Lorentz covariant matrix elements. Finite-time processes and inter-frame
experiments of the sort considered by us here are generally avoided, because
it is assumed there is no significant novel physics involved. An important
factor contributing to this train of thought is that the scattering limit
makes calculations based on Feynman diagrams relatively straightforward.
Such a simplification does not happen for finite-time and inter-frame
processes. A well-motivated approach to finite time, localized quantum field
theory was developed by Schwinger \cite{SCHWINGER:1969}, but it remains an
exception and most approaches to quantum field theory tend to avoid the
topic.

We now consider the implications of the relativity principle and ask the
following question: if according to the relativity principle frames $%
\mathcal{F}$ and $\mathcal{F}^{\prime }$ are \textquotedblleft just as good
as each other\textquotedblright , why does the quantum horizon $B$ appear to
distinguish between the two?

A little thought soon resolves the question. If the relativity principle is
valid, then there must be a symmetry between the two frames. There is no
doubt that a quantum signal can be prepared at $P$ and received at $%
P^{\prime }$, if $P^{\prime }$ is in or on the forwards lightcone with
vertex $P.$ Quantum causality rules out the transmission of a quantum signal
from $P^{\prime }$ to $P$, and the transmission of a signal from $Q$ to $%
Q^{\prime }$. But no principle forbids the possibility of a physical signal
being sent from $Q^{\prime }$ to $Q$, if $Q$ is in the forwards lightcone
with vertex $Q^{\prime }$. Indeed, symmetry demands such a possibility.

It is convenient at this point to set the origin of spatial and temporal
coordinates in both frames $\mathcal{F}$ and $\mathcal{F}^{\prime }$ at the
quantum horizon, so that now $Q$ and $P$ are simultaneous in $\mathcal{F}$
at time $t=0$ and $Q^{\prime }$ and $P^{\prime }$ are simultaneous in $%
\mathcal{F}^{\prime }$ at time $t^{\prime }=0$, as shown in Figure $14$. The
event horizon is now labelled $O$.

\begin{figure}[t]
\centerline{\includegraphics[width=5.0in]{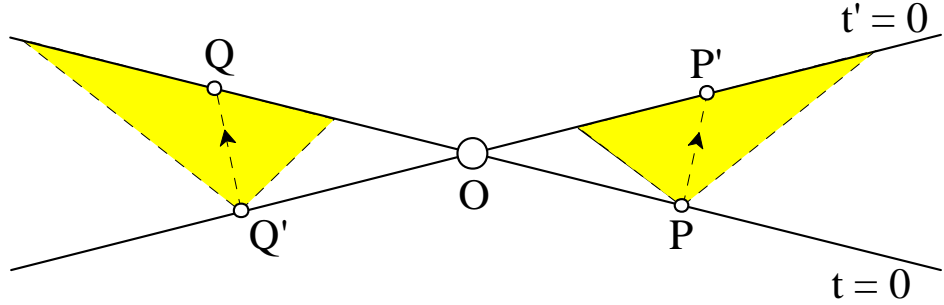}}
\caption{$O$ is the quantum horizon. Shaded regions are forwards lightcones.
Signals from $P$ to $P^{\prime}$ and from $Q^{\prime}$ to $Q$ are shown with
subluminal transmission speeds; this does not alter the overall conclusions.}
\end{figure}

It is clear from this diagram that the Lorentz transformation (\ref{1118})
leaves out much important information about how information can pass between
the two frames, particularly in the case of genuine inter-frame experiments
of the sort discussed here. Indeed, a much better physical description would
be to regard $P$ and $Q^{\prime }$ as ESDs belonging to an initial
Heisenberg net which sends signals to a subsequent Heisenberg net containing
ESDs $P^{\prime }$ and $Q$. For each of these nets, their ESDs are scattered
over some spacelike hypersurface in Minkowski spacetime, but neither
hypersurface is associated with a single inertial frame. Note that ESDs $P$
and $Q$ are regarded as at rest in the same inertial frame, and similarly
for $P^{\prime}$ and $Q^{\prime}$; there is therefore significant contextual
information held by the observer which is not indicated in Figures $13$ and $%
14$ and cannot be ignored.

Some work has already been done on some issues related to entangled states
and quantum horizons in such inter-frame experiments \cite{J2006P}. Our
conclusions support the QDN view that it is not physically meaningful to
talk about the preparation of quantum states, entangled or not, without
reference to any context of subsequent observation. It is the choice of test
apparatus which determines whether a state of an SUO should be regarded as
entangled or not. For instance, in any discussion involving quantum
information and black hole physics, problems will inevitably arise whenever
quantum states are discussed without due regard for the equipment used to
test them. For this reason, all discussions of quantum mechanics across
event horizons or wavefunctions for the universe without due reference to
observers and their equipment should be avoided. As we have shown above, it
is insufficient even in SR to talk about Lorentz transformations without any
reference to the apparatus involved in inter-frame experiments. When these
are taken properly into account, the equations of special relativity have to
be interpreted much more carefully and according to quantum principles.

\

\ \

\begin{center}
\textbf{\Large PART\ IV: Commentary}
\end{center}

\

It is our belief that the above examples demonstrate the viability of QDN as
an alternative approach to QM. However, much remains to be done to establish
further our intuition that QDN should be capable of doing everything that
SQM can do and perhaps more. We are currently working our way through
various other quantum optics experiments to confirm the viability and useful
of our approach in that area. It is clear that at the one-photon level, the
formalism gives an economical method of dealing with quite large scale
networks. There may be problems with more complicated two-photon networks
because the $X$ term in expression (\ref{econ}) vanishes only for sure when
it is applied to the void state. However, a computer algebra approach would
be able to deal with the complexities which arise on that score.

The QDN\ approach holds some promise of having applicability to other areas
of quantum physics apart from quantum optics. We shall report developments
in those area elsewhere. There are a number of points about QDN which it is
appropriate to comment on here.

\section{Hamiltonians and Lagrangians}

SQM is generally characterized by the specification of a Hamiltonian
operator, from which the dynamical evolution of state vectors can be
determined. Frequently, such an operator is obtained from a classical model
based on standard classical mechanical principles.

It will be evident from our review that there has been no mention of
Hamiltonians or Lagrangians, so it looks as if that aspect of SQM has
nothing in common with QDN. This would be a misleading deduction, however.
Under those circumstances where there were many signal degrees of freedom in
an experiment, and with an appropriate discussion of how discrete time could
be represented in terms of a continuum approximation, we expect QDN can be
made to look more like SQM, particularly in those cases where the rank of
the Heisenberg net remains constant. Our discussion of path summations
suggests that QDN should look like the conventional Feynman path integral
formulation of SQM in the appropriate limits.

Hamiltonians and Lagrangians are just convenient ways of encoding acquired
contextual information about a dynamical system, i.e., the previously
discovered rules of the dynamics. The fact is that what matters in both SQM
and QDN are the transition amplitudes. Once these are known, then either
formalism should give good results. Where they differ is in how those
amplitudes are obtained. SQM has certain standard procedures for obtaining
the transition amplitudes from a given Hamiltonian or Lagrangian, such as
evaluating path integrals or solving partial differential equations.

We regard Hamiltonians and the SQM formalism associated with them as useful
weapons in the description of vast parts of the quantum universe, but there
is no theorem which says that these are all the weapons we need. There might
not be such a thing as a Hamiltonian or Lagrangian for the universe, the
finding of which has been the dream of particle physicists for decades. What
leads to our worry about such a concept is that in practice (i.e., as they
are actually used), Hamiltonians are not absolute, but contextual. Each
model requires its own Hamiltonian, and changes in the apparatus generally
impose changes in that Hamiltonian, such as happens when external
electromagnetic fields are introduced. The QDN view is that Hamiltonians
arise only as and when the circumstances of an experiment dictate. In other
words, a Hamiltonian without any associated concept of observation is a
metaphysical concept with no physical value.

With the current state of development of QDN dynamics, we may need on
occasion to use results from SQM in order to find correct expressions for
semi-unitary operators in QDN. This should not be regarded as much different
in principle to the use of classical mechanical Hamiltonians as templates
for Hamiltonian operators SQM. The important point about our formalism is
that it gives a different conceptual basis to the meaning of the
calculations.

\section{Quantum counterfactuality}

Counterfactuals are true statements in classical logic, such as
\textquotedblleft $P$ implies $Q$\textquotedblright, made under the
circumstance that it is known that the premise $P$ is in fact false.
Counterfactuals are widely used in ordinary life and form the basis of the
classical world view, but they should be treated with great care or avoided
when in comes to discussions of quantum processes. In the classical world
view, things which did not, or could not, happen are sometimes assigned
significance and truth values which they do not merit. For example, the
classical world view would circumvent Heisenberg's uncertainty principle
with an argument such as the following: \textquotedblleft We have just
measured the position of this particle with absolute precision. However, if
we had chosen instead to measure its momentum, we would have established
that dynamical variable with absolute precision. Therefore, this proves that
a particle can have precise position and precise momentum
simultaneously.\textquotedblright

We call the principle that \textquotedblleft \emph{if something has not been
observed, then it is irrelevant}\textquotedblright\ the Heisenberg-Peres
principle, or \emph{quantum counterfactuality}. QDN adheres to quantum
counterfactuality by a strict adherence to the basic principle that only
what the observer knows and does with their apparatus has significance. This
has bearing on the old debate between the block universe model of spacetime
and the process time view. QDN is committed to the latter. Real observers
exist as one-off processes in the physical universe, and their experiments
are run serially in process time. Given unique information, there is only
one preferred basis associated with any given apparatus, and it is not
permitted to discuss other bases as if they had physical significance
without explaining carefully what this might mean. This requires
acknowledging the use of counterfactuals, and brings us to the next point,
the issue of symmetries.

\section{Quantum symmetries}

SQM makes great use of symmetry arguments, involving such symmetries as
rotational and translational invariance, as if the associated
transformations could be done at any time. Such arguments are usually
vindicated retrospectively by the results obtained in the laboratory, but it
has to be pointed out that in general, there are invariably a number of
hidden assumptions left unstated in such discussions. These invariably clash
with what actually happens in the laboratory. Any given individual run of an
experiment involves only one realization of the apparatus, and generally,
observers cannot change their apparatus in the middle of a run without a
great deal of physical consequences. For example, in a Stern-Gerlach
experiment, the main magnetic field has to lie in one direction only. If the
outcomes of other runs with the magnetization axis lying along different
directions are related by symmetries, then clearly what is involved are
comparisons of outcomes of different apparatus at different times and
places. This will in general involve the use of counterfactuals in a way
consistent with quantum principles. We do not have the space to comment on
this point further, save to say that it is our belief that QDN principles
provide a sound basis for such discussions.

\section{Final comments}

Our experience with the formalism described here leads us to have some
confidence that it has more useful things to say about quantum physics than
discussed here. It gives a different perspective on the significance of
quantum amplitudes and allows for a wider form of discussion than is usual
in SQM. We have had no opportunity here to discuss some important topics
such as mixed states or what a physically correct approach to
\textquotedblleft quantum gravity\textquotedblright\ might involve, save to
say that in both cases, we envisage a true novelty: uncertainty about the
actual apparatus and not just about its labstates \cite{J2005B}. What this
means is that at a given time, an observer might not know for sure what sort
of Heisenberg net they would be using in the future, and could give only
probability estimates for a range of possibilities. It is our intuition that
the real world, in which observers and apparatus themselves are created and
destroyed by dynamical processes, might be modelled one day by some enhanced
version of quantized dynamical networks.

\section*{Acknowledgements}

I am indebted to a number of people who were of value and assistance to me
during the time this work was developing. I warmly thank Lino Buccheri,
Metod Saniga, Mark Stuckey and Vito DiGes\`{u}, all of whom I first met in
Palermo in 1999 and subsequently became colleagues in the study of Time. I
am particularly grateful to Michel Planat of Besan\c{c}on, who was of great
assistance at a critical point. Without his help this article would never
have been written. My former students Jon Eakins and Jason Ridgway-Taylor
were constant in their support and were excellent collaborators. Most of
all, I am greatly indebted to Dr. K. K. Phua of World Scientific for
inviting me to write this review.

\end{document}